\documentclass[journal]{IEEEtran}
\usepackage{array}
\usepackage{url}
\usepackage{cite,graphicx,multirow,multicol,setspace,amsmath,wrapfig}%,subcaption}
\usepackage{tabulary,amssymb,amsfonts,adjustbox,booktabs,qtree,tikz-qtree,synttree,wrapfig}
\usepackage[flushleft]{threeparttable}
\usepackage{makecell,eqnarray}
\usepackage{cuted,sidecap}
\usepackage[linesnumbered,ruled]{algorithm2e}
\usepackage{algorithmic,textcomp,xcolor}
\usepackage{pifont}
\usepackage{commath}
\usepackage{rotating,forest}
\usepackage[switch]{lineno}
\usepackage{slashbox}
\usepackage{placeins}
\usepackage{amsthm}
\newtheorem{remark}{Remark}
\usepackage{tree-dvips}
\usepackage{tabularx}
\usepackage{qtree}
\usepackage{enumitem}
\graphicspath{ {Images/} }

\makeatletter
\let\MYcaption\@makecaption
\makeatother

\usepackage[font=footnotesize]{subcaption}

\makeatletter
\let\@makecaption\MYcaption
\makeatother

\def\BibTeX{{\rm B\kern-.05em{\sc i\kern-.025em b}\kern-.08em
    T\kern-.1667em\lower.7ex\hbox{E}\kern-.125emX}}
\usepackage{amsthm}
\theoremstyle{definition}
\makeatletter
\newcommand{\algorithmfootnote}[2][\footnotesize]{%
  \let\old@algocf@finish\@algocf@finish% Store algorithm finish macro
  \def\@algocf@finish{\old@algocf@finish% Update finish macro to insert "footnote"
    \leavevmode\rlap{\begin{minipage}{\linewidth}
    #1#2
    \end{minipage}}%
  }%
}
\usepackage{tikz}
\usetikzlibrary{shapes,arrows,decorations.markings,calc,positioning}
\newdimen\nodeDist

\begin{document}
% \linenumbers
\title{Quantification of Non-stationary Power Quality Events: A New Index Based on $\ell_p$-norm of Energy}

\author{Faizal~Hafiz,~\IEEEmembership{}
        Chirag~Naik,~\IEEEmembership{}
        Davide~La Torre~\IEEEmembership{}
        and~Akshya~Swain,~\IEEEmembership{Senior Member,~IEEE}
        % <-this % stops a space
        
\thanks{F. Hafiz and D. La Torre are with SKEMA Business School, Université Côte d’Azur, Sophia Antipolis, France. Emails: faizal.hafiz@skema.edu; davide.latorre@skema.edu; C. Naik is with S.N. Patel Institute of Technology, Gujarat Technological University, India. Email: chirag.naik@snpitrc.ac.in; A. Swain is with the Department of Electrical, Computer \& Software Engineering, The University of Auckland, New Zealand. E-mail: a.swain@auckland.ac.nz\\

This article has been accepted for publication in \textit{IEEE Transactions on Systems, Man, and Cybernetics: Systems} and it is available at DOI: 10.1109/TSMC.2024.3480457\\

\textcopyright 2024 IEEE. Personal use of this material is permitted. Permission from IEEE must be obtained for all other uses, in any current or future media, including reprinting/republishing this material for advertising or promotional purposes, creating new collective works, for resale or redistribution to servers or lists, or reuse of any copyrighted component of this work in other works.}
% \thanks{C. Naik is with S.N. Patel Institute of Technology, Gujarat Technological University, India. Email: chirag.naik@snpitrc.ac.in}
% \thanks{A. Swain is with the Department of Electrical, Computer \& Software Engineering, The University of Auckland, New Zealand. E-mail: a.swain@auckland.ac.nz}% <-this % stops a space
}

% \markboth{}%
% {Hafiz \MakeLowercase{\textit{et al.}}: Quantification of Non-stationary Power Quality Events: A New Index Based on $\ell_p$-norm of Energy}%IEEE TRANSACTIONS ON SYSTEMS, MAN, AND CYBERNETICS: SYSTEMS


\maketitle

\begin{abstract}
The present study proposes a new index to quantify the severity of non-stationary power quality (PQ) disturbance events. In particular, the severity of PQ events is estimated from their energy distribution in temporal-frequency space. The index essentially measures the $\ell_p$-norm between the energy distributions of an event and the nominal voltage signal. The efficacy of the new index is demonstrated considering a wide class of major non-stationary PQ events, including sag, swell, interruptions, oscillatory transients, and simultaneous events. The results of this investigation, with simulated, real and experimental data, convincingly demonstrate that the proposed index is generic, monotonic, easy to interpret, and can accurately quantify the severity of non-stationary events.
\end{abstract} % ASD signficance 


\ifCLASSOPTIONpeerreview
\begin{center} \bfseries EDICS Category: 3-BBND \end{center}
\fi
\IEEEpeerreviewmaketitle

\section{Introduction}

\IEEEPARstart{T}he proliferation of power electronics converters and consumer electronics, the emergence of smart grid technologies, and the push towards sustainable distributed generation in recent years are some of the contributing factors to the deteriorating quality of power supply. In essence, any departure from the ideal sinusoid voltage denotes a degradation in the quality of the power supply and is often referred to as a Power Quality (PQ) disturbance event. Such events can have detrimental effects ranging from loss of data and production time to equipment failure~\cite{Lu:Shen:2007,Bollen:2000,IEEE-519:1993,IEEE:1159,Mcgranaghan:Roettger:2002,Caramia:Carpinelli:2009}. The consequent remedial response primarily depends on effective severity quantification and monitoring of power quality~\cite{IEEE:1159,Caramia:Carpinelli:2009,Mcgranaghan:Roettger:2002,heydt:1998,Caicedo:Agudelo:2023}. %This study, therefore, focuses on quantification of PQ events. %In particular, PQ indices (\textit{e.g.}, Total Harmonic Distortion~\cite{}) serve as essential tool to easily estimate and interpret the severity of events. Salmeron:Herrera:2009

In response to the growing concern about the quality of power supply, the severity quantification of PQ events has received significant attention; albeit, most of these efforts have primarily been limited to stationary events (\textit{e.g.}, harmonics). For instance, the conventional PQ indices such as Total Harmonic Distortion (THD), Total Demand Distortion (TDD), and Distortion Index (DIN) are defined in the frequency domain. These indices can easily be determined using Fourier analysis and are easy to interpret. Hence, they are highly acceptable for quantifying stationary harmonics and have been used to determine acceptable thresholds in PQ standards\cite{Caramia:Carpinelli:2009,IEEE-519:1993}. However, the conventional indices can only quantify the stationary events (\textit{e.g.}, harmonics) and cannot be extended for non-stationary events as the temporal information associated with such events is lost in the Fourier analysis~\cite{Jaramillo:Heydt:2000,Mariscotti:2000,Morsi:Hawary:2007,Morsi:El-Hawary:2009,Morsi:Hawary:2010}. It is worth highlighting that, in practice, \emph{non-stationary} events, such as voltage \emph{sags}, can constitute a significant portion of PQ events~\cite{Bollen:2000,Upadhya:Singh:2022}. Further, sags, which are limited to only few cycles in duration, can often have economic impacts similar to complete outages~\cite{Mcgranaghan:Roettger:2002}. Despite a relatively higher frequency of occurrence and financial losses associated with non-stationary events, the severity quantification of such events has received relatively less attention. This has been the motivation behind the present investigation.%=-02§ which aims to quantify non-stationary events., \textit{e.g.}, see the indices for power system resilience in~\cite{Talukder:Ibrahim:2021}.

Most non-stationary events, such as sag, swell, and transients, can be characterized by deviation in \textit{magnitude}, \textit{frequency}, and \textit{duration}~\cite{IEEE:1159,Bollen:2000,Bollen:2005}. Thus, it is desirable that an ideal PQ index captures the deviations from any combination of these defining parameters. Quantifying the event under consideration based on both temporal and frequency-domain behavior is essential to achieve such sensitivity to the defining parameters. Further, it is desirable that an ideal PQ index have a \textit{monotonic} relationship with the event severity, \textit{i.e.}, an increase in severity should be captured by a higher index value and vice versa. For instance, a voltage sag with a $50\%$ reduction in magnitude for a duration of $20$ cycles is relatively more severe than a sag with a $30\%$ magnitude reduction for $\frac{1}{2}$ cycle duration. This should translate into a higher index value for the former sag event.

Such a quantification exercise is vital to estimating the expected severity of events, which serves as the basis of PQ remedial responses from consumers and utilities. From a consumer perspective, the severity estimation is pivotal to selecting, designing, and deploying remedial strategies, as their economic viability hinges on the severity of events~\cite{Mcgranaghan:Roettger:2002}. On the other hand, the estimation of severity (via \emph{quantification} indices) and the \emph{frequency of occurrence} (via \emph{classification}, \textit{e.g.}, see~\cite{Caicedo:Agudelo:2023,Hafiz:Swain:IET:2018}) enables utilities to isolate the source of PQ events, select and shape an appropriate remedial response as well as provide consumer guidelines for better ride-through performance. However, to the best knowledge of the authors, the quantification of non-stationary events by a generalized PQ index is yet to be fully explored. While a few existing studies explicitly address this issue~\cite{Shin:Powers:2006,Andreotti:Bracale:2009,ferreira:2015,Olivia:Rosa:2018,Naik:Kundu:2013}, their applicability may be limited in practice due to various factors such as the need to estimate/segment the fundamental component~\cite{Shin:Powers:2006,Andreotti:Bracale:2009}, lack of interpretability~\cite{ferreira:2015,Olivia:Rosa:2018,Naik:Kundu:2013}, and absence of multi-scale analysis~\cite{ferreira:2015,Olivia:Rosa:2018} which is crucial to simultaneous events like \emph{sag-with-transients}, as discussed in detail in Section~\ref{sec:literature_review}.

% Old Lit-review was here - now moved to Section II

% Proposed approach
Given that the quantification of non-stationary events remains an open issue, this study aims to develop an effective yet easy-to-interpret index to quantify major non-stationary events such as sag, swell, and oscillatory transients, to bridge this gap. The core idea is to quantify the severity of an event with respect to the nominal voltage at the Point of Common Coupling (PCC), albeit in a compact functional space. To this end, it is argued that the energy of wavelet coefficients or Temporal-Frequency Energy Distribution (TFED) of events can act as an \textit{indicator} of event severity. Following this argument, a generalized PQ index is proposed, which essentially quantifies the deviation in non-stationary events with respect to the nominal voltage in terms of the $\ell_p$-norm of corresponding TFEDs. The key contributions of this investigation are as follows:
\begin{itemize}
    \item It is demonstrated that the temporal-frequency energy distribution of wavelet coefficients can serve as a severity indicator of a given non-stationary event. Accordingly, a normalized similarity index based on $\ell_p-$norm is proposed to quantify the deviation in TFED of a given event. The proposed index is non-negative and bounded in $[0,1]$, and, therefore, is easy to interpret. Further, it is established that the proposed index varies \emph{monotonically} with the severity of events, through both theoretical and empirical analysis on synthetic and real events. 
    
    \item The efficacy of the index is evaluated considering a comprehensive and diverse PQ event database containing a wide degree of severity that conforms to the IEEE Std. 1159~\cite{IEEE:1159}. This is further augmented by a comparative evaluation with the \emph{wavelet deviation index}~\cite{Naik:Kundu:2013}, on a set of experimental and real-life disturbance recordings available from the University of Cádiz (UoC)~\cite{SagDatabase} and the EPRI national repository of PQ events~\cite{EPRIDatabase}.
 
    \item The sensitivity of the proposed index to mother wavelet selection is extensively investigated by considering 70 distinct wavelets from three distinct wavelet families. This analysis convincingly demonstrates that the proposed index retains the desired monotonic behavior irrespective of the selected wavelet. %Subsequently, the index behavior across different wavelets is estimated by a parametric regression models, which maps the characteristic parameters of events (\textit{e.g.}, \emph{depth} and \emph{duration} of sags) with the index values.
\end{itemize}

The rest of the article is organized as follows: We begin by reviewing the existing PQ indices and highlighting possible limitations in Section~\ref{sec:literature_review}. The PQ event dataset and temporal-frequency decomposition being considered are discussed in Section~\ref{s:Prelim}. The proposed index, its analytical properties, and the integration of \emph{preferences} are discussed next in Section~\ref{s:Proposed}. This is followed by a comprehensive evaluation of the proposed indices in Section~\ref{s:Res} (\emph{non-stationary events}), Section~\ref{s:ResSimEvents} (\emph{simultaneous events}), and Section~\ref{s:ResRealEvents} (\emph{comparative evaluation over real sags and transients}). Finally, we discuss issues related to wavelet selection (Section~\ref{s:wavsense}) and possible applications from utility perspectives in Section~\ref{s:utilityapp}.

%-----------------------------------------------------------
% \vspace{-4mm}
\section{Related Works} %: : Severity Estimation of Non-stationary Events
\label{sec:literature_review}

Most existing investigations into severity estimation of non-stationary events can broadly be categorized into either \emph{reformulated indices}~\cite{Jaramillo:Heydt:2000,Mariscotti:2000,Morsi:Hawary:2007,Morsi:El-Hawary:2009,Morsi:Hawary:2010,Galeano:2015,Biswal:Dash:2012,Naik:kundu:2012,urbina:2017,Yu:Zhao:2021,Thirumala:Umarikar:2015,Thirumala:Jain:2017,Coelho:Dantas:2023} or \emph{deviation indices}~\cite{Shin:Powers:2006,Andreotti:Bracale:2009,ferreira:2015,Olivia:Rosa:2018,Naik:Kundu:2013}. We briefly review some of the relevant investigations from both categories, along with possible limitations in the following subsections.

%-----------------------------------
\vspace{-3.5mm}
\subsection{Reformulation of Conventional Indices}
\label{subsec:litrevA}

The earlier attempts to accommodate non-stationary events primarily focused on the reformulation of the conventional PQ indices (\textit{e.g.}, THD, DIN, k-factor), as their behavior was well understood. To this end, various time-frequency information extraction approaches have been considered which include but are not limited to Short-term Fourier transform (STFT), wavelets, Stockwell-Transform (ST), and their extensions~\cite{Jaramillo:Heydt:2000,Mariscotti:2000,Morsi:Hawary:2007,Morsi:El-Hawary:2009,Morsi:Hawary:2010,Galeano:2015,Biswal:Dash:2012,Naik:kundu:2012,urbina:2017,Yu:Zhao:2021,Thirumala:Umarikar:2015,Thirumala:Jain:2017,Coelho:Dantas:2023}. For instance, STFT is considered in~\cite{Jaramillo:Heydt:2000}, which essentially applies FFT to a moving fixed-length time segment of a given signal to extract time-frequency information. The caveat here is the need for careful balancing of the window-length, which determines the trade-off in time and frequency resolution~\cite{Mariscotti:2000}. These limitations can be alleviated by a variable analysis window associated with wavelets, which is leveraged in several existing redefined PQ indices~\cite{Morsi:Hawary:2007,Morsi:El-Hawary:2009,Morsi:Hawary:2010}.%\textit{e.g.}, see

Several recent investigations focused on augmenting wavelet analysis to enhance the time-frequency information extraction. For instance, Urbina-Salas \emph{et al.}~\cite{urbina:2017} augmented wavelet analysis with single-sideband modulation (SSM) and Hilbert transform (HT) to calculate instantaneous PQ indices. The rationale here is to reduce spectral leakage by using SSM which shifts the fundamental frequency and its harmonics to the center of wavelet-filters. The HT is applied at the output of each wavelet-subband, providing an analytical signal which is used to calculate redefined instantaneous PQ indices. A combination of SSM and HT was also considered in~\cite{Yu:Zhao:2021} to augment wavelet analysis, with an emphasis on \emph{inter-harmonics}. A two-stage decomposition was proposed; the first stage estimates and isolates the \emph{inter-harmonics} using SSM and time-invariant undecimated wavelets. The next stage processes the remaining signal by applying frequency shifting wavelets followed by HT to calculate instantaneous PQ indices. The issue of wavelet spectrum leakage was also investigated in Thirumala \emph{et al.}~\cite{Thirumala:Umarikar:2015,Thirumala:Jain:2017}, where Empirical Wavelet Transform (EWT) was applied to reformulate the PQ indices. This approach essentially segments the Fourier spectrum based on the significant frequencies in the measured signal and adapts scaling and wavelet functions accordingly to reduce spectrum leakage. However, the length of the measurement window directly influences the Fourier spectrum, its segmentation, and subsequent adaptive filter design. This may limit the applications to rapid non-stationary events. Apart from wavelets, a combination of ST and HT was also considered in~\cite{Kaushik:Mahela:2020} to determine two instantaneous PQ indices. While these indices were found to be effective for classification, their application to severity estimation may be hindered by the lack of interpretability as the indices are not bounded. Further, the empirical weight associated with the index (see, TLI in~\cite{Kaushik:Mahela:2020}) may need careful adjustment to suit a particular class of event.

It is worth emphasizing that while the reformulated indices successfully address the simultaneous time-frequency information extraction requirements, the lack of a single and easy to interpret index often limits their practical applications. For instance, the severity estimation from a utility perspective is often approached as a single generic index, whose values can be used to identify the specific parts of a distribution grid where mitigation is required~\cite{Herath:Gosbell:2005}. The need for an easy to interpret, generic index can also be extended to the consumer perspective of severity of estimation. This is further illustrated by a recent investigation by Bracale \emph{et al.}~\cite{Bracale:Caramia:2021}, which focuses on disturbance modeling of electric arc furnaces. Such models are designed and optimized using a generic deviation based PQ index for non-stationary events~\cite{Shin:Powers:2006}, which does not rely on reformulation as will be discussed in the next subsection. % which can broadly be classified as a deviation index, 

%-----------------------------------
\vspace{-3mm}
\subsection{Deviation Indices}
\label{subsec:litrevB}

The primary notion behind the deviation indices is to estimate the severity as a direct function of departure from an \emph{ideal} component or signal. This can be accomplished by either decomposing a given event into the fundamental and the remaining spectral components~\cite{Shin:Powers:2006,Andreotti:Bracale:2009} or by relying on an external \emph{ideal} reference signal~\cite{ferreira:2015,Olivia:Rosa:2018,Naik:Kundu:2013}, \textit{e.g.}, the fundamental sinusoid or the nominal voltage, as discussed in the following:

The severity estimation approach in~\cite{Shin:Powers:2006,Andreotti:Bracale:2009} relies on the decomposition of the given event into the \emph{fundamental} and \emph{disturbance} components. The objective here is to derive several indices by comparing the time-frequency localized energy of the fundamental component with that of the remaining spectral components of the event. In particular, Shin \emph{et al.}~\cite{Shin:Powers:2006} suggested reduced interference distributions to estimate time-frequency distributions (TFD) of the fundamental and the disturbance components. However, the fundamental component was estimated via a curve-fitting procedure assuming that the fundamental frequency remains stable, which may not hold due to load variations and frequency regulation. To overcome this issue, Andreotti \emph{et al.}~\cite{Andreotti:Bracale:2009} suggested an adaptive prony method (APM) to calculate the same indices but replaced TFD with a parametric model that represents events as a linear combination of a pre-fixed number of complex exponentials. The subsequent event model directly provides access to the fundamental component, and thereby, dispenses the need for the curve-fitting based estimation. In particular, APM dynamically adapts the analysis time-window to deal with non-stationary events through an iterative procedure that terminates upon reaching a user-defined fitting-error threshold. However, the minimum length of such a window is restricted to properly estimate the fundamental component, which may limit the ability to handle short duration non-stationary events. Further, the estimated event model can be sensitive to the initial parameter settings and the a priori specified error threshold. 

In contrast to~\cite{Shin:Powers:2006,Andreotti:Bracale:2009}, the severity estimation can directly be approached as a deviation from an external \emph{ideal} signal (\textit{e.g.}, nominal PCC voltage or a pure sinusoidal voltage) in a higher-order compact space, thus eliminating the need to estimate the fundamental component in the given event. This can be achieved by a combination of compact signal representations and a deviation metric, \textit{e.g.}, \emph{principal curves}~\cite{Ferreira:Seixas:2013,ferreira:2015},  \emph{higher-order-cumulants}~\cite{Gerek:2006,Rosa:Gonzalez:2009,Rosa:Agustin:2012,aguera:2011,Rosa:Agustin:2013,Olivia:Rosa:2018,Olivia:2022}, and \emph{wavelet analysis}~\cite{Naik:Kundu:2013}, as discussed in the following: 

Ferreira \emph{et al.}~\cite{ferreira:2015,Ferreira:Seixas:2013} explored the application of Principal Curves (PC), a nonlinear extension of principal components, as a compact representation and for the extraction of distinguishing features. In particular, in~\cite{Ferreira:Seixas:2013} a representative PC was extracted using the k-segmentation algorithm~\cite{Verbeek:Vlassis:2002} for different classes of events and nominal voltage. Subsequently, the squared Euclidean distance of a given event to a particular representative PC is used to decide its membership to the corresponding disturbance. While this approach was found to be an effective direct event \emph{classifier}, the distance between the representative PCs of fundamental frequency events was found to be negligible~\cite{Ferreira:Seixas:2013}, \textit{i.e.}, it is not trivial to distinguish among \emph{sag}, \emph{swell}, and \emph{interruptions} in the proposed feature space. Further investigation in~\cite{ferreira:2015} suggested the use of only one representative PC derived from an \emph{ideal} voltage signal, which served as the reference. Subsequently, the severity index is calculated as the squared Euclidean distance between a given event and its projection on an \emph{ideal} PC. It is easy to follow that this index is non-negative with the zero minimum value (indicating the absence of any disturbance), but it is theoretically unbounded, and, therefore, it is not easy to interpret. Further, unlike wavelet analysis, PCs lack multi-scale information extraction, which can limit their applications to simultaneous events (\textit{e.g.}, \emph{sag-with-transients}). Such simultaneous events often contain information on multiple scales and can be very challenging to quantify as will be discussed with illustrative examples in Section~\ref{s:ResSimEvents}.% A deviation from such `ideal' voltage behavior is considered to be a disturbance measure.

Extensive investigations into applications of \emph{high-order statistics} have also been explored as an alternative compact representation in~\cite{Gerek:2006,Rosa:Agustin:2013,Rosa:Gonzalez:2009,aguera:2011,Rosa:Agustin:2012,Olivia:Rosa:2018,Olivia:2022}. These investigations are rooted in the fact that the ideal voltage behavior can be modeled as a Gaussian process, and \emph{higher-order cumulants} can detect non-Gaussian behavior, which is typically associated with PQ events~\cite{Gerek:2006,Rosa:Agustin:2013}. For example, \cite{Gerek:2006,Rosa:Gonzalez:2009} proposed local extrema of higher-order cumulants up to $4^{th}-order$ (\textit{i.e.}, \emph{variance}, \emph{skewness}, and \emph{kurtosis}) in a pre-fixed time interval, beginning with the event inception, as distinguishing features of events. These concepts are further explored in~\cite{Rosa:Agustin:2012,aguera:2011}, where the \emph{higher-order cumulants} of a nominal voltage are demonstrated to be $\{variance, \ skewness, \ kurtosis\} = \{ 0.5, \ 0, \ -1.5 \}$. A departure from these known values in a particular pair of cumulants can be linked to a particular class of PQ event~\cite{Olivia:Rosa:2018} and can also be used to estimate characteristic parameters of sags and swells~\cite{aguera:2011}. An obvious extension of this desirable property is to determine a new severity index which is essentially a sum of deviations in cumulants corresponding to a given event and the \emph{ideal} sinusoidal voltage, see~\cite{Olivia:Rosa:2018} for details. This index shares some of the mathematical properties with the PC-based index in~\cite{ferreira:2015}, \textit{i.e.}, non-negativity, the minimum zero value corresponding to event absence. The index is, however, theoretically unbounded, with no definitive maximum value, as it depends on the extent and severity of the PQ disturbances.

Naik and Kundu~\cite{Naik:Kundu:2013} proposed the energy of wavelet coefficients as the compact representation. The authors, in particular, focused on the severity estimation of \emph{transients}, and, therefore, proposed a weighted sum of energy deviations. The assigned weight progressively decreases from the higher to lower scales, which makes the proposed index relatively more sensitive to energy deviations in higher scales. While such an approach is effective for \emph{transients}, it leads to reduced sensitivity to lower scales, introducing a non-monotonic behavior when fundamental frequency events are considered, \textit{e.g}, \emph{sags}, as will be shown in Section~\ref{subsec:realsag}.

Finally, it is worth noting that the notion of deviation from a reference voltage/signal has been explored in other areas of PQ analysis, \textit{e.g.}, A waveshape change detection approach in~\cite{Bastos:Santoso:2020} proposed \emph{similarity scores} to compare to consecutive waveform cycles to identify the inception of PQ events. Hu \emph{et al.}~\cite{Hu:Yang:2023} proposed a goodness-of-fit based statistical hypothesis test, which evaluates whether the probability distributions of a test and reference voltages are equal, and subsequently estimates the duration of \emph{sags}.

To summarize, while a few investigations address the issue of quantification of non-stationary events, several aspects of this issue are yet to be explored, which include but are not limited to,
\begin{itemize}[itemindent=0pt,left=0pt]
    \item \emph{Monotonicity}: A monotonic index behavior with event severity is key to the wide-scale acceptability of any index. However, this aspect is seldom explicitly addressed in the existing investigations~\cite{Shin:Powers:2006,Andreotti:Bracale:2009,ferreira:2015,Olivia:Rosa:2018,Naik:Kundu:2013}, which typically build on a few illustrative events/case studies. 
    \item \emph{Simultaneous Events}: While such events (\textit{e.g.}, \emph{sag-with-transients}) are fairly common phenomena~\cite{Bollen:2000}, this aspect of the severity estimation is typically not explored in detail in~\cite{Shin:Powers:2006,Andreotti:Bracale:2009,ferreira:2015,Olivia:Rosa:2018,Naik:Kundu:2013}. Such scenarios present unique estimation challenges for which access to multi-scale information is crucial, as will be discussed in Section~\ref{s:ResSimEvents}.
    \item \emph{Interpretability}: As noted earlier, the deviation/distance metrics used to quantify the differences in a given event and the ideal signal are often unbounded. Consequently, it is not trivial to interpret the existing indices and determine meaningful thresholds. Such a step is crucial to understanding event severity in terms of end-consumer impacts, as will be demonstrated in Section~\ref{s:utilityapp}.
\end{itemize}

%-----------------------------------------------------------
\vspace{-3mm}
\section{Preliminaries}
\label{s:Prelim}
%--------------------------------------------------------------
%                       PQ Events
%--------------------------------------------------------------
% 
\begin{table}[!b]
    \centering
    \caption{Parametric Models of Power Quality Events$^\dagger$}
	\label{t:events}
	\begin{adjustbox}{max width=0.47\textwidth} 
	    \begin{threeparttable}
	   % \small 
	    \begin{tabular}{c c} 
			\toprule
			\bfseries PQ Event & \bfseries Model \\ [0.8ex]%& \bfseries Parameters$^{\dagger}$
			\midrule
% 			Pure Sinusoid & $ v_1(t) = \alpha \sin(\omega t)$ & $0.9<\alpha<1.1$\\ [1.0ex]
% 			\midrule	
			Sag$^\ddagger$ & \makecell{$v(t) = \Big[1-\alpha\{u(t_1)-u(t_2)\} \Big] \times \sin(\omega t)$\\[1.5ex] $0.1\leq\alpha\leq0.9$, $t_1<t_2$, $0.5T\leq t_2-t_1 \leq30T$} \\ [0.7ex]
			\midrule
			Swell$^\ddagger$ & \makecell{$v(t) = \Big[1+\alpha\{u(t_1)-u(t_2)\} \Big] \times \sin(\omega t)$ \\[1.5ex] $0.1\leq\alpha\leq0.8$, $t_1<t_2$, $0.5T\leq t_2-t_1 \leq30T$} \\ [0.7ex]
			\midrule
			Interruption$^\ddagger$ & \makecell{$v(t) = \Big[1-\alpha\{u(t_1)-u(t_2)\} \Big] \times \sin(\omega t)$ \\[1.5ex] $0.9 <\alpha\leq1.0$, $0.5T\leq t_2-t_1 \leq30T$} \\ [1.2ex]
			\midrule
			\makecell{Oscillatory\\ Transient} & \makecell{$v(t) = \sin(\omega t) + u_{t_1} \beta e^{\gamma t_2} \sin(\omega_{tr}t)$ \\[1.2ex] \makecell{$-125\leq\gamma\leq -25$, $1\leq\beta\leq 4$, $400\leq f_{tr} \leq 4000$ \\ $\omega_{tr}=2\pi f_{tr}$, $0.3\leq t_1 \leq 0.9$, $t_2 =  (t-t_1)u_{t_1}$} } \\  [1.7ex]
			\midrule
			\makecell{Sag with\\ Transient$^\ddagger$} & \makecell{$v(t) = [\{1-\alpha(u(t-t_1)-u(t-t_2))\} \times  \sin(\omega t)] $ \\ $ + u_{t_1} \beta e^{\gamma t_2} \sin(\omega_{tr}t)$\\[1.5ex] $ 0.1 \leq \alpha \leq 0.9 $, $-125\leq\gamma\leq -25$, $1\leq\beta\leq 4$, $0.3\leq t_1 \leq 0.9$\\ $ 0.5T \leq t_2-t_1 \leq 30T $, $t_2 =  (t-t_1)u_{t_1}$ \\ $400\leq f_{tr} \leq 4000$} \\  [1.2ex]
			\midrule
			\makecell{Swell with\\ Transient$^\ddagger$} & \makecell{$v(t) = [\{1+\alpha(u(t-t_1)-u(t-t_2))\} \times \sin(\omega t)] $\\ $ + u_{t_1} \beta e^{\gamma t_2} \sin(\omega_{tr} t)$ \\[1.2ex] $ 0.1 \leq \alpha \leq 0.8 $, $ -125 \leq \gamma \leq -25 $, $ 1 \leq \beta \leq 4 $, $0.3\leq t_1 \leq 0.9$ \\ $ 0.5T \leq t_2-t_1 \leq 30T $, $t_2 =  (t-t_1)u_{t_1}$ \\ $400\leq f_{tr} \leq 4000$} \\ [1.2ex]
			\bottomrule
		\end{tabular}
		\begin{tablenotes}
        \item $^{\dagger}$ For each event duration $T = 40 \ cycles$; fundamental frequency $f = 50 \ Hz$; $\omega=2\pi f$; $\omega_{tr}=2\pi f_{tr}$; sampling frequency $F_s = 10 \ kHz$; $^\ddagger$ $u(t)=1$ if $t>0$; otherwise $u(t)=0$; duration of \textit{sag} and \textit{swell} is denoted by `$t_d$' and is given by: $t_d=t_2-t_1$. 
        \end{tablenotes}
        \end{threeparttable}
		\end{adjustbox}
		\vspace{-4mm}
\end{table}
%--------------------------------------------------------------
%----------------------------------------------------------
\subsection{Non-stationary PQ Events}
\label{s:PQEvents}

This study mainly focuses on four non-stationary PQ events, which include voltage sags, voltage swells, momentary interruptions, and oscillatory transients. Such non-stationary events are often difficult to quantify as this requires information from both the temporal and frequency domains. Further, the possibilities of simultaneous non-stationary events (\textit{e.g., sag with transient}) within a measurement window can not be ruled out. A comprehensive PQ event database, which includes a wide variety of single and simultaneous non-stationary events, is, therefore, crucial to validate any PQ index. To this end, we consider both parametric models and real-life measurements of PQ events, as discussed in the following.

In this study, the IEEE Std. 1159~\cite{IEEE:1159} is being followed, which defines various PQ events in terms of magnitude, frequency, and duration. Based on these definitions, parametric models are being used to simulate non-stationary events of interest, as shown in Table \ref{t:events}. Each event is simulated for the duration of $40$ cycles at the fundamental frequency of $50\ Hz$ and is sampled at $10 \ kHz$. In addition, the measurement noise is emulated by adding zero-mean white Gaussian noise of ($45 \ dB$ SNR) to each simulated event.

Further, the simultaneous non-stationary events, such as \textit{sag with transient}, often arise from auto re-close operations of circuit-breakers~\cite{Bollen:2000}. Such simultaneous events are often relatively difficult to quantify due to distinct defining parameters for each constitute event, \textit{e.g.}, \textit{sag with transient} is characterized by magnitude ($\alpha$) and duration ($t_d$) of \textit{sag} as well as magnitude ($\beta$), duration ($\gamma$) and frequency ($f_{tr}$) of an \textit{oscillatory transient} (see Table~\ref{t:events}). Therefore, this study considers two simultaneous events (\emph{sag with transient} and \emph{swell with transient}) to validate the proposed index's efficacy further. Note that the parametric models are being used to synthesize both single and simultaneous events, as seen in Table~\ref{t:events}. Moreover, real voltage sags and experimental transients are also considered for further validation, which will be discussed in Section~\ref{s:ResRealEvents}.
%%%========================================
\begin{table}[!b]
  % \vspace{-5mm}
  \centering
  \caption{DWT Decomposition}
  \label{t:DWT}%
  \begin{adjustbox}{max width=0.3\textwidth}
    \begin{tabular}{ccc}
    \toprule
    \makecell{\textbf{Decomposition} \\ \textbf{Level}} & \makecell{\textbf{Frequency} \\ \textbf{Band}} & \makecell {\textbf{Frequency} \\ \textbf{Content} (\boldmath{$Hz$)}} \\
    \midrule
    1     & $\mathcal{B}_1$    & $2500 - 5000$ \\
    2     & $\mathcal{B}_2$    & $1250 - 2500$ \\
    3     & $\mathcal{B}_3$    & $625 - 1250$ \\
    4     & $\mathcal{B}_4$    & $312.5 - 625$ \\
    5     & $\mathcal{B}_5$    & $156.20 -  312.25$ \\
    6     & $\mathcal{B}_6$    & $78.10 - 156.20$ \\[0.5ex]
    \midrule
    \multirow{2}{*}{7}     & $\mathcal{B}_7$ (detail)    & $39.05 - 78.10$ \\
         & \makecell{$\mathcal{B}_8$ (approximation)}    & $0 - 39.05$ \\
    \bottomrule
    \end{tabular}%
    
%     \begin{tabular}{ccccccccc}
%     \toprule
%     \makecell{\textbf{Decomposition}\\ \textbf{Level}} & 1 & 2 & 3 & 4 & 5 & 6 & \multicolumn{2}{c}{7} \\
%     \midrule
%     \makecell{\textbf{Frequency}\\ \textbf{Band}} & $\mathcal{B}_1$ & $\mathcal{B}_2$ & $\mathcal{B}_3$ & $\mathcal{B}_4$ & $\mathcal{B}_5$ & $\mathcal{B}_6$ & \makecell{$\mathcal{B}_7$\\ (detail)} & \makecell{$\mathcal{B}_8$\\ (approximation)} \\
%     \midrule
%     \makecell{\textbf{Frequency}\\ \textbf{Content} ($\boldsymbol{Hz}$)} & $2500 - 5000$ & $1250 - 2500$ & $625 - 1250$ & $312.5 - 625$ & $156.20 - 312.25$ & $78.10 - 156.20$ & $39.05 - 78.10$ & $0 - 39.05$ \\
%     \bottomrule
% \end{tabular}

\end{adjustbox}
\end{table}%
%%%=======================================================

% It is worth emphasizing that parametric models are crucial for validating the PQ index, as such models allow for precise control of event characteristics. For instance, the IEEE Std. 1159 characterizes sag in terms of a $10\%$ to $90\%$ reduction in magnitude ($\alpha$) and $0.5$ cycle to $30$ cycles variation in duration ($t_d$). Such variations in sag can easily be emulated by varying the magnitude $\alpha$ and/or duration ($t_2-t_1$) in the parametric model of sag (see Table~\ref{t:events}). The parametric model thus allows for curating a comprehensive PQ event database and extensive evaluation of the proposed PQ index.

% Finally, real voltage sags and experimental transients are also considered for further validation. The details of these events and the corresponding evaluation of the proposed index are discussed in Section~\ref{s:ResRealEvents}.

%-------------------------------------------------------
\vspace{-3mm}
\subsection{Temporal Frequency Decomposition of PQ Events}
\label{s:DWT}

The core idea of the proposed index is to quantify the deviations in PQ events through \textit{time-frequency} analysis, as will be discussed in Section~\ref{s:Proposed}. It is, therefore, pertinent to briefly discuss the time-frequency decomposition being considered. This study, in particular, considers DWT for temporal-frequency analysis owing to its relatively lower computational complexity over other approaches such as Stockwell Transform~\cite{Hafiz:Swain:IET:2018}. This is crucial to cover a broad frequency spectrum of PQ events, varying from DC to several MHz~\cite{IEEE:1159}. A detailed treatment on DWT is beyond the scope of this study and can be found elsewhere~\cite{Daubechies:1992}. 

% A multi-resolution analysis by DWT can be viewed as a decomposition of a given signal by an iterative filter bank for a pre-fixed number of \textit{levels}. In particular, the signal is passed through a pair of half-band filters at each decomposition level  and then down-sampled by a factor of $2$. The outputs from \textit{high-pass} and \textit{low-pass} filters are referred to as \textit{detail} and \textit{approximation}. At each level, the \textit{detail} serves as the process output, whereas the \textit{approximation} is passed to the next level for further decomposition. This procedure is iterated till a pre-fixed decomposition level is reached. A detailed treatment of this subject is beyond the scope of this study and can be found elsewhere~\cite{Daubechies:1992}. 

An application of DWT involves a judicious selection of base wavelets and the number of decomposition levels~\cite{Daubechies:1992,Song:Xia:2017}. In this study, mid-order \texttt{symlets} (\textit{e.g.}, \texttt{sym4} or \texttt{sym6}) are used on the basis of a detailed wavelet sensitivity analysis, which will be discussed in Section~\ref{s:wavsense}. Further, the depth of decomposition is determined using the following rule of thumb,
%%%===================================
\begin{equation}
    \label{eq:ThumbRule}
    \frac{f_s}{2^{D+1}} \leq f \leq \frac{f_s}{2^D}
\end{equation}
%%%===================================
where, `$f$' denotes the \textit{frequency of interest} ($f=50 \ Hz$);`$f_s$' denote the sampling frequency ($f_s = 10 kHz$) and `$D$' gives the number of decomposition levels. Following this thumb rule, the decomposition level in this study is determined to be $7$, \textit{i.e.}, $D=7$. This gives a total of $8$ frequency bands after decomposition, including $7$ \textit{details} ($\mathcal{B}_1, \mathcal{B}_2, \dots, \mathcal{B}_7$) and the approximation at the $7^{th}$ level ($\mathcal{B}_8$), as shown in Table~\ref{t:DWT}.

% The rationale behind the proposed approach is to quantify the non-stationary PQ events using the corresponding temporal-frequency energy distribution. It is argued that, for a given class of non-stationary events (\textit{e.g., sag} or \textit{transient}), such energy distribution varies with the \textit{severity} of the event. The core idea of the proposed approach is, therefore, to treat temporal-frequency energy distribution as an \textit{indicator} of event severity.
%---------------------------------------------------------
\section{Proposed Index}
\label{s:Proposed}

\begin{figure}[!t]
    \centering
    \begin{subfigure}{0.21\textwidth}
        \includegraphics[width=\textwidth]{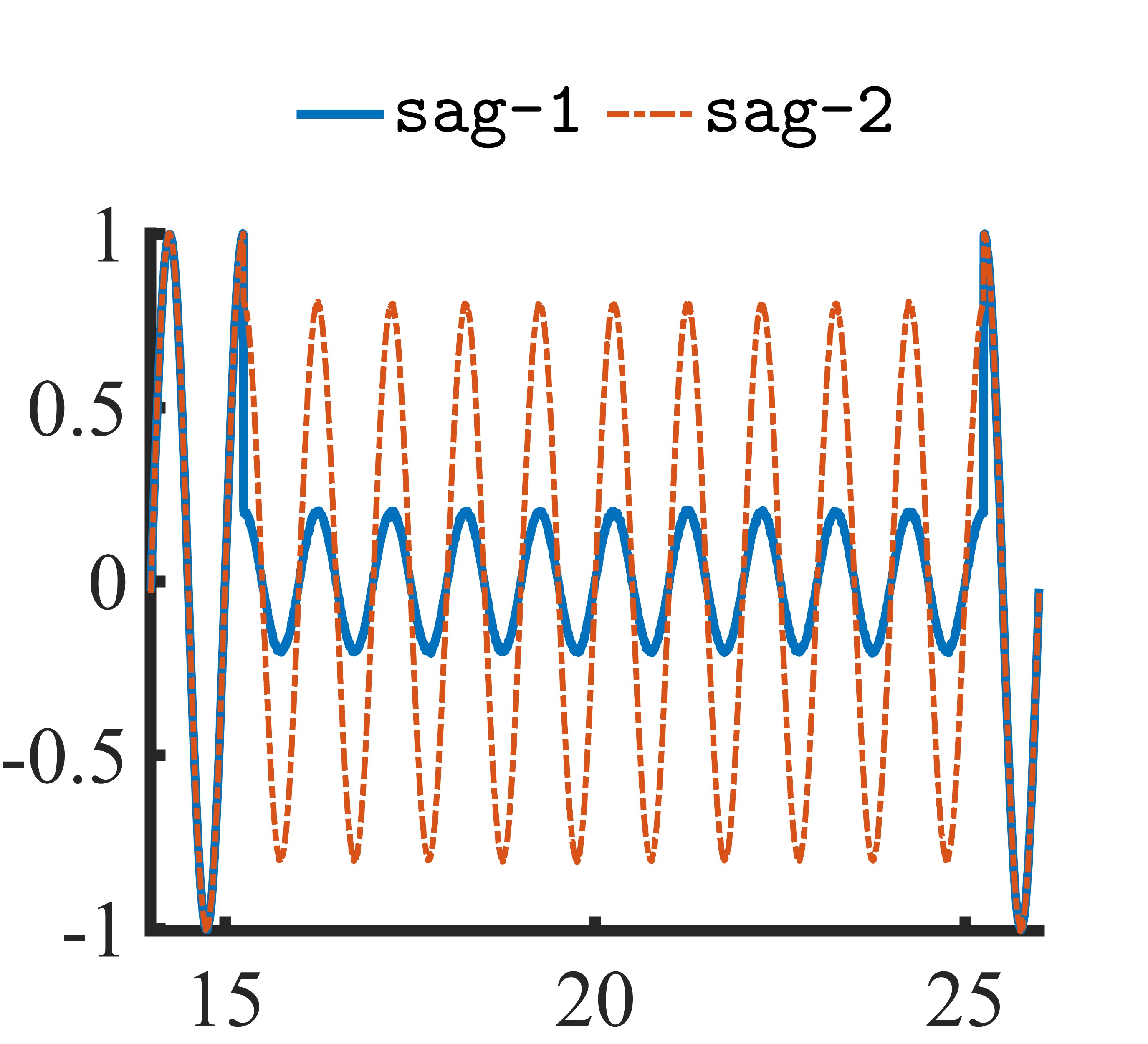}
        \caption{Illustrative \texttt{sags}}% with $\{ \alpha, t_d \} =\{ 0.8, 10 \ cycles\}$}
        \label{f:sag1}
    \end{subfigure}
    \hfill
    \begin{subfigure}{0.21\textwidth}
        \includegraphics[width=\textwidth]{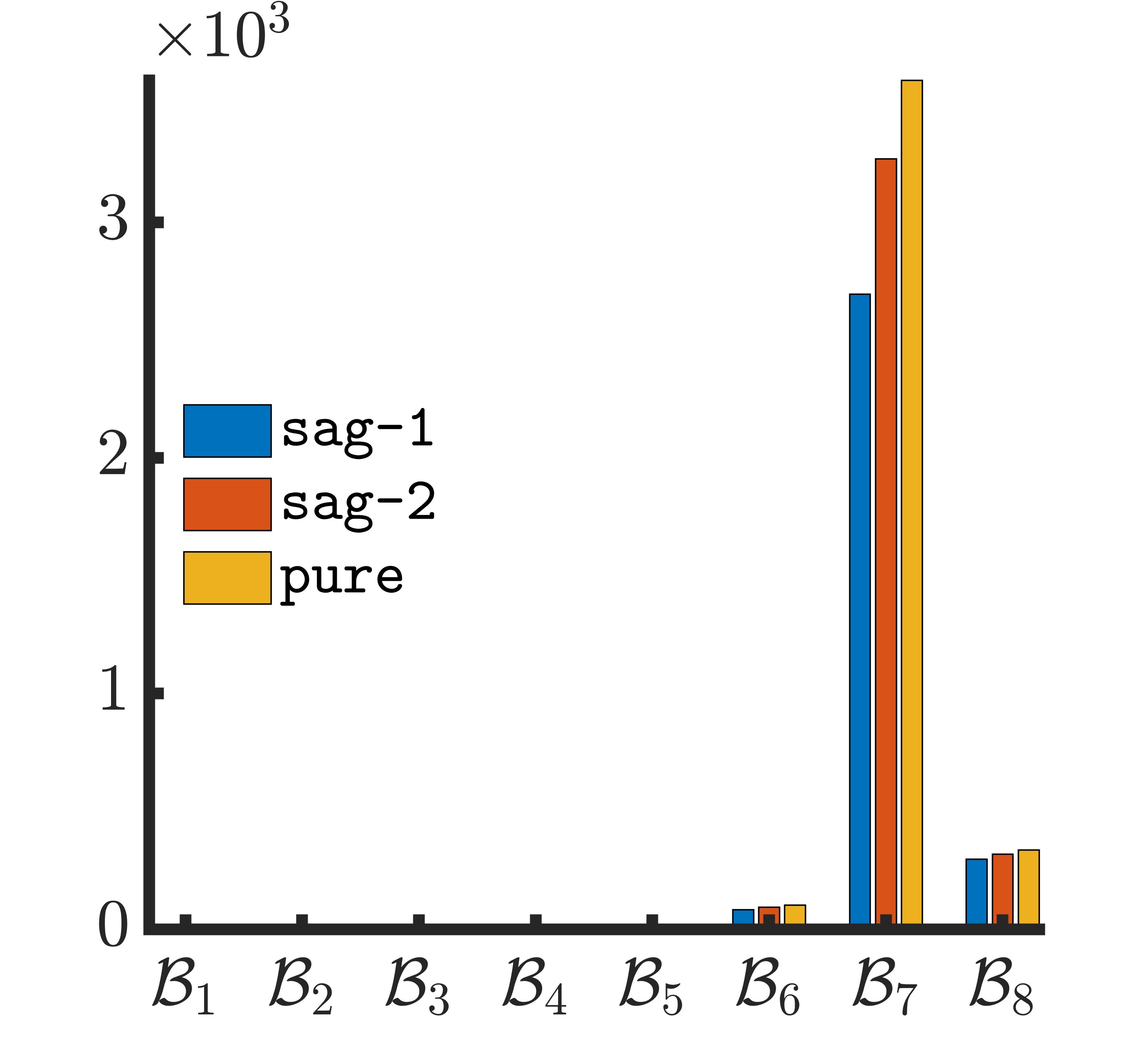}
        \caption{Energy distribution of \textit{sags}}
        \label{f:SagBar}
    \end{subfigure}    
    \caption{Illustrative \texttt{sag} events and the corresponding energy of wavelet coefficients. The \texttt{x} and \texttt{y-axes} in Fig.~\ref{f:sag1} respectively denote \texttt{time} (cycles) and \texttt{magnitude} (pu). \texttt{Pure} denotes an ideal 1 (pu) and $50 \ Hz$ sinusoidal voltage signal. The \texttt{x} and \texttt{y-axes} in Fig.~\ref{f:SagBar} respectively denote \textit{frequency bands} and the wavelet coefficient \textit{energy}. $\mathcal{B}_1, \mathcal{B}_2, \dots, \mathcal{B}_8$ denote frequency bands after $7^{th}$ level decomposition as shown in Table~\ref{t:DWT}.}
    \label{f:SagExample}
    % \vspace{-6mm}
\end{figure}

Let $\mathcal{S}$ denote the severity of a particular class of PQ event, which can be approached as a function of defining `traits' of the event class. Most of such traits are well defined in the IEEE Std. 1159~\cite{IEEE:1159}. For instance, the severity of the fundamental frequency events such as \textit{sag}, \textit{swell}, and \textit{interruption} can be thought of as a function of the corresponding \emph{magnitude deviation/depth} ($\alpha$) and \emph{duration} ($t_d$),
%------------------------------------
\begin{align}
    \mathcal{S} \leftarrow \mathcal{F}(\alpha,t_d)% \quad \mathcal{S}_{hf} = \mathcal{F}(\beta,\gamma, f_{tr})
\end{align}
%--------------------------------------
The goal of a PQ index is then to estimate the unknown function $\mathcal{F}(\cdot)$ to quantify the severity of a given event. In particular, this study proposes an index based on the temporal-frequency energy distribution of an event to estimate $\mathcal{F}(\cdot)$. It is argued that such energy distribution varies with the event's severity for a given class of non-stationary events (\textit{e.g., sag} or \textit{transient}). Therefore, the core idea of the proposed approach is to treat temporal-frequency energy distribution as an \textit{indicator} of event severity. To illustrate this further, consider two \textit{sag} events with different degrees of severity, as shown in Fig.~\ref{f:sag1}. These events are generated using the following specifications in the parametric model of \textit{sag} (see Table~\ref{t:events}):
\begin{itemize}
    % \smallskip
    \item \texttt{sag-1}: $\{ \alpha_1, t_{d,1} \} =\{ 0.8 \ pu, 10 \ cycles \}$. %$80\%$ drop in the magnitude with duration of 10 cycles, \textit{i.e.},
    % \smallskip
    \item \texttt{sag-2}: $\{ \alpha_2, t_{d,2} \} =\{ 0.2 \ pu, 10 \ cycles\}$. %$20\%$ drop in the magnitude with duration of 10 cycles, \textit{i.e.},
    % \smallskip
\end{itemize}

While the durations of both sags are equal, the magnitude drop with \texttt{sag-1} is relatively higher ($80\%$). Accordingly, \texttt{sag-1} is relatively more severe than \texttt{sag-2}, \textit{i.e.}, $\mathcal{S}_1>\mathcal{S}_2$. Both the events are decomposed to the $7^{th}$ level using DWT (see Section~\ref{s:DWT}), and the energy of consequent \textit{detail} and \textit{approximation} is determined. Fig.~\ref{f:SagBar} shows the energy distribution of these \textit{sag} events along with a pure sinusoidal voltage signal ($\{1 \ pu, 50 \ Hz\}$). It is worth noting that \textit{sags} are essentially fundamental frequency events that translate into skewed energy distribution towards the fundamental frequency band (\textit{i.e.}, $\mathcal{B}_7$ band, see Table~\ref{t:DWT}). Further,  the \textit{sag} events are characterized by a momentary reduction in voltage magnitude, which translates into energy reduction in the $\mathcal{B}_7$ band in comparison to the pure sinusoidal voltage, as seen in Fig.~\ref{f:SagBar}. It is interesting to see that the reduction of the $\mathcal{B}_7$ band energy is relatively higher with $sag-1$ in comparison to $sag-2$ (see Fig.~\ref{f:SagBar}), which conforms to the known severity of these events.

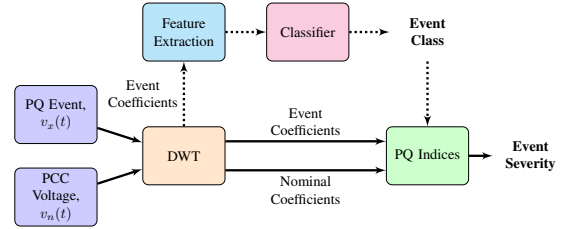
\begin{figure}[!t]
\centering
\begin{adjustbox}{max width=0.41\textwidth}
% Define block styles
\tikzstyle{block} = [rectangle, draw, fill=white, 
    text width=5em, text centered, rounded corners, minimum height=4em,line width=0.02cm]%blue!20
\tikzstyle{block1} = [rectangle, draw, fill=white, 
    text width=6.5em, text centered, rounded corners, minimum height=4em,line width=0.02cm]%blue!20
\tikzstyle{invisible_block} = [rectangle, draw=white, fill=white, text width=4.5em, text centered, rounded corners, minimum height=4em,line width=0.02cm]%blue!20
\tikzstyle{line} = [draw, -latex']
\tikzstyle{cloud} = [draw, ellipse,fill=white, node distance=4.8cm,
    minimum height=3em, line width=0.02cm]%red!20

\begin{tikzpicture}[auto]
    \centering
    % % Place nodes
    \node[block,fill=blue!20] at (0, 1)   (event) {PQ Event, $v_x(t)$};
    \node [block,fill=blue!20] at (0,-1) (pure) {PCC Voltage, ${v}_{n} (t)$};
    \node [block,fill=orange!20] at (3.1,0) (prep) {DWT};
    % \node [block1, fill=orange!20] at (7.25,0) (fe) {Energy\\ Determination};
    \node [block,fill=green!20] (PQI) at (9,0) {PQ Indices}; %(10.5,0)
    \node [invisible_block] at (11.5,0) (BOTTOM END) {\textbf{Event\\Severity}};

    \node [block,fill=cyan!25] at (3.1,3) (FE) {Feature\\Extraction};
    \node [block,fill=magenta!25] at (6.1,3) (ML) {Classifier};
    \node [invisible_block] at (9,3) (END) {\textbf{Event\\Class}};
    
    % % % Draw edges
    \path [line,ultra thick] (event) -- (prep);
    \path [line,ultra thick] (pure) -- (prep);
    \path[line,ultra thick] ([yshift=-10pt]prep.east) -- node[align=center, below] {Nominal\\ Coefficients} ([yshift=-10pt]PQI.west);
    \path[line,ultra thick] ([yshift=10pt]prep.east) -- node[align=center, above] {Event\\ Coefficients} ([yshift=10pt]PQI.west);
    \path [line,ultra thick] (PQI) -- (BOTTOM END);

    \path[line,ultra thick,dotted] (prep.north) -- node[align=center, left] {Event\\ Coefficients} (FE.south);
    \path [line,ultra thick, dotted] (FE) -- (ML);
    \path [line,ultra thick,dotted] (ML) -- (END);
    \path [line,ultra thick,dotted] (END.south) -- (PQI);

\end{tikzpicture}  
\end{adjustbox}
\caption{The proposed framework for quantification of non-stationary events.}
\label{f:IndexBlock}
% \vspace{-6mm}
\end{figure}

The illustrative example thus convincingly demonstrate that the \textit{temporal-frequency} energy distribution of the events can be used as an \emph{indicator} of severity. The next step of the \textit{index} development is to determine a convenient single-valued function of such energy distribution similar to Total Harmonic Distortion (THD). Since the objective of the disturbance index is to determine the deviation in events with respect to the nominal sinusoid, the $\ell_p$-norm between the energy distribution of events and the nominal PCC voltage is proposed as the disturbance index. Fig.~\ref{f:IndexBlock} outlines the overall framework of the proposed approach. In the following, the steps involved in determining the proposed index are discussed in detail.

%%%========================================
%%%=======================================================
%%%===========================================================
\vspace{-2mm}
%%%=========================================================

%------------------------------------------------------------
\vspace{-3mm}
\subsection{Energy Norm Index}
\label{s:ENI}
Let $v_{x}(t)$ and $v_{n}(t)$ respectively denote the PQ event under consideration and the reference nominal voltage at the Point of Common Coupling (PCC), in an appropriate \emph{per-unit system}. Let the total number of samples in the given event be equal to `$N$', and the total number of decomposition levels be equal to `$D$'. Subsequently, the number of wavelet coefficients at a particular decomposition level is given by,

%--------------------------------------
\vspace{-2mm}
\begin{small}    
\begin{equation}
    % \small
    \label{eq:ns}
    N_k=N/(2\times k), \quad k \in [1,D]
\end{equation}
\end{small}
%------------------------------   
For the given event $v_{x}(t)$, the energy of the \textit{approximation} and \textit{detail} coefficients can be given by,

%------------------------------------
\vspace{-2mm}
\begin{small}    
\begin{equation}
    \label{eq:e}
    e_{x,k} = \sum \limits_{i=1}^{N_k} |d_i|^2, \quad k \in [1,D], \qquad e_{x,D+1} = \sum \limits_{i=1}^{N_D} |a_i|^2
    %   e_x^a = \sum \limits_{i=1}^{N_D} |a_i|^2, \quad e_{x,k}^d = \sum \limits_{i=1}^{N_k} |d_i|^2, \ \ k \in [1,D]
\end{equation}
\end{small}
%--------------------------------
where, `$d_i$' and `$a_i$' respectively denote the \textit{detail} and \textit{approximation} coefficients; $e_{x,k}$ denotes the energy of the detail coefficients at the `$k^{th}$' decomposition level, and $e_{x,{D+1}}$ denotes the energy of the approximation coefficients.

Subsequently, the temporal-frequency energy distribution (TFED) of the PQ disturbance event $v_{x}(t)$ is given by,

%------------------------------------------------
\vspace{-2mm}
\begin{small}  
\begin{equation}
\label{eq:ex}
    {\mathcal{E}}_{x} =\begin{bmatrix} \overbrace{\begin{matrix} e_{x,1} & e_{x,2} & \dots & e_{x,D} \end{matrix}}^{details} & \overbrace{\begin{matrix} e_{x,D+1} \end{matrix}}^{approximation} \end{bmatrix} 
\end{equation}
\end{small}
%----------------------------------------------
Similarly, the energy distribution vector of `$nominal$' PCC voltage is given by,
%---------------------------------------------   
\vspace{-2mm}
\begin{small}  
\begin{equation}
\label{eq:epure}
    {\mathcal{E}}_{n} = \begin{bmatrix} e_{n,1} & e_{n,2} & \dots & e_{n,D} & e_{n,D+1} \end{bmatrix}
\end{equation}
\end{small}
%---------------------------------------------

Once the energy distribution vectors ${\mathcal{E}}_x$ and ${\mathcal{E}}_{n}$ are determined, the disturbance index can be quantified by calculating the $\ell_p$-norm as follows: 
%-------------------------------------------------------------
\begin{align}
\label{eq:edi}
    \rm{Index} = ||{\mathcal{E}}_{x} - {\mathcal{E}}_{n}||_p
\end{align}
%-------------------------------------------------------------
An increasing severity will lead to a higher deviation in the energy distribution of the event, which will translate into an increase in the $\ell_p$-norm based index. 

Note that while the $\ell_p$-norm based index in~(\ref{eq:edi}) can capture the severity of PQ events, it is often difficult to interpret results in terms of absolute values associated with the energy norms. It is, therefore, desirable to bound and normalize the index for ease of interpretation. This is achieved by normalizing the index with respect to the sum of the $\ell_p$-norms of ${\mathcal{E}}_{x}$ and ${\mathcal{E}}_{n}$; which bounds the index from above by a constant, as follows: 
%------------------------------------------------------------
\begin{align}
\label{eq:edinormalized1}
    {\mathcal{ENI}}={||{\mathcal{E}}_{x}-{\mathcal{E}}_{n}||_p}\bigg{/}{\sqrt{||{\mathcal{E}}_{x}||_p^2 + ||{\mathcal{E}}_{n}||_p^2}}
\end{align}
%------------------------------------------------------------
where, $\mathcal{ENI}$ denotes the proposed Energy Norm Index. It is worth noting that this approach is similar to normalized similarity indices, which have been used in image and signal processing to measure how much two images or signals differ from each other; see, for instance~\cite{vrscay2011,Fang:Fang:2017,brunet}. Next, we examine some of the properties of $\mathcal{ENI}$:
\vspace{-3mm}

\ \\
{\bf Proposition 1.} The energy norm index for any given event $v_x(t)$ is bound in $[0,1]$.
\begin{proof}
It is easy to prove that the maximum value of  $\mathcal{ENI}$ over the positive orthant is $1$. In fact, by plugging $\mathcal{E}_{x}=0$ we can easily prove that $\mathcal{ENI} = {||\mathcal{E}_{n}||_p}\big{/}{\sqrt{||\mathcal{E}_{n}||_p^2}}=1$. 

Furthermore, for any $\mathcal{E}_{x}$, we have:
%------------------------------------------------------------
\begin{equation}
    \mathcal{ENI}={||\mathcal{E}_{x}-\mathcal{E}_{n}||_p}\Big{/}{\sqrt{||\mathcal{E}_{x}||_p^2 + ||\mathcal{E}_{n}||_p^2}} \le  1
\end{equation}
%------------------------------------------------------------
which shows that the maximum is attained when $\mathcal{E}_{x}=0$.
\end{proof}
\begin{remark}
It is also true that $\mathcal{ENI} = 0$ if and only if 
$\mathcal{E}_{x}= \mathcal{E}_{n}$ and that $\mathcal{ENI}$ is a symmetric index. However, in general, the triangular property is not satisfied, which makes $\mathcal{ENI}$ not to be a metric. 
\end{remark}
\ \\ %\smallskip
{\bf Proposition 2.} For any given event, $\mathcal{ENI}$ is increasing with respect to the variable $e_{x,j}$ over the interval $[e_{n,j},+\infty)$ if and only if, $\displaystyle{1 - \frac{e_{n,j}}{e_{x,j}} \ge \frac{||\mathcal{E}_{x}- \mathcal{E}_{n}||_p^2}{||\mathcal{E}_{x}| _p^2 + ||\mathcal{E}_{n}||_p^2}}$

\begin{proof}
The definition of $\mathcal{ENI}$ is, 
\begin{equation*}
\mathcal{ENI} = {||\mathcal{E}_{x}-\mathcal{E}_{n}||_p}\bigg{/}{\sqrt{||\mathcal{E}_{x}||_p^2 + ||\mathcal{E}_{n}||_p^2}} 
\end{equation*}
where $\mathcal{E}_{x}=(e_{x,j})$ and $\mathcal{E}_{n}=(e_{n,j})$, 
$j\in 1...D+1$, $e_{x,j},e_{n,j}\ge 0$. Then,
\begin{align*}
\frac{d \mathcal{ENI}^2}{d e_{x,j}}  = 
\frac{2 (e_{x,j}-e_{n,j}) (||\mathcal{E}_{x}||_p^2 + ||\mathcal{E}_{n}||_p^2)-2 ||\mathcal{E}_{x}- \mathcal{E}_{n}||_p^2 e_{x,j}}{(||\mathcal{E}_{x}||_p^2 + ||\mathcal{E}_{n}||_p^2)^2}
\end{align*}

By simple algebra one can show that $\mathcal{ENI}$ is increasing over the interval $[e_{n,j},+\infty)$ if and only if the following condition holds,
\begin{equation}
\label{eq:cond1}
1 - \frac{e_{n,j}}{e_{x,j}} \ge \frac{||\mathcal{E}_{x}- \mathcal{E}_{n}||_p^2}{||\mathcal{E}_{x}| _p^2 + ||\mathcal{E}_{n}||_p^2}
\end{equation}
which concludes the proof. 
\end{proof}

Note that whenever the opposite condition,
\begin{equation}
\label{eq:cond2}
1 - \frac{e_{n,j}}{e_{x,j}} \le \frac{||\mathcal{E}_{x}- \mathcal{E}_{n}||_p^2}{||\mathcal{E}_{x}| _p^2 + ||\mathcal{E}_{n}||_p^2}
\end{equation}

is satisfied over the interval $[e_{n,j},+\infty)$ then the index $\mathcal{ENI}$ is decreasing. As a particular case, let us also notice that over the interval $(-\infty,e_{n,j}]$ the above condition is always satisfied (the LHS is negative while the RHS is positive) and, therefore, the index $\mathcal{ENI}$ is always decreasing. \hfill \qedsymbol

%%%===========================================
\begin{figure}[!t]
    \centering
    \includegraphics[width=0.48\textwidth]{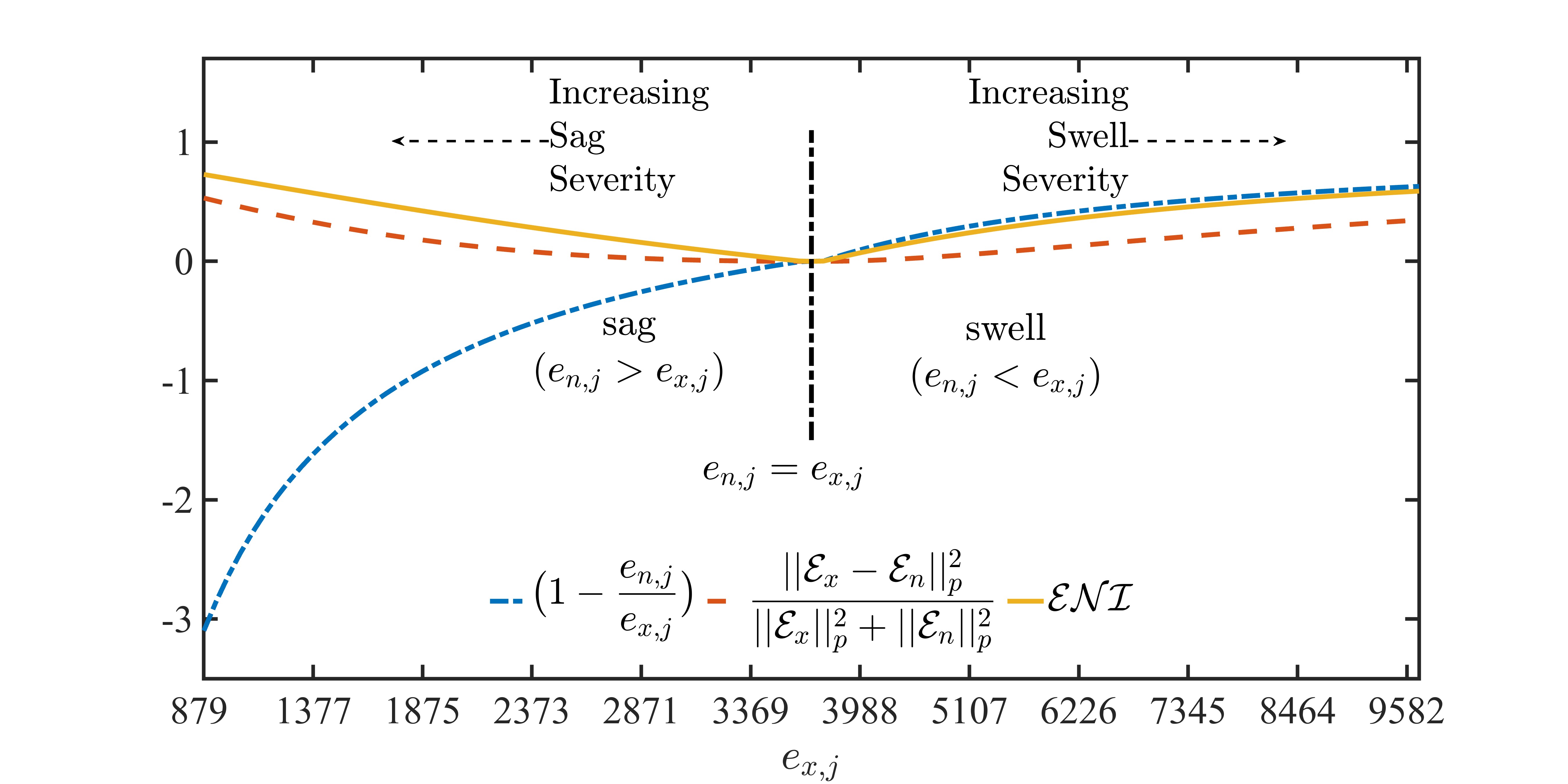}
    \caption{Variations in the monotonicity conditions and $\mathcal{ENI}$ for $sag$ and $swell$ events. $e_{n,j} = 1995$ denotes the nominal  energy. All variations in $sag$ and $swell$ events are contained in the regions ($e_{n,j}>e_{x,j}$) and ($e_{n,j}<e_{x,j}$). } 
    \label{f:proof}
    % \vspace{-2mm}
\end{figure}
%%%=============================================

It is worth emphasizing that the monotonicity condition in either (\ref{eq:cond1})~or~(\ref{eq:cond2}) is satisfied for PQ events. To understand this further, consider all possible severity variations of $sag$ and $swell$ as per the IEEE Std. 1159, \textit{i.e.}, $t_d\in[0.5, \ 30]$; $\alpha\in[0.1,0.9]$ ($sag$) and $\alpha\in[0.1,0.8]$ ($swell$), see Table~\ref{t:events}. For these variations, we examine the monotonicity conditions for the \emph{fundamental} frequency band ($\mathcal{B}_7$, Table~\ref{t:DWT}), since the energy of both $sag$ and $swell$ is predominantly contained in this band, \textit{i.e.}, $e_{x,j} = e_{x,7}$. Fig.~\ref{f:proof} shows the variations in $\mathcal{ENI}$, the monotonicity conditions, and $e_{x,j}$ corresponding to these severity variations. As expected, these results show two distinct regions corresponding to $sag$ and $swell$, which are separated by the nominal event ($e_{n,j}=e_{x,j}$), as follows:
\begin{itemize}
    \item Region-1 ($e_{n,j}>e_{x,j}$): This region includes all $sag$ events and satisfies the monotonicity condition in~(\ref{eq:cond2}). In this region, an \emph{increase} in $e_{x,j}$ indicates a \emph{reduction} in the severity of $sag$. This is reflected by the corresponding reduction in $\mathcal{ENI}$. 
    \item Region-2 ($e_{n,j}<e_{x,j}$): This region includes all $swell$ events and satisfies the monotonicity condition in~(\ref{eq:cond1}). Here, an \emph{increase} in $e_{x,j}$ is associated with an \emph{increase} in the severity of $swell$, which is captured by the increase in $\mathcal{ENI}$.
\end{itemize}
Detailed analysis of various parametric variations and their impacts on the event severity and the proposed indices will be discussed in Section~\ref{s:Res}~and~\ref{s:ResSimEvents}. 

%---------------------------------------------
\vspace{-4mm}
\subsection{Preference Integration}
\label{s:Pref}

In many practical scenarios, the quantification of severity may be subjected to the preferences of a Decision Maker (DM). For instance, the DM may be interested in monitoring transient events in a particular frequency band and, therefore, prefer a higher index sensitivity in the frequency bands of interest. A similar scenario will be investigated later in Section~\ref{subsec:nonmonoSimE}. It is worth emphasizing that such preferences are not integrated into $\mathcal{ENI}$, \textit{i.e.}, the deviation in each frequency band is given equal `importance'. The following two approaches are, therefore, proposed to reformulate/augment $\mathcal{ENI}$ and integrate the preferences of DM: (1) Weighted Index (\textit{reformulation}) and (2) Segmented Index (\textit{augmentation}). %These approaches are discussed in the following:

%---------------------------------------------
\smallskip
\subsubsection{Weighted Index}
\label{s:WNI}
One of the possible approaches to accommodate distinct preferences of the DM is to reformulate the proposed index as the Weighted energy Norm Index ($\mathcal{WNI}$), which is given by,

%-----------------------------------------------------
\vspace{-3mm}
\begin{small}
\begin{align}
\label{eq:edinormalized2}
    \mathcal{WNI} & ={||W_x-W_n||_p}\bigg{/}{\sqrt{||W_x||_p^2 + ||W_n||_p^2}}\\
    \text{with,} \quad W_x &= W\odot \mathcal{E}_{x}, \quad W_n = W\odot\mathcal{E}_{n} \nonumber
\end{align}
\end{small}
%----------------------------------------------------
where, $W = \begin{bmatrix} w_1, & w_2, & \dots, & w_{D+1}\end{bmatrix}$; $w_i\in[0,1]$ denotes the preference weight for the $i^{th}$ frequency band such that $\sum \nolimits_{i=1}^{D+1} w_i = 1$; and $\odot$ denotes the Hadamard product. Such weights can be determined by quantifying the preferences of DM through \emph{multiplicative preference relations} approach~\cite{Zhang:Chen:2004}. For sake of brevity, we discuss this approach through an illustrative example in the following and refer to~\cite{Zhang:Chen:2004} for a detailed treatment on preference formulation and integration:

The multiplicative preference relations approach translates and quantifies the preferences of DM following the steps outlined in Algorithm~\ref{al:pref}. In particular, this approach requires two inputs from the DM: \emph{ordered rankings} ($O$) of the frequency bands and the intensity for such rankings $\mathcal{I}\in[1,9]$.

To understand this process, consider that the DM prefers a higher index sensitivity to transient events. The first step is to encode such a preference in ordered rankings as follows:
%-------------------------------------------------
\begin{equation}
    \label{eq:preforder}
    O = \begin{bmatrix} O_{1}, O_{2}, \dots, O_{(D+1)} \end{bmatrix} = \begin{bmatrix} 1, 2, \dots, (D+1) \end{bmatrix}
\end{equation}
%-------------------------------------------------
These rankings indicate that the order of preference ($O$) is progressively \emph{decreasing} from the \emph{highest} (the first \textit{detail}, $\mathcal{B}_1$, Table~\ref{t:DWT}) to the last frequency band (\textit{approximation}, $\mathcal{B}_{D+1}$). Next, the DM selects the \emph{preference intensity}, $\mathcal{I} \in [1,9]$; where $\mathcal{I} = 1$ denotes \emph{`indifference'} and $\mathcal{I} = 9$ indicates an \emph{`extreme prejudice'} towards a particular frequency band~\cite{Zhang:Chen:2004}.

For simplicity, assume that the number of wavelet decomposition levels $D=2$. Further, let the intensity of preference ordering in (\ref{eq:preforder}) be $9$, \textit{i.e.}, $\mathcal{I} =9$, then the preference weight for each frequency band is determined as follows:

%----------------------------------------------------------
\vspace{-2mm}
\begin{small}  
\begin{align*}
    % \label{eq:prefExample1}
    O & = \begin{bmatrix} O_{1}, & O_{2}, & O_3 \end{bmatrix} = \begin{bmatrix} 1, & 2, & 3 \end{bmatrix}, \quad \text{and} \quad  \mathcal{I} =9, \quad \text{gives,} \\
    & \begin{bmatrix} 
        \tau_{1,1} & \tau_{1,2} & \tau_{1,3}\\ 
        \tau_{2,1} & \tau_{2,2} & \tau_{2,3}\\
        \tau_{3,1} & \tau_{3,2} & \tau_{3,3}
      \end{bmatrix} = 
      \begin{bmatrix} 
        1 & \sqrt{9} & 9\\
        \frac{1}{\sqrt{9}} & 1 & \sqrt{9}\\
        \frac{1}{9} & \frac{1}{\sqrt{9}} & 1
      \end{bmatrix}\\%\nonumber
      W & = \frac{\begin{bmatrix} w_1, & w_2, & w_3 \end{bmatrix}}{\sum_{i=1}^{3} w_i} = \begin{bmatrix} 0.6923,  & 0.2308, & 0.0769 \end{bmatrix}
\end{align*}
\end{small}
%-------------------------------------------------------
Similarly, if the DM is interested in mid-frequency bands, then the preference rankings can be given by:  $O = \begin{bmatrix} 2, & 1, & 3 \end{bmatrix}$, and the corresponding weights are determined to be: $ W = \begin{bmatrix} 0.2308, & 0.6923, & 0.0769 \end{bmatrix}$.

%-------------------------------------------------
%------          Pseudo Code: MTD
%----------------------------------------------
% \vspace{-3mm}
\begin{algorithm}[!t]
    \footnotesize
    \SetKwComment{Comment}{*/ \ \ \ }{}
    % \algorithmfootnote{$\mathcal{I} \in [1,9]$ denotes the intensity of DM's preferences; where $\mathcal{I} = 1$ denotes \emph{`indifference'} and $\mathcal{I} = 9$ denotes an \emph{`extreme prejudice'} of the DM towards a particular frequency band}
     Specify the frequency band rankings, $O = \begin{bmatrix} O_{1}, \dots, O_{(D+1)} \end{bmatrix}$\\
     \BlankLine
     Select the preference intensity, $\mathcal{I} \in [1,9]$\\
    \BlankLine
     \For{i = 1 to $(D+1)$} 
     {  
         \For{j = 1 to $(D+1)$ \nllabel{line:mtd1}}
         {
           $\delta_O = \frac{O_j - O_i}{D}, \qquad
           \tau_{i,j} = \mathcal{I}^{\delta_O}$
         }\nllabel{line:mtd2} % end of j
         $w_i = \displaystyle \Big( \prod \limits_{j=1}^{D+1} \tau_{i,j} \Big)^{1/(D+1)}$ \nllabel{line:mtd3}
     }% end of i
    \Comment*[h] {priority weights}
    $W = \begin{bmatrix} w_1 & w_2 & \dots & w_{D+1} \end{bmatrix}\big/\sum_{i=1}^{D+1} w_i$ 
\caption{Preference Articulation}
\label{al:pref}
\end{algorithm}
% \vspace{-5mm}
% %----------------------------------------------------

%--------------------------------------------------------
% \smallskip
% \vspace{-6mm}
\subsubsection{Segmented Index}
\label{s:LNI}
The rationale behind this approach is to calculate additional index over segmented energy distributions. To this end, the inherent filtering capabilities of DWT are exploited. In particular, the energy distribution of events over $(D+1)$ frequency distribution is segmented to segregate the energy in the fundamental and the remaining frequency bands (see Table~\ref{t:DWT}), as follows:

%-----------------------------------------------------
\vspace{-2mm}
\begin{small}  
\begin{align}
% \small
\label{eq:filtered}
  \mathcal{E}_{x}^{lf} & = \begin{bmatrix} e_{x,D} & e_{x,(D+1)} \end{bmatrix}, \quad  \mathcal{E}_{n}^{lf} & = \begin{bmatrix} e_{n,D} & e_{n,(D+1)} \end{bmatrix}
\end{align}
\end{small}
%-----------------------------------------------------
$\mathcal{E}_{x}^{lf}$ and $\mathcal{E}_{n}^{lf}$ respectively denote the low-pass (\textit{fundamental}) energy distributions of the event $v_{x}$ and the nominal PCC voltage, $v_n(t)$. 

Subsequently, the Low-pass energy Norm Index ($\mathcal{LNI}$) is determined to gauge the deviation only in the lower frequency bands, as follows:

%--------------------------------------------------------
\vspace{-2mm}
\begin{small}  
\begin{equation}
% \small
\label{eq:filteredindex}
      \mathcal{LNI} = {||{\mathcal{E}}_{x}^{lf}-{\mathcal{E}}_{n}^{lf}||_p}\bigg{/}{\sqrt{||{\mathcal{E}}_{x}^{lf}||_p^2 + ||{\mathcal{E}}_{n}^{lf}||_p^2}}
\end{equation}
\end{small}
%-------------------------------------------------------
%%%=============================================
% \vspace{-6mm}
\subsection{Index Selection}
\label{s:indexselect}

Since three different variants of the energy norm index are being proposed, it is pertinent to briefly discuss the selection of an index appropriate to a prevailing scenario. We begin by noting that the proposed indices are designed to work in tandem with a PQ event classifier. The nature of a given event (\textit{e.g.}, $sag$ or $transient$) is assumed to be known \textit{apriori} through a companion classifier, as shown in Fig.~\ref{f:IndexBlock}. 

Further, among all proposed energy norm indices, $\mathcal{ENI}$ is suitable for all scenarios in which the DM does not have any particular preference. In comparison, $\mathcal{WNI}$ allows more flexibility and can be adapted to accommodate distinct preferences by adjusting the preference weights, $W$ (see Algorithm~\ref{al:pref}). It is worth noting that $\mathcal{ENI}$ is a particular case of $\mathcal{WNI}$, which indicates an equal preference to all frequency bands, \textit{i.e.}, $\mathcal{WNI} = \mathcal{ENI}$ when all preference weights are equal $w_1=w_2=\dots=w_{D+1}$.

\begin{figure}[!t]
\centering
% \footnotesize
% \small
\begin{adjustbox}{max width=0.5\textwidth}%max height=0.5\textheight
% \tikzset{grow'=right,level distance=32pt}
% \Tree [.\textbf{Preference?} [.(yes) $\mathcal{WNI}$ ]
%         [.(no) [.\textbf{Simultaneous?} [.(yes) {$\mathcal{LNI}$ and $\mathcal{ENI}$ or $\mathcal{WNI}$} ] [.(no) [.\textbf{Transient?} [.(yes) {$\mathcal{ENI}$ or $\mathcal{WNI}$} ] [.(no) {$\mathcal{ENI}$} ]] ] ] 
%               ].VP ]

\begin{forest}
  for tree={
    grow=east, % direction of growth
    draw, % draw the nodes
    edge={-latex}, % arrow for edges
    rounded corners, % rounded corners for nodes
    align=center, % center-align text
    anchor=west, % anchor nodes to the west
    calign=center, % center-align children
    parent anchor=east, % attach parent to the east
    child anchor=west, % attach child to the west
    l sep+=5pt, % sibling separation
    s sep+=5pt, % level separation
    font=\normalsize, % set font size
    inner sep=3pt, % inner separation
    outer sep=3pt, % outer separation
  }
  [\textbf{Preference?}
    [yes [\(\mathcal{WNI}\)]]
    [no
      [\textbf{Simultaneous?}
        [yes [\(\mathcal{LNI}\) and \(\mathcal{ENI}\) or \(\mathcal{WNI}\)]]
        [no
          [\textbf{Transient?}
            [yes [\(\mathcal{ENI}\) or \(\mathcal{WNI}\)]]
            [no [\(\mathcal{ENI}\)]]
          ]
        ]
      ]
    ]
  ]
\end{forest}

\end{adjustbox}
\caption{Selection of an appropriate energy norm index}
\label{f:tree}
% \vspace{-4mm}
\end{figure}
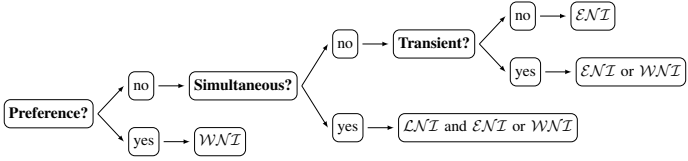

The segmented index, $\mathcal{LNI}$, is particularly designed to estimate the contribution of low frequency events (\textit{e.g.}, $sag$ or $swell$) to the overall severity of simultaneous events (\textit{e.g.}, $sag \ with \ transient$). This index augments the severity information provided by $\mathcal{ENI}$ and is particularly well suited for non-monotonic severity variations in simultaneous events, which will be discussed at length in Section~\ref{subsec:nonmonoSimE}. Fig.~\ref{f:tree} depicts a simple decision tree for selecting an appropriate index. Algorithm~\ref{al:propIndex} outlines the steps involved in the determination of the proposed indices. Note that, in the following, all indices are reported in terms of \textit{percentage} for sake of simplicity.

% \textcolor{blue}{
Further, the proposed index quantifies the deviation of an event from the \emph{reference voltage} using wavelet coefficients. While an ideal sinusoidal voltage could serve as the \emph{reference}, we adopt the \emph{nominal} PCC voltage, $v_n$, as it naturally includes measurement noise and typical magnitude and frequency variations. Accordingly, any substantial \emph{deviation} from $v_n$ indicates the presence of an event. The energy distribution, $\mathcal{E}_n$, over several cycles of $v_n$ (determined by the measurement window) is measured and stored \textit{a priori}, which serves as the \emph{reference energy distribution}.
% }

%%%==================================================
%------------------------------------------------------

%--------------------------------------------------
%             Pseudo Code:Proposed Approach
%--------------------------------------------------
\begin{algorithm}[!t]
    \footnotesize
    \SetKwInOut{Input}{Input}
    \SetKwInOut{Output}{Output}
    \SetKwComment{Comment}{*/ \ \ }{}
    \Input{PQ Event, $v_x (t)$; and nominal energy distribution, $\mathcal{E}_{n}$}
    \Output{PQ Indices: $\mathcal{ENI}$, $\mathcal{LNI}$ and $\mathcal{WNI}$}
    \BlankLine
    \Comment*[h] {Energy distribution of the event}\\
    \BlankLine
    Decompose $v_x(t)$ for $D$ levels\nllabel{line:1}\\
    \BlankLine
    $\mathcal{E}_{x} \leftarrow \varnothing$\\
    \For{k = 1 to $D$} 
        { \BlankLine
           \Comment*[h] {Energy of details}\\%coefficients
           $e_{x,k} = \sum \limits_{i=1}^{N_k} |d_i|^2$, \qquad $\mathcal{E}_{x} \leftarrow \{ \mathcal{E}_{x} \cup e_{x,k} \} $\\
          \BlankLine
          \BlankLine
         \Comment*[h] {Energy of approximation}\\ %coefficients\\
          \If{$k=D$}
             {$e_{x,D+1} = \sum \limits_{i=1}^{N_D} |a_i|^2$, \qquad $\mathcal{E}_{x} \leftarrow \{ \mathcal{E}_{x} \cup e_{x,D+1} \}$\\
             }
        } 
     \BlankLine
     %\Comment*[h] {Calculation of $\mathcal{ENI}$}\\
    %  \BlankLine
     ${\mathcal{ENI}} \leftarrow {||{\mathcal{E}}_{x}-{\mathcal{E}}_{n}||_p}\bigg{/}{\sqrt{||{\mathcal{E}}_{x}||_p^2 + ||{\mathcal{E}}_{n}||_p^2}}$\\
     \BlankLine
     \Comment*[h]{Segmented Indices}\\
     \BlankLine
      \If{segmentation required}
      {
         \BlankLine
         \Comment*[h]{Segment event energy distributions}\\
         \BlankLine
         $\mathcal{E}_{x}^{lf} \leftarrow \begin{bmatrix} e_{x,D}, & e_{x,(D+1)} \end{bmatrix}, \quad \mathcal{E}_{n}^{lf} \leftarrow \begin{bmatrix} e_{n,D}, & e_{n,(D+1)} \end{bmatrix}$
        %  \BlankLine
        %  \Comment*[h] {Determine $\mathcal{LNI}$}\\
         \BlankLine
          $\mathcal{LNI} \leftarrow {||{\mathcal{E}}_{x}^{lf}-{\mathcal{E}}_{n}^{lf}||_p}\bigg{/}{\sqrt{||{\mathcal{E}}_{x}^{lf}||_p^2 + ||{\mathcal{E}}_{n}^{lf}||_p^2}}$\\
      }
      \BlankLine
      \BlankLine
      \Comment*[h]{Accommodating Preferences}\\
      \BlankLine
      \If{preference exists}
      {
          \BlankLine
          Determine preference weights ($W$) as per Algorithm~\ref{al:pref}\\
          \BlankLine
          $W_x \leftarrow W\odot \mathcal{E}_{x}, \qquad W_n \leftarrow W\odot\mathcal{E}_{n}$\\
        %   \BlankLine
        %   \Comment*[h]{Weighted Index}\\
          \BlankLine
          $\mathcal{WNI} \leftarrow {||W_x-W_n||_p}\bigg{/}{\sqrt{||W_x||_p^2 + ||W_n||_p^2}}$
      }

\caption{Determination of Proposed Index}
\label{al:propIndex}
\end{algorithm}
% \vspace{-2mm}
%--------------------------------------------------

\section{Evaluation over Single Non-stationary Events}
\label{s:Res}

% In the following, the efficacy of the proposed index, $\mathcal{ENI}$, is rigorously determined. To this end, 
% The evaluation of $\mathcal{ENI}$ is being carried out in three steps. First, major non-stationary events including \textit{sag, swell, interruption} and \textit{oscillatory transient} are being considered in Section~\ref{s:ResSgItSwl} and~\ref{s:ResOscTr}. The effects of multi-stage \textit{sag} events are discussed in Section~\ref{s:ResMultiStage}. Finally, $\mathcal{ENI}$ is further validated by considering the experimental \textit{sag} event dataset in Section~\ref{s:ResRealSag}. %The simultaneous non-stationary events are considered next in Section~\ref{s:ResSimEvents}.

%--------------------------------------------------------
\subsection{Sag, Interruption and Swell}
\label{s:ResSgItSwl}

The first part of this investigation focuses on fundamental frequency non-stationary events including \textit{sag}, \textit{interruption}, and \textit{swell}. It is easy to follow that the severity of these events, denoted by $\mathcal{S}_{f}$, is a function of the  corresponding \textit{magnitude deviation} ($\alpha$) and \textit{duration} ($t_d$), \textit{i.e.},

%------------------------------------------
\vspace{-2mm}
\begin{small}    
\begin{equation}
    \label{eq:lowseverity}
     \mathcal{S}_{f} = \mathcal{F}(\alpha,t_d)   
\end{equation}
\end{small}
%------------------------------------------
An increase in $\alpha$ and/or $t_d$ translates into an increase in severity of these classes of events; \textit{see} Table~\ref{t:events} and~\cite{IEEE:1159}.

Let $\mathcal{N}$ denote a group of a particular class of low frequency events with the following parametric variations:
%------------------------------------------
\begin{equation}
    \label{eq:lfvar}
     (\alpha_i \le \alpha_{i+1}) \wedge (t_{d,i} \le t_{d,i+1}), \qquad i\in [1,\mathcal{N}-1]
\end{equation}
%------------------------------------------
%%---- Index Plots-------------------------------------------
\begin{figure}[!b]
\centering
\small
\begin{subfigure}{.23\textwidth}
  \includegraphics[width=\textwidth]{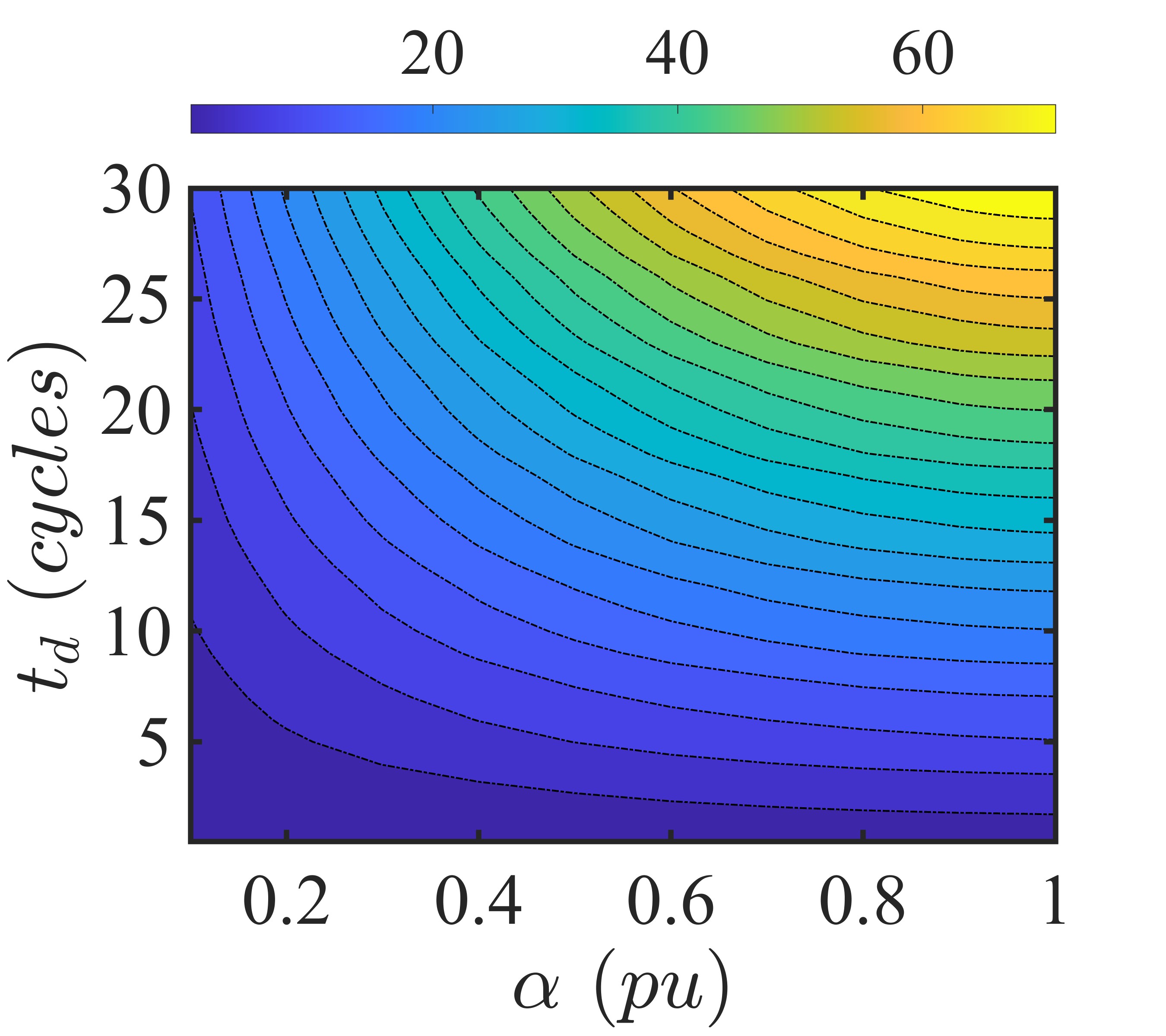}
  \caption{Sags + Interruptions}
  \label{f:sag}
\end{subfigure}
\hfill
% \hspace{0.1\textwidth}
\begin{subfigure}{.23\textwidth}
  \includegraphics[width=\textwidth]{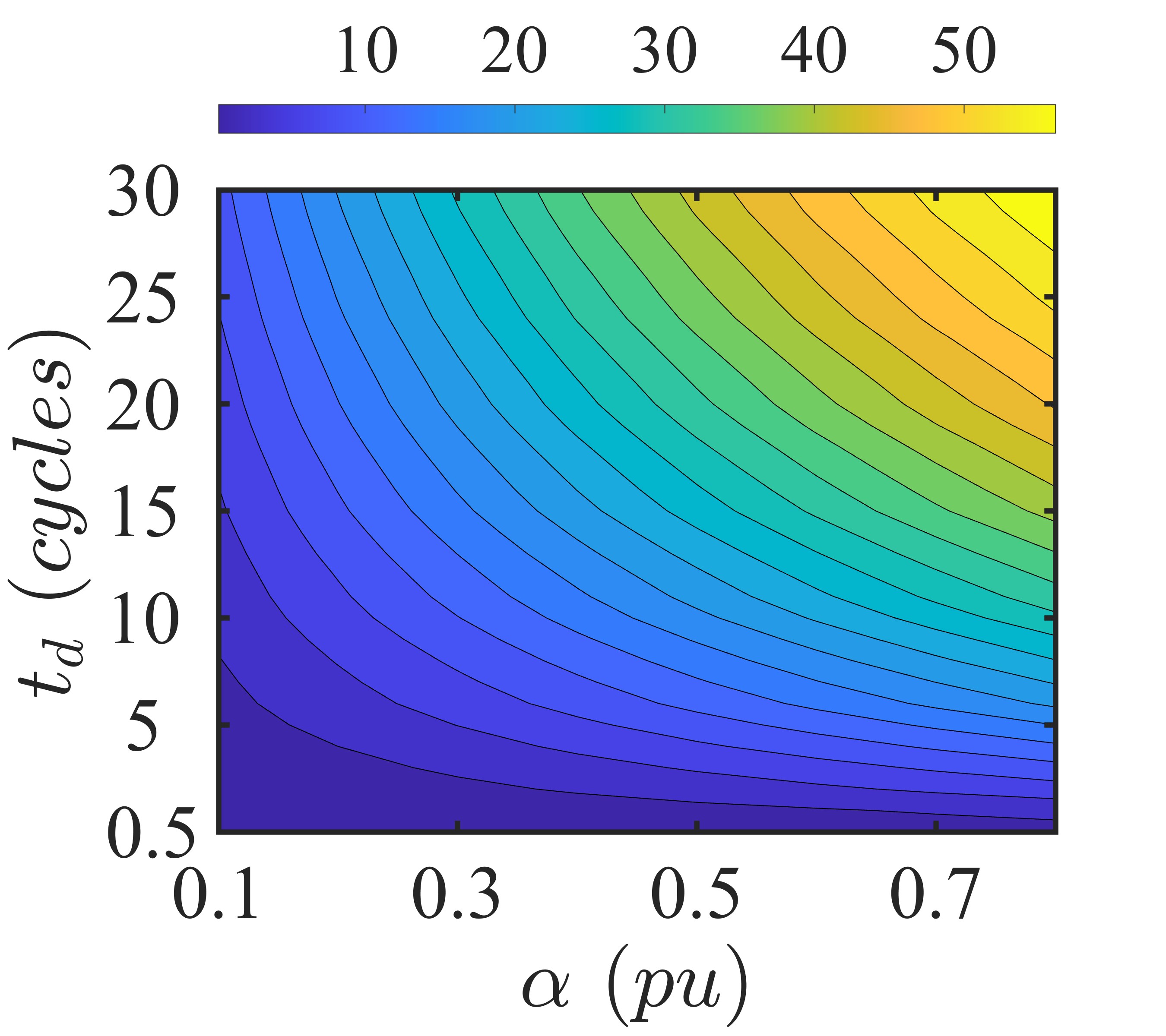}
  \caption{Swells}
  \label{f:swell}
\end{subfigure}
\caption{Variations in $\mathcal{ENI}$ for Sags, Interruptions, and Swells. $\mathcal{ENI}$ is determined using $\ell_2-$norm.}
\label{f:enilow}
\vspace{-5mm}
\end{figure}
% \FloatBarrier
%--------------------------------------------------------
The severity of these events can then be ranked as follows:

%------------------------------------------
\vspace{-2mm}
\begin{small}   
\begin{equation}
    \mathcal{S}_{f,1} \le \mathcal{S}_{f,2} \le \dots \le \mathcal{S}_{f,\mathcal{N}}
\end{equation}
\end{small}
%------------------------------------------
It is, therefore, safe to infer that an increase in $\alpha$ and/or $t_d$ should reflect in a monotonic increase of a PQ index and vice-versa. This forms the basis of the remainder of this test.

% %--------------------------------------------
% \begin{wrapfigure}[13]{r}{0.24\textwidth}
%    \centering
%    \includegraphics[width=0.239\textwidth]{surf_EDI_L2_norm_Swl}
%    \caption{Variations in $\mathcal{ENI}$ for $swell$.}
%   \label{f:swell}
% \end{wrapfigure}
% %-----------------------------------------
A total of 250 events with a distinct degree of severity are generated for each class of low frequency disturbance to evaluate the efficacy of $\mathcal{ENI}$. These events are generated as per the scenario given by~(\ref{eq:lfvar}) and with the following variations in the parameters: the \textit{duration} ($t_d$) of all events is varied from $\frac{1}{2} \ cycle$ to $30 \ cycles$, while the \textit{magnitude} ($\alpha$) is varied in the range of $[0.1, 0.9]$ (\textit{sag}), $[0.91, 1]$ (\textit{interruption}) and $[0.1, 0.8]$ (\textit{swell}), following the guidelines of the IEEE Std. 1159~\cite{IEEE:1159}. For each event, $\mathcal{ENI}$ is determined following the steps outlined in Algorithm~\ref{al:propIndex}. The outcomes of this test are depicted in Fig.~\ref{f:sag} (\textit{sag} and \textit{interruption}) and Fig.~\ref{f:swell} (\textit{swell}). These results clearly show a monotonic increase in $\mathcal{ENI}$ with increase in $\alpha$ and/or $t_d$. %for all class the events.% These results, thus, establish the ability of $\mathcal{ENI}$ to quantify the deviation from the reference PCC voltage in a monotonic fashion. 

%--------------------------------------------------------
\vspace{-3.5mm}
\subsection{Oscillatory Transients}
\label{s:ResOscTr}

The severity of oscillatory transients is primarily dependent on the following three parameters: \textit{magnitude} ($\beta$), \textit{decay time constant} ($\gamma$) and \textit{frequency} ($f_{tr}$), \textit{see} Table~\ref{t:events} and~\cite{IEEE:1159}. Accordingly, the severity of transient events can be thought of as a function of these parameters:

%------------------------------------------
\begin{small}
    \begin{equation}
    \label{eq:transseverity}
    \mathcal{S}_{tr} = \mathcal{F}_{tr}(\beta, \ \gamma, \ f_{tr})
\end{equation}
\end{small}
%------------------------------------------
Consider a group of total $\mathcal{N}$ number of transient events of a particular frequency ($f_{tr}$) with the following parametric variations:

%------------------------------------------
\begin{small}
\begin{equation}
    \label{eq:transvar}
    (\beta_i \le \beta_{i+1}) \wedge (\gamma_i \le \gamma_{i+1}), \qquad i\in [1,\mathcal{N}-1]
\end{equation}
\end{small}
%------------------------------------------
Note that $\gamma$ is always set to less than zero ($\gamma<0$) to emulate the exponential transient decay (see Table \ref{t:events}). An increase in $\gamma$, therefore, translates into a longer duration of the transient. Accordingly, an increase in $\beta$ and/or $\gamma$ leads to an increase in the severity of transient events. Hence, for the parametric variations given by~(\ref{eq:transvar}), the events can be arranged in the increasing order of severity,

%------------------------------------------
\vspace{-2mm}
\begin{small}   
\begin{equation}
    \mathcal{S}_{tr,1} \le \mathcal{S}_{tr,2} \le \dots \le \mathcal{S}_{tr,\mathcal{N}}
\end{equation}
\end{small}
%------------------------------------------

\begin{figure}[!t]
\begin{subfigure}{.23\textwidth}
  \includegraphics[width=\textwidth]{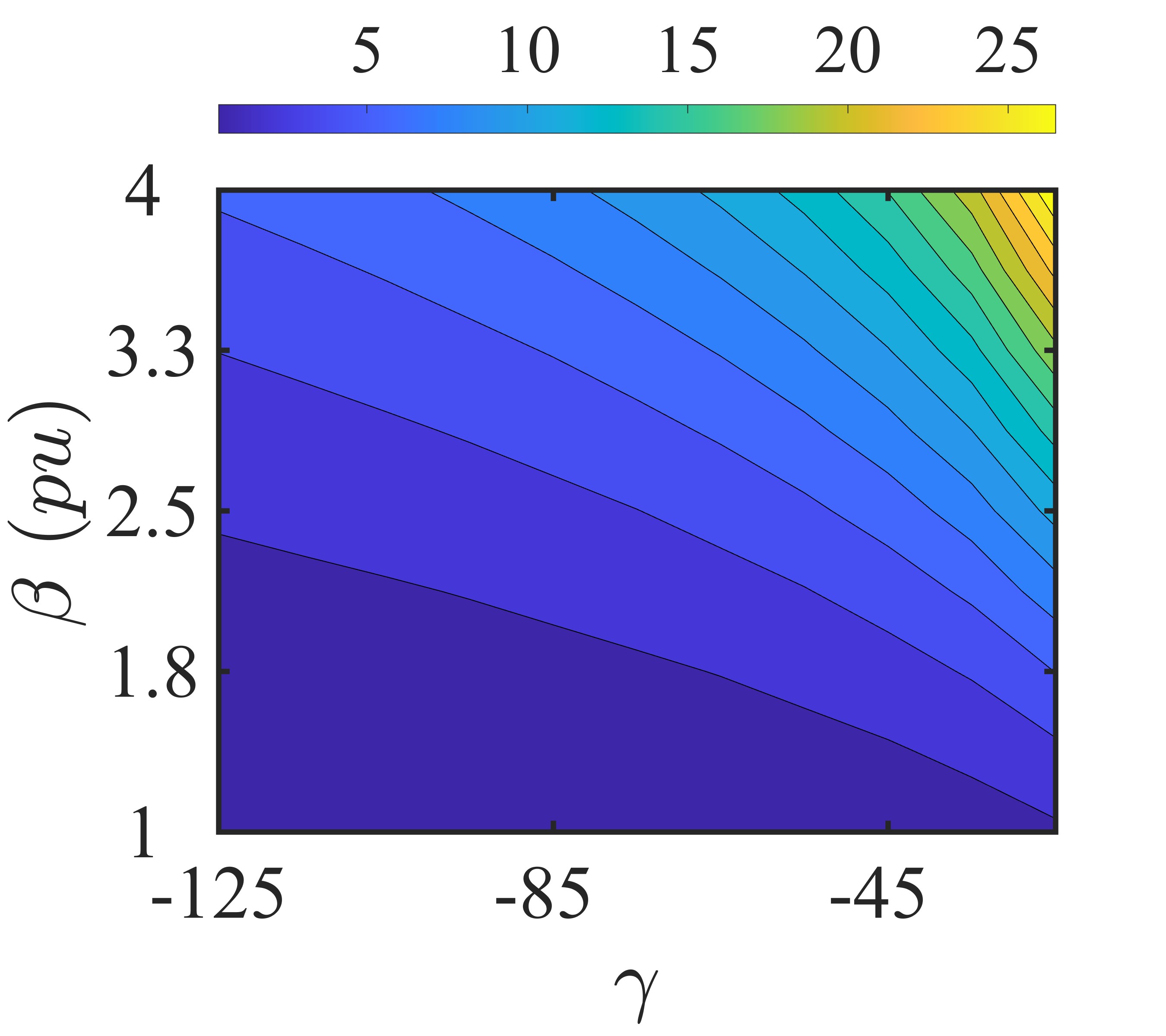}
  \caption{$\mathcal{ENI}$}
  \label{f:tranENI}
\end{subfigure}%
\hfill
\begin{subfigure}{.23\textwidth}
  \includegraphics[width=\textwidth]{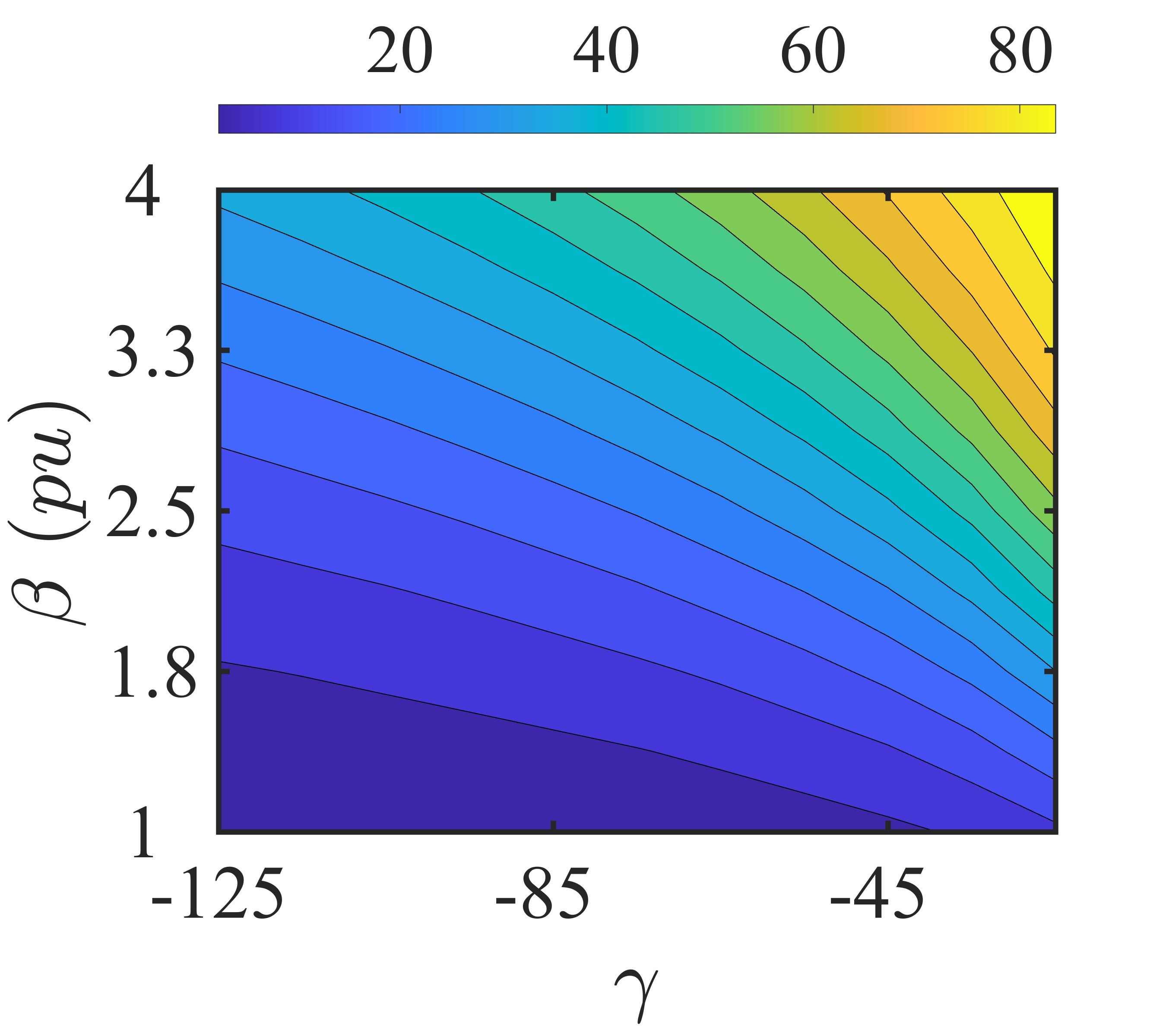}
  \caption{$\mathcal{WNI}$}
  \label{f:tranWNI}
\end{subfigure}
\caption{Variations in energy norm indices for transients with $f_{tr}=4 \ kHz$.}
\label{f:restran}
% \vspace{-5.7mm}
\end{figure}

This part of the investigation aims to determine whether the proposed norm indices can capture monotonic variations in $\mathcal{S}_{tr}$ in such scenarios. To this end, for a particular transient frequency $f_{tr}$, a total of 100 events with a distinct degree of severity are generated as per~(\ref{eq:transvar}) with the following variations in the parameters: $\beta \in [1,4]$ and $\gamma \in [-125,-25]$. The frequency of transients is varied from $400 \ Hz$ to $4000 \ kHz$. The severity of these events is determined following the step outlined in Algorithm~\ref{al:propIndex}. The consequent variations in the proposed indices are shown in Fig.~\ref{f:restran} for transient frequency of $f_{tr}=4000 \ Hz$. The results obtained for the other frequencies do not differ significantly and therefore are omitted here.
%%---- Index Plots------------------------
%-----------------------------------------

The results in Fig.~\ref{f:tranENI} clearly indicate that the variations in $\mathcal{ENI}$ are monotonic for the parametric variations given by~(\ref{eq:transvar}). However, it is worth noting that $\mathcal{ENI}$ is relatively less sensitive to transient variations. For instance, $\mathcal{ENI}$ varies from $0.28\%$ (minimum) to $73\%$ (maximum) for the sag events (see~Fig.~\ref{f:sag}). In contrast, the maximum value of $\mathcal{ENI}$ for the transient is limited to approximately $30\%$ (see~Fig.~\ref{f:tranENI}). This decrease in sensitivity can be explained by the fact that $\mathcal{ENI}$ is being calculated over a non-overlapping measurement window of 40 cycles. The duration, and thereby the energy, of the transient event is relatively small compared to the fundamental frequency component over such duration of the measurement cycle. This can be alleviated by reducing the measurement window, say from $40$ to $10 \ cycles$. 

The other alternative is to use the weighted energy norm index, $\mathcal{WNI}$. In particular, the preferences are adjusted to be more sensitive to the higher frequency bands (see Section~\ref{s:WNI}) to increase the sensitivity of $\mathcal{WNI}$ towards transients. The weighted index is determined following the steps outlined in Algorithms~\ref{al:pref} and~\ref{al:propIndex}. Fig.~\ref{f:tranWNI} shows the consequent variations in $\mathcal{WNI}$ for the same parametric variations in transient events. The results show that $\mathcal{WNI}$ is both monotonic and sensitive to transient events; the minimum and maximum values of $\mathcal{WNI}$ are $2.5\%$ and $90\%$, respectively.

% In this study, a total of 1700 transient instances of distinct severity are being generated with the following variations in the parameters: $\beta \in [1,4]$, $\gamma \in [50,500]$ and $f_{tr} \in [0.4, 4] \ kHz$. The severity of these event is determined following the step outlined in Algorithm~\ref{al:propIndex}. The consequent variations in $\mathcal{ENI}$ are shown in Fig.~\ref{f:tran1} ($f_{tr}=800 \ Hz$), Fig.~\ref{f:tran2} ($f_{tr}=2000 \ Hz$) and Fig.~\ref{f:tran3} ($f_{tr}=4000 \ Hz$). The outcome of this test are also shown for few combinations of $\{ \beta, \gamma\}$ in Table~\ref{t:Tr}, for further clarity. The results obtained for the remaining parameter combinations do not differ significantly, and therefore are not shown here. The results in Fig.~\ref{f:tran1}-\ref{f:tran3} and Table~\ref{t:Tr} clearly show that $\mathcal{ENI}$ can effectively quantify the severity of oscillatory transients. This further establishes the monotonic property of $\mathcal{ENI}$.

% the monotonicity of the proposed index $\mathcal{EDI}$ can monotonically quantify the severity of oscillatory transients.

%%%====================================================
\section{Evaluation over Simultaneous Events}
\label{s:ResSimEvents}

This part of the study considers simultaneous events, which are often relatively difficult to quantify due to the distinct nature of the constituent events. In particular, two classes of simultaneous events are being considered: \textit{sag with transients} and \textit{swell with transients}. Both classes of simultaneous events are synthesized using the parametric models in Table~\ref{t:events}. The severity of these events is controlled by the following five parameters: $\alpha$ (\textit{magnitude of sag/swell}), $t_d = (t_2-t_1)$ (\textit{duration of sag/swell}), $\beta$ (\textit{peak magnitude of transient}), $\gamma$ (\textit{decay time constant of transient}) and $f_{tr}$ (\textit{frequency of transient}).

The overall severity of the simultaneous events can be thought of as a function of the severity of both low (\textit{sag/swell}) and high frequency (\textit{transient}) constituent events:

%------------------------------------------
\vspace{-2mm}
\begin{small}   
\begin{align}
% \small
    \label{eq:simsev}
      \mathcal{S}_o & = \mathcal{G}(\mathcal{S}_{f}, \mathcal{S}_{tr}), \\
    \text{where, } \mathcal{S}_{f} & = \mathcal{F}(\alpha,t_d) \quad \mathcal{S}_{tr} = \mathcal{F}_{tr}(\beta,\gamma, f_{tr}) \nonumber
\end{align}
\end{small}
%------------------------------------------
Further, as discussed earlier, $\mathcal{S}_f$ monotonously increases with an increase in $\alpha$ and/or $t_d$. Similarly, for a particular transient frequency $f_{tr}$, $\mathcal{S}_{tr}$ is monotonically dependent on magnitude ($\beta$) and decay constant ($\gamma$).

It is worth emphasizing that based on the variations in the severity of constituent events, there exist four distinct possibilities between the overall ($\mathcal{S}_o$) and constituent severities ($\mathcal{S}_f$ and $\mathcal{S}_{tr}$), as shown in Table~\ref{t:Rel}. These possibilities can be grouped into either \textit{monotonic} or \textit{non-monotonic} scenarios, as will be discussed in the following subsections. 

%%%====================================================
\begin{table}[!t]
    \centering
	\caption{Relationship between Overall and Constituent Severities}
	\label{t:Rel}
	\begin{adjustbox}{max width=0.46\textwidth} 
	\begin{threeparttable}
	\small 
   
    \begin{tabular}{c|cc}
    \cmidrule{2-3}  \backslashbox{\makecell{\textbf{Severity of}\\\textbf{Transient} \boldmath$\ \mathcal{S}_{tr}$}}{\makecell{\textbf{Severity}\\\textbf{Sag/Swell}$ \ \mathcal{S}_f$}} & \makecell{\textbf{Increase}\\\textbf{in} \boldmath$\mathcal{S}_{f}$,\\[0.7ex] \boldmath$(\alpha_i \le \alpha_{i+1})\wedge$\\ \boldmath$(t_{d,i} \le t_{d,i+1})$} & \makecell{\textbf{Decrease}\\\textbf{in} \boldmath$\mathcal{S}_{f}$,\\[0.7ex] \boldmath$(\alpha_i \ge \alpha_{i+1})\wedge$\\\boldmath$ (t_{d,i} \ge t_{d,i+1})$} \\
    \midrule 
    \makecell{\textbf{Increase in} \boldmath$\mathcal{S}_{tr}$, \\[0.7ex] \boldmath$(\beta_i \le \beta_{i+1}) \wedge (\gamma_i \le \gamma_{i+1})$} & \makecell{\textit{monotonic}\\\textit{increase in}\\ $\mathcal{S}_o$} & \textit{non-monotonic}, $\mathcal{S}_o$\\[0.5ex]
    \midrule
    \makecell{\textbf{Decrease in} \boldmath$\mathcal{S}_{tr}$,\\[0.7ex] \boldmath$(\beta_i \ge \beta_{i+1}) \wedge (\gamma_i \ge \gamma_{i+1})$} & \textit{non-monotonic}, $\mathcal{S}_o$ & \makecell{\textit{monotonic}\\\textit{decrease in}\\$\mathcal{S}_o$} \\
    \bottomrule
    \end{tabular}%
   
    % \begin{tablenotes}
    %     \small
    %     \item $^\dagger$
    % \end{tablenotes}
    \end{threeparttable}
	\end{adjustbox}
	% \vspace{-5mm}
\end{table}
%%%===================================================
%%%===========================================
\begin{figure}[!t]
    \centering
    \begin{subfigure}{0.238\textwidth}
        \includegraphics[width=0.9\textwidth]{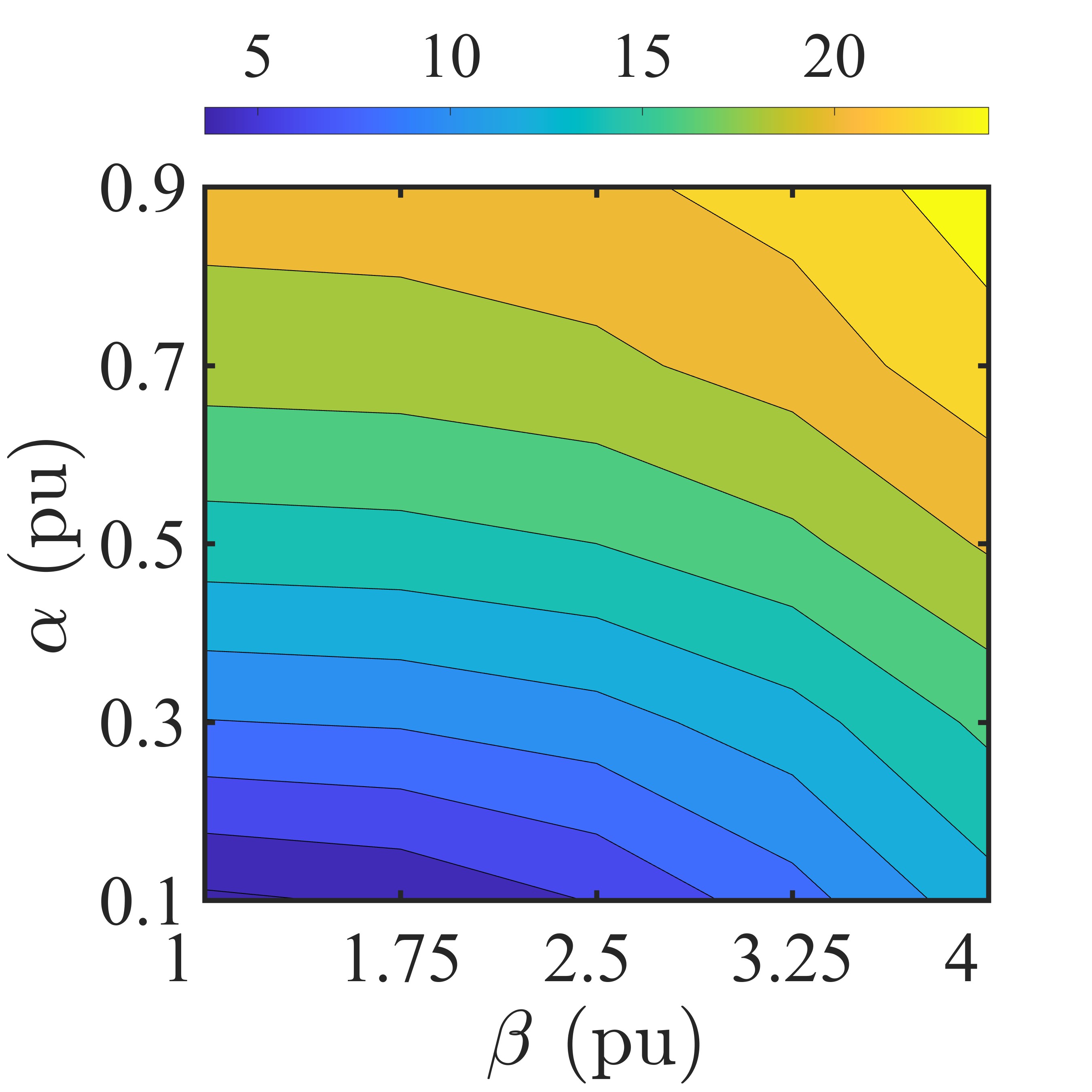}
        \caption{with $\{t_d, \gamma \} = \{10 \ cycles, -55\}$}
        \label{f:sim_ab}
    \end{subfigure}
    \hfill
    \begin{subfigure}{0.238\textwidth}
        \includegraphics[width=0.9\textwidth]{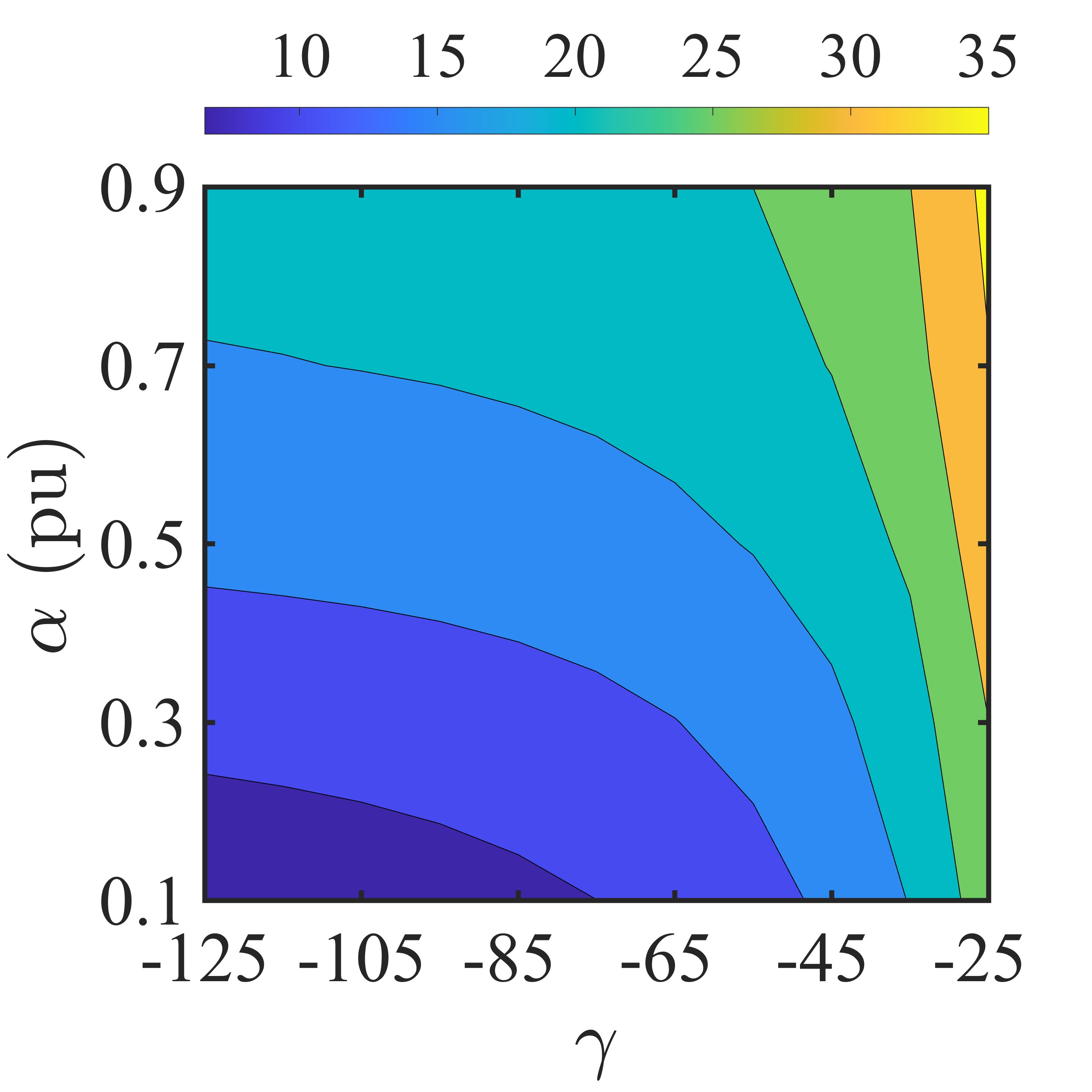}
        \caption{with $\{t_d, \beta \} = \{10 \ cycles, 4 \}$}
        \label{f:sim_ag}
    \end{subfigure}
    \hfill
    \begin{subfigure}{0.238\textwidth}
        \includegraphics[width=0.9\textwidth]{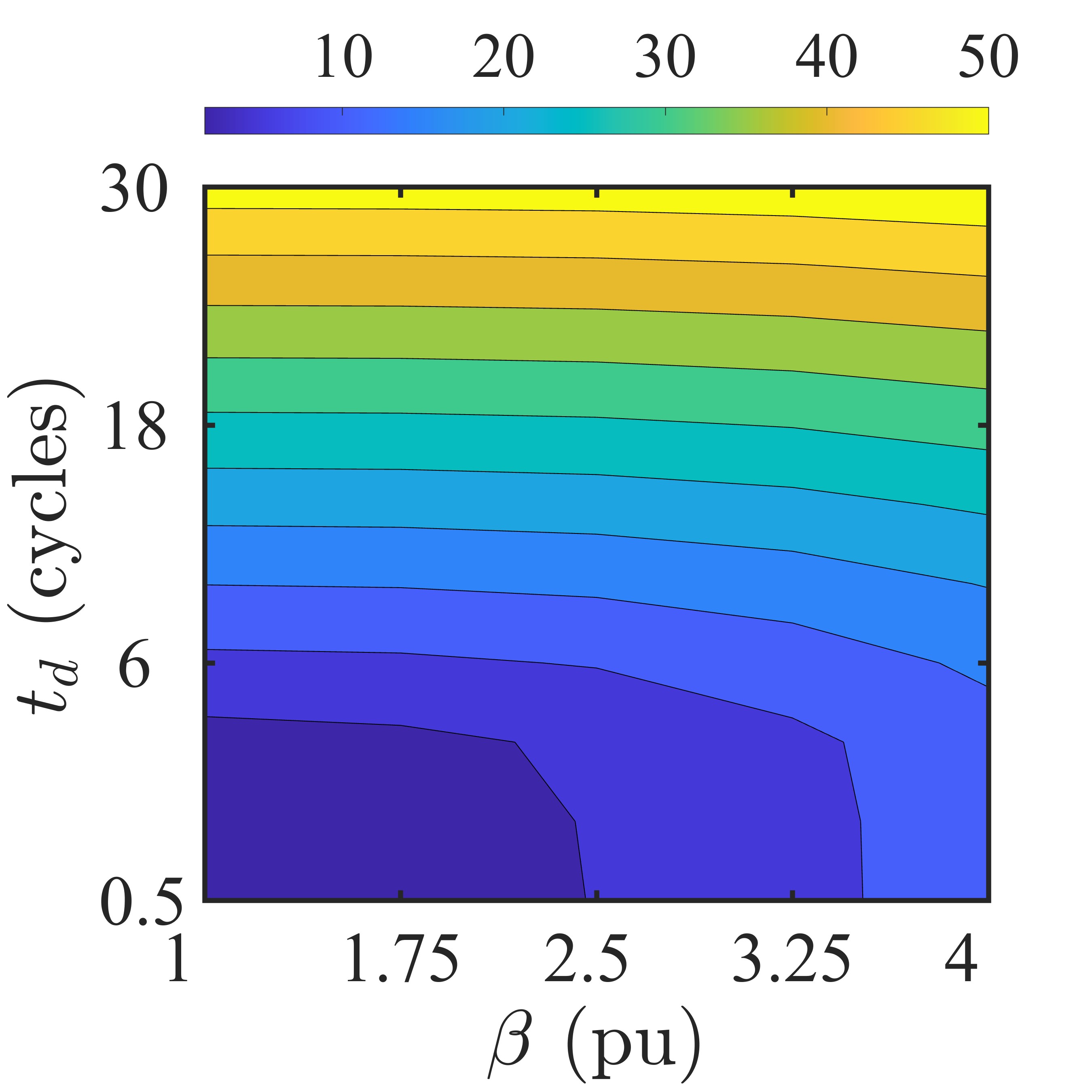}%swT_EDI.jpg
        \caption{with $\{\alpha, \gamma \} = \{0.5, -55\}$}
        \label{f:sim_tb}
    \end{subfigure}
    \hfill
    \begin{subfigure}{0.238\textwidth}
        \includegraphics[width=0.9\textwidth]{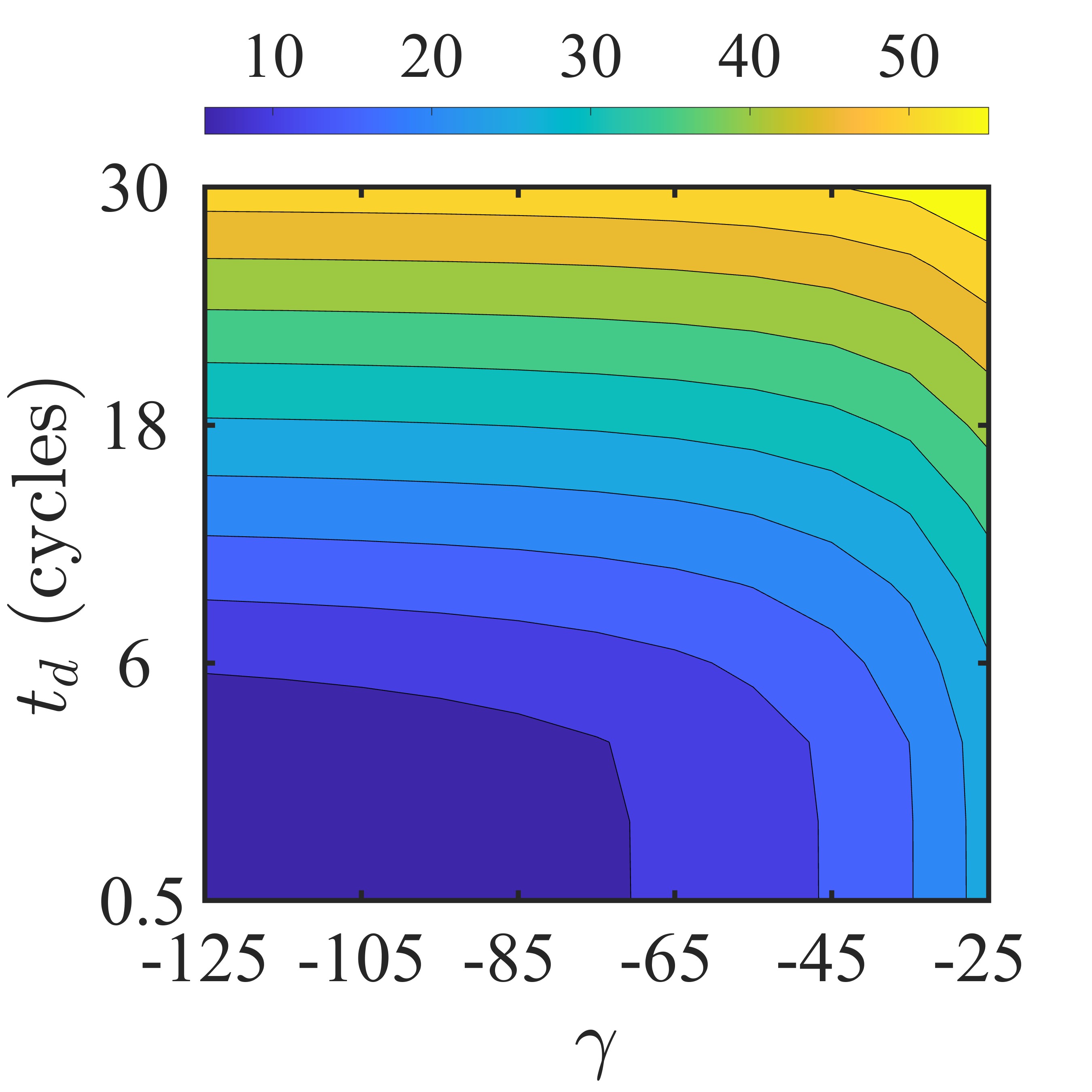}%
        \caption{with $\{\alpha, \beta \} = \{0.5, 4 \}$}
        \label{f:sim_tg}
    \end{subfigure}
    \caption{Variations in $\mathcal{ENI}$ for sag with transients under monotonic scenario.} % The frequency of oscillatory transients is fixed to $4000 \ Hz$. %and decay time constant is set to 50, \textit{i.e.}, $f_{tr} = 800 \ Hz$ and $\gamma = 50$
    \label{f:simE}
% \vspace{-5.6mm}
\end{figure}
%%%=============================================
% %-----------------------------------------------------
\vspace{-3mm}
\subsection{Monotonic Scenario}
\label{subsec:monoSimE}

In this scenario, the severity of both the fundamental ($\mathcal{S}_f$) and the higher frequency ($\mathcal{S}_{tr}$) constituent events monotonically increases or decreases. For a particular transient frequency ($f_{tr}$), this scenario can be represented by either of the following parametric variations (see~Table~\ref{t:events}~and~\ref{t:Rel}): 

%--------------------------------------------------
\vspace{-2mm}
\begin{small}
\begin{align}
    \label{eq:simparvarmonoup}
    (\alpha_i \le \alpha_{i+1}) \wedge (t_{d,i} \le t_{d,i+1}) \wedge  (\beta_i \le \beta_{i+1}) \wedge (\gamma_i \le \gamma_{i+1})\\
    \label{eq:simparvarmonodown}
    (\alpha_i \ge \alpha_{i+1}) \wedge (t_{d,i} \ge t_{d,i+1}) \wedge  (\beta_i \ge \beta_{i+1}) \wedge (\gamma_i \ge \gamma_{i+1})
\end{align}
\end{small}
%--------------------------------------------------
where, $i \in [1,\mathcal{N}-1]$ and $\mathcal{N}$ denotes the total number of simultaneous events under consideration.

For the parametric variations given by~(\ref{eq:simparvarmonoup}), $\mathcal{S}_f$ and $\mathcal{S}_{tr}$ monotonically increase. Hence, the overall severity of simultaneous events can be ranked as,

%------------------------------------------
\vspace{-2mm}
\begin{small}   
\begin{equation}
% \small
    \mathcal{S}_{o,1} \le \mathcal{S}_{o,2} \le \dots \le \mathcal{S}_{o,\mathcal{N}}
\end{equation}
\end{small}
%------------------------------------------

Similarly, a monotonic decrease in $\mathcal{S}_o$ is obtained with the variations given by~(\ref{eq:simparvarmonodown}), as both $\mathcal{S}_f$ and $\mathcal{S}_{tr}$ monotonically decrease in this scenario. 

These scenarios are emulated by varying the following four parameters: $\alpha \in [0.1 \ pu, \ 0.9 \ pu]$, $t_d \in [\frac{1}{2} \ cycle, \ 30 \ cycles]$, $\beta\in[1 \ pu, \ 4 \ pu]$, and $\gamma\in[-125,-25]$. Without the loss of generality, the frequency of the transient events is fixed to $4000\ Hz$ to emulate the transient arising from switching of capacitor banks. Further, white Gaussian noise of $45 \ dB$ SNR is added to each event to simulate the measurement noise. A total of $1250$ events are generated through these parametric variations for both sag with transients and swell with transients.

The proposed index, $\mathcal{ENI}$, is determined for all events following the steps outlined in Algorithm~\ref{al:propIndex}. The objective here is to determine the impact of variations in parameters, $\{\alpha, t_d, \beta, \gamma \}$, on $\mathcal{ENI}$. The results of this evaluation are shown in Fig.~\ref{f:simE}. For ease of interpretation, these results are discussed in four stages; each stage demonstrates the impact of variations in a particular pair of parameters, while fixing the remaining two parameters, as follows:  
\begin{itemize}[noitemsep, left=0pt]
    \item \textit{Impact of $\{\alpha, \beta\}$}: Fig.~\ref{f:sim_ab} shows the influence of $\alpha$ and $\beta$ on $\mathcal{ENI}$, when the time duration of both low and high frequency constituent events is fixed, \textit{i.e.}, $\{t_d, \gamma \} = \{10 \ cycles, -50\}$.
    \item \textit{Impact of $\{\alpha, \gamma\}$}: Fig.~\ref{f:sim_ag} shows the variations in $\mathcal{ENI}$ with respect to $\alpha$ and $\gamma$, while the sag duration and the transient magnitude are fixed, \textit{i.e.}, $\{t_d, \beta \} = \{10 \ cycles, 4 \ pu\}$.
    \item \textit{Impact of $\{t_d, \beta\}$}: Fig.~\ref{f:sim_tb} depicts the impact of $t_d$ and $\beta$, while $\alpha$ and $\gamma$ are respectively set to $0.5 \ pu$ and $-50$.
    \item \textit{Impact of $\{t_d, \gamma\}$}: Fig.~\ref{f:sim_tg} shows the influence of the duration of the constituent events ($t_d$ and $\gamma$), while their magnitudes are fixed, \textit{i.e.}, $\alpha=0.5 \ pu$ and $\beta = 4 \ pu$.
\end{itemize}

The results in Fig.~\ref{f:simE} clearly indicate a monotone behavior of $\mathcal{ENI}$ for all possible pairs of parameters of sag with transient. Note that similar behavior is also observed for swell with transient events; hence the corresponding results are omitted here for brevity. %Similar observations can also be inferred, from Table \ref{t:SimE}, for \textit{swell with transients} events. It is thus clear that $\mathcal{EDI}$ can easily quantify severity arising from variation in any parameter of the simultaneous events. 
%%%===================================================
\begin{table*}[!t]
    \centering
	\caption{Indices for Sag with Transients (Non-monotonic Scenario)}
	\label{t:simNonmono1}
	\begin{adjustbox}{max width=0.7\textwidth} 
	\begin{threeparttable}
	\small 
   
    \begin{tabular}{c|cc|c|cc|cc}
    \toprule
    \textbf{Remark} & \makecell{\textbf{Parameters,}\\[0.5ex]\boldmath$\{\alpha, \ t_d, \  \beta, \ \gamma\}^\dagger$} & \makecell{\textbf{Known}\\\textbf{Severity}} & \makecell{\textbf{Overall},\\[0.5ex]\boldmath{$\mathcal{ENI}$}} & \makecell{\boldmath{$\mathcal{LNI}$}} & \makecell{\boldmath{$\mathcal{WNI}$}} & \textbf{Observation} & \makecell{\textbf{Estimated}\\\textbf{Severity}} \\[1ex]
    \midrule
    \multirow{2}{*}{\makecell{Event\\Pair-1}} & \makecell{$\{0.5, \ 10, \ 1, \ -55\}$} & \makecell{$\mathcal{S}_{f,1} > \mathcal{S}_{f,2}$}& 15.10 & 15.08 & 15.70 & $\mathcal{LNI}_1>\mathcal{LNI}_2$ & $\rm{Sg.-1 > Sg.-2}$\\[1ex]
    & \makecell{$\{0.3, \ 10, \ 4, \ -55\}$} & $\mathcal{S}_{tr,1} < \mathcal{S}_{tr,2}$ & 16.49 & 9.85  & 55.82 & $\mathcal{WNI}_1<\mathcal{WNI}_2$ &$\rm{Tr.-1 < Tr.-2}$ \\[1ex]
    \midrule
    \multirow{2}{*}{\makecell{Event\\Pair-2}} & $\{0.3, \ 6, \ 3.25,\ -75\}$ & \makecell{$\mathcal{S}_{f,1} < \mathcal{S}_{f,2}$} & 8.60  & 5.88  & 30.24 & $\mathcal{LNI}_1<\mathcal{LNI}_2$ & $\rm{Sg.-1 < Sg.-2}$\\[1ex]
    & $\{0.5, \ 6, \ 1, \ -125\}$ & $\mathcal{S}_{tr,1} > \mathcal{S}_{tr,2}$  & 8.91  & 8.91  & 9.10 & $\mathcal{WNI}_1>\mathcal{WNI}_2$ &$\rm{Tr.-1 > Tr.-2}$ \\
    \bottomrule
    \end{tabular}%
    \begin{tablenotes}
        \small
        \item $^\dagger$ $t_d$ gives sag duration in cycles; the frequency of oscillatory transients is fixed to $f_{tr} = 2 \ kHz$.
    \end{tablenotes}
    \end{threeparttable}
	\end{adjustbox}
	\vspace{-4mm}
\end{table*}
%%%=============================================
%%%=============================================
\vspace{-3mm}
\subsection{Non-monotonic Scenario}
\label{subsec:nonmonoSimE}

Under this scenario, the changes in the severity of constituent events are not in agreement. For instance, consider the following scenario:

%--------------------------------------------------
\vspace{-2mm}
\begin{small}
\begin{equation}
% \small
    \label{eq:simparvarnonmono1}
     (\alpha_i \ge \alpha_{i+1}) \wedge (t_{d,i} \ge t_{d,i+1}) \wedge (\beta_i \le \beta_{i+1}) \wedge (\gamma_i \le \gamma_{i+1})
\end{equation}
\end{small}
%--------------------------------------------------
here, $\mathcal{S}_f$ decreases while $\mathcal{S}_{tr}$ increases over the given group of events. Similarly, consider the other parameter scenario where $\mathcal{S}_f$ is increasing while $\mathcal{S}_{tr}$ decreases:

%--------------------------------------------------
\vspace{-2mm}
\begin{small}
\begin{equation}
% \small
    \label{eq:simparvarnonmono2}
    (\alpha_i \le \alpha_{i+1}) \wedge
    (t_{d,i} \le t_{d,i+1}) \wedge (\beta_i \ge \beta_{i+1}) \wedge (\gamma_i \ge \gamma_{i+1})
\end{equation}
\end{small}
%--------------------------------------------------

The quantification of overall severity, $\mathcal{S}_o$, under such non-monotonic scenarios is not straightforward as it can often be subjective depending on the preference of the Decision Maker (DM). For instance, the impacts of higher frequency events are often considered relatively more severe~\cite{heydt:1998}; under such a scenario, DM may be biased to make the index more sensitive to transients. 

To simulate this scenario, a pair of simultaneous events is synthesized corresponding to the parameter variations in~(\ref{eq:simparvarnonmono1})~and~(\ref{eq:simparvarnonmono2}). The detailed parameter settings are shown in Table~\ref{t:simNonmono1}. It is worth noting that preferences are not integrated into $\mathcal{ENI}$. For a given pair of non-monotonic simultaneous events, $\mathcal{ENI}$ will be higher for the event with a higher overall deviation in the energy distribution; for instance, see the first event pair in Table~\ref{t:simNonmono1}. Further, the impact of relatively short duration transients on the overall energy distribution can be lower than the companion sag/swell event. It is, therefore, natural to expect that $\mathcal{ENI}$ would be more sensitive to constituent sag/swell for relatively shorter duration transients, \textit{e.g.}, see the second pair of events in Table~\ref{t:simNonmono1}.

To overcome this issue, calculation of two additional energy norm indices, $\mathcal{LNI}$ and $\mathcal{WNI}$, is recommended. The rationale is to estimate both constituent severities, $\mathcal{S}_f$, and $\mathcal{S}_{tr}$. This additional information can aid the DM in evaluating the overall severity, $\mathcal{S}_o$. In particular, $\mathcal{LNI}$ estimates $\mathcal{S}_f$, \textit{i.e.}, the severity of $sag/swell$ (see~Section~\ref{s:LNI}). In contrast, $\mathcal{WNI}$ is used to estimate $\mathcal{S}_{tr}$ by assigning a higher priority to the higher frequency bands, see~Section~\ref{s:WNI}. 

The values of $\mathcal{LNI}$ and $\mathcal{WNI}$ for non-monotonic simultaneous events are shown in Table~\ref{t:simNonmono1}. The results indicate that $\mathcal{WNI}$ is relatively more sensitive to transients, even for shorter duration transients, \textit{e.g.}, see the results for the second pair in Table~\ref{t:simNonmono1}. Similarly, $\mathcal{LNI}$ can correctly estimate the severity of low frequency events. For instance, consider the second pair of events in Table~\ref{t:simNonmono1}; for these events known severity of constituent sag is $\mathcal{S}_{f,2} > \mathcal{S}_{f,1}$. This is correctly estimated by $\mathcal{LNI}$, \textit{i.e.}, $\mathcal{LNI}_2 = 8.9\%$, which is higher than the same for the first event, $\mathcal{LNI}_1=5.8\%$ (see Table~\ref{t:simNonmono1}). Note that similar results are also obtained for swell with transients, which are omitted here due to space constraints.

%======================================================
\section{Comparative Evaluation over Real Events}
\label{s:ResRealEvents} 

%=============================================
\subsection{Real Disturbances}
\label{subsec:realsag}

For further validation of the proposed approach, comparative evaluations are carried out on several real PQ disturbances gathered from the following two public databases: real sag recordings in the University of Cádiz (UoC) database~\cite{SagDatabase}; and the EPRI national repository of PQ events~\cite{EPRIDatabase}. The UoC database contains 13 distinct \emph{sag} events, which were gathered from a low voltage distribution network (300 V, 50 Hz) at a sampling frequency of $f_s = 20 \ kHz$, see~\cite{aguera:2011,SagDatabase} for details. Further, nine real sag events are collected from two EPRI assets which are part of North American medium voltage utility networks ($7.2$ kV, 60 Hz): Asset ID - 112678 ($f_s \approx 30 \ kHz$)  and 101159 ($f_s \approx 7 \ kHz$); see~\cite{EPRIDatabase} for details.

%%%===========================================
\begin{figure}[!b]
    \centering
    \begin{subfigure}{0.227\textwidth}
        \includegraphics[width=\textwidth]{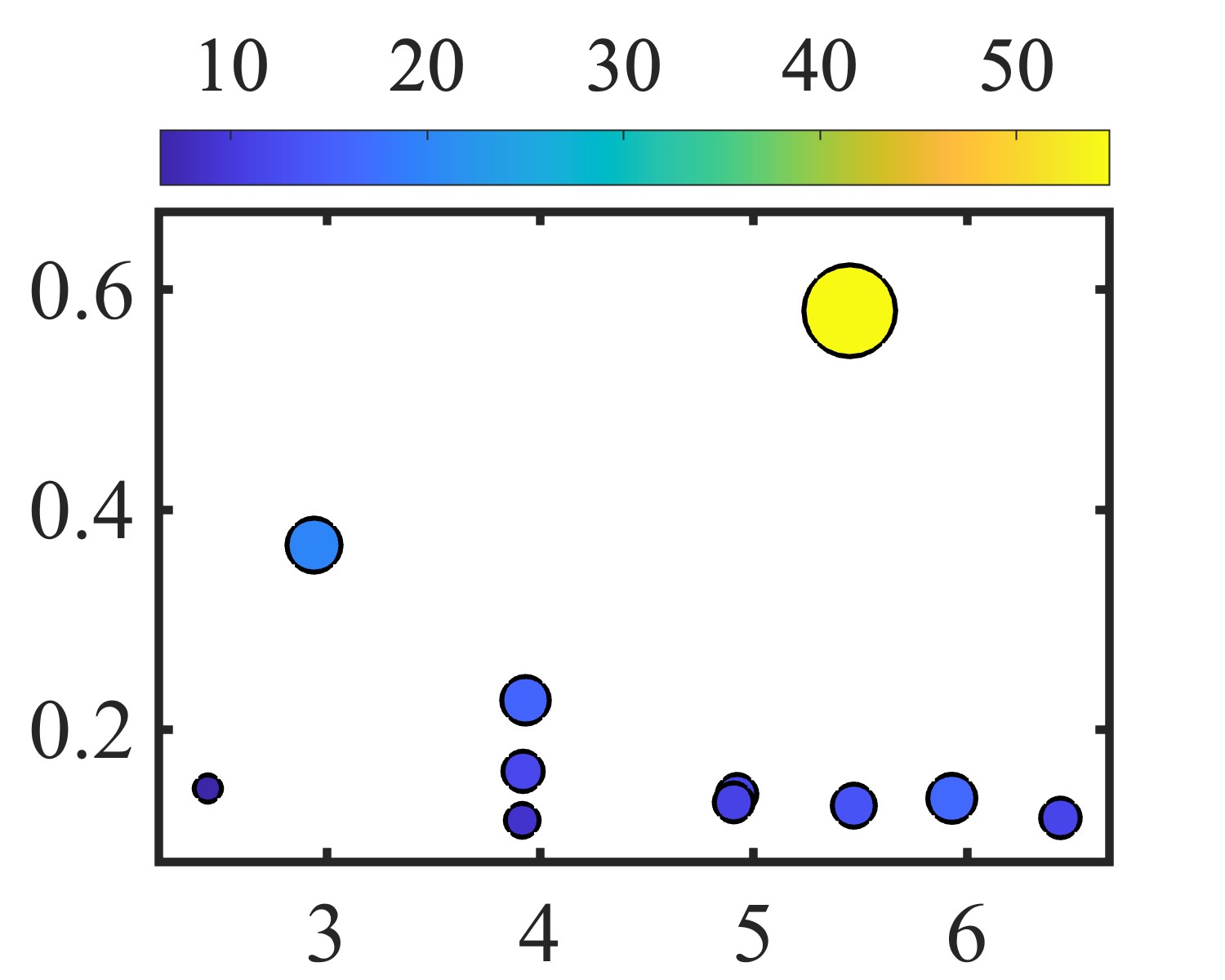}%
        \caption{UoC Sags $+ \ \mathcal{ENI}$}
        \label{f:UoC+ENI}
    \end{subfigure}
    \hfill     
    \begin{subfigure}{0.227\textwidth}
        \includegraphics[width=\textwidth]{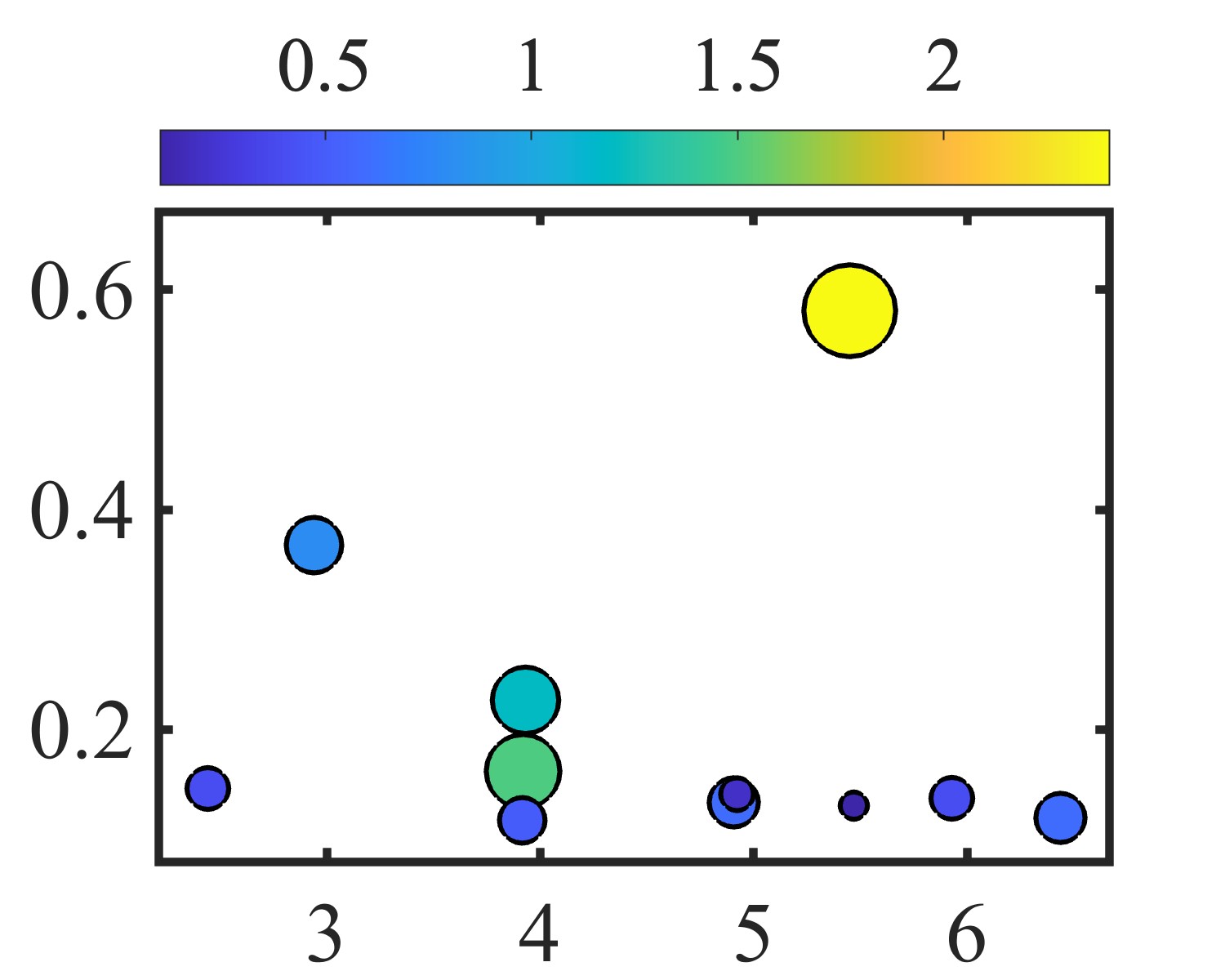}%
        \caption{UoC Sags $+ \ \Delta WD_e$}
        \label{f:UoC+WDe}
    \end{subfigure}     
    \hfill 
    \begin{subfigure}{0.227\textwidth}
        \includegraphics[width=\textwidth]{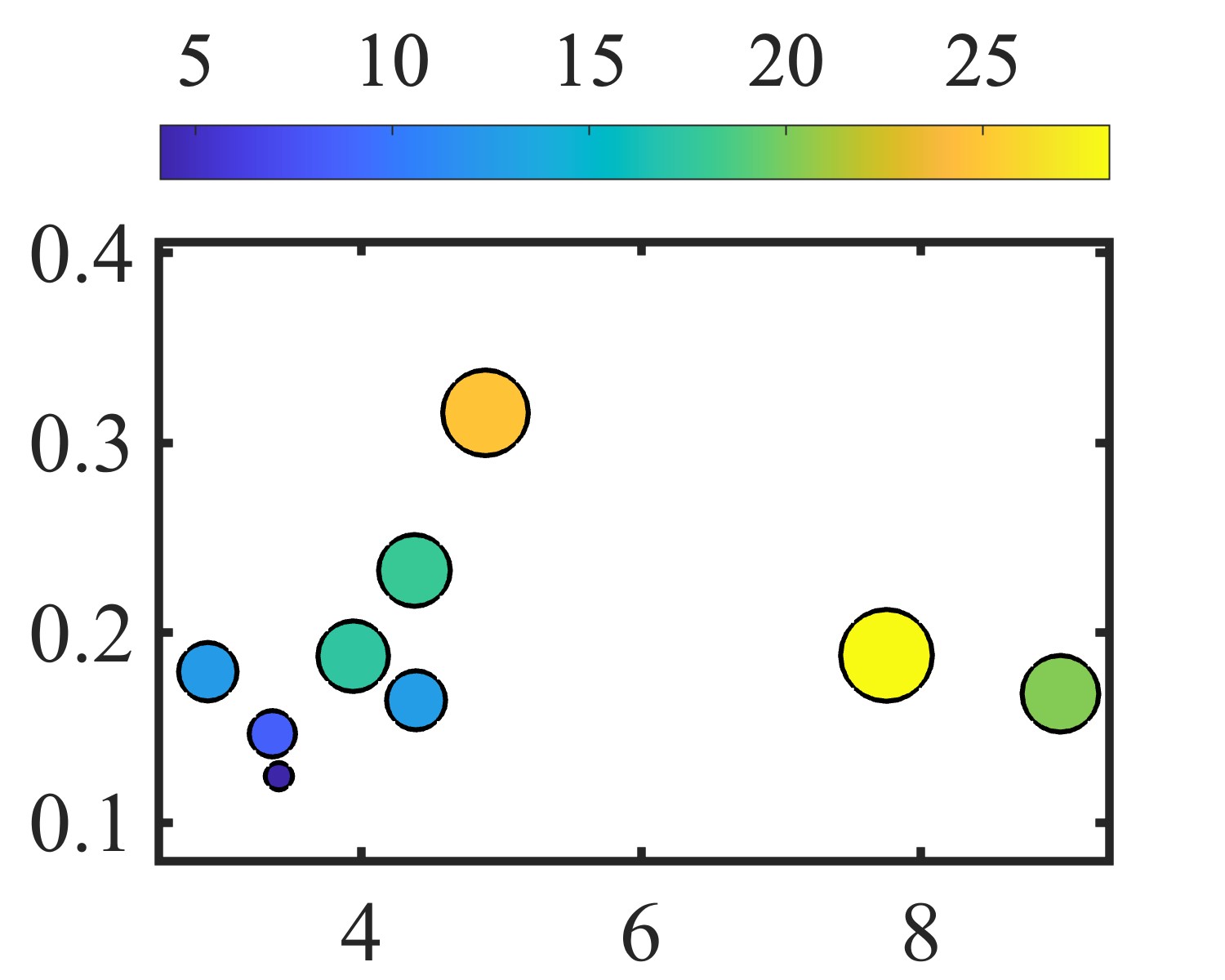}%
        \caption{EPRI Sags $+ \ \mathcal{ENI}$}
        \label{f:EPRI+ENI}
    \end{subfigure}
    \hfill     
    \begin{subfigure}{0.227\textwidth}
        \includegraphics[width=\textwidth]{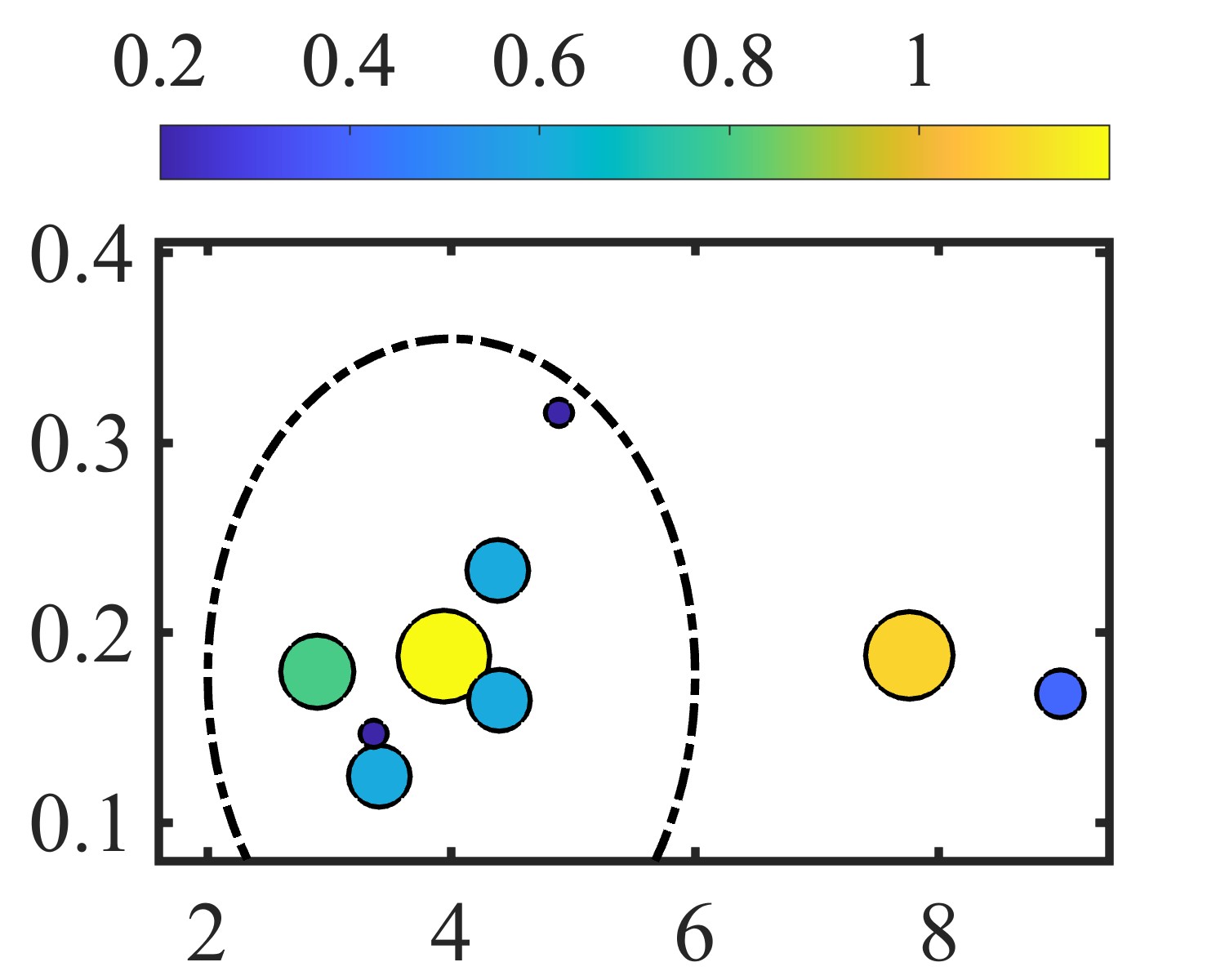}%
        \caption{EPRI Sags $+ \ \Delta WD_e$}
        \label{f:EPRI+WDe}
    \end{subfigure}
     
    \caption{Variation in indices for real sag events in the UoC~\cite{SagDatabase} and EPRI~\cite{EPRIDatabase} databases. \texttt{x} and \texttt{y-axes} respectively denote the \texttt{duration}, $t_d \ (cycles)$ and the \texttt{sag magnitude}, $\alpha \ (pu)$. The values indices are denoted by the radius and the color of circles; a larger radius and bright yellow color of a circle denotes a higher value. The values of $\Delta WD_e$ inside the black dotted line in Fig.~\ref{f:EPRI+WDe} illustrate its \emph{non-monotonic} behavior.} 
    \label{f:realSags}
    %\textcolor{blue}{Variation in indices for real sag events in the EPRI database: EPRI Asset 112678: 502	618	734	1027	1075	1134	531	626	698, 	101159: 48 (a) $\mathcal{ENI}$ (b) $\Delta WD_e$}
    % \vspace{-5mm}
\end{figure}
%%%=============================================

Further, while the conventional PQ indices such as THD are well-established, the lack of temporal information associated with such indices often limits their applicability to non-stationary events. Therefore, we consider the \emph{wavelet deviation index}~\cite{Naik:Kundu:2013}, $\Delta WD_e$, for comparative evaluation purposes. $\Delta WD_e$ was specifically designed to quantify the severity of high-frequency PQ disturbances such as \emph{transients} and \emph{harmonics} and can accommodate non-stationary events as it is also defined based on wavelet coefficients.

%-----------------------------------------
% \vspace{-2mm}
\begin{table}[!b]
  \centering
  \caption{Illustrative Example of Non-monotonic behavior of $\Delta WD_e$ \label{t:realSag}}
  \begin{adjustbox}{max width=0.49\textwidth}  
  \begin{threeparttable}
    \begin{tabular}{ccccc}
    \toprule
    \multirow{2}{*}{\textbf{Event }} & \multicolumn{2}{c}{\textbf{Esimated Parameters}} & \multirow{2}{*}{\boldmath $\mathcal{ENI}$} & \multirow{2}{*}{\boldmath $\Delta WD_e$} \\
    \cmidrule{2-3}  & \boldmath $\alpha \ (pu)$ & \boldmath $t_d \ (cycles)$ &       &  \\
    \midrule
    
   \texttt{Sag-1} (\texttt{UoC}) & 0.16  & 3.92  & 11.98 & 1.48 \\[0.5ex]
   \texttt{Sag-2} (\texttt{UoC}) & 0.23  & 3.93  & 15.99 & 1.21 \\[0.5ex]
    \cmidrule{2-5}       & \multicolumn{2}{c}{\makecell{\textbf{Known Severity:}\\ \texttt{Sag-1} $<$ \texttt{Sag-2}}} & \makecell{\textbf{Estimation:}\\\texttt{Sag-1} $<$ \texttt{Sag-2}} & \makecell{\textbf{Estimation:}\\\texttt{Sag-1} $>$ \texttt{Sag-2}} \\[1.2ex]
    \midrule
    \texttt{Sag-3} (EPRI: 698$^\dagger$) & 0.12  & 3.41  & 4.06  & 0.64 \\[0.5ex]
    \texttt{Sag-4} (EPRI: 626$^\dagger$) & 0.15  & 3.36  & 8.52  & 0.19 \\[0.5ex]
    \cmidrule{2-5}       & \multicolumn{2}{c}{\makecell{\textbf{Known Severity:}\\ \texttt{Sag-3} $<$ \texttt{Sag-4}}} & \makecell{\textbf{Estimation:}\\\texttt{Sag-3} $<$ \texttt{Sag-4}} & \makecell{\textbf{Estimation:}\\\texttt{Sag-3} $>$ \texttt{Sag-4}} \\[0.5ex]
    
    \bottomrule
    \end{tabular}%
    \begin{tablenotes}
        \small
        \item $^\dagger$ Real sag recorded at the EPRI Asset 112678~\cite{EPRIDatabase}
    \end{tablenotes}
 \end{threeparttable}
 \end{adjustbox}
\end{table}%
% \vspace{-5mm}
%-----------------------------------------

We begin by normalizing real sag events in both databases into per unit (pu), where the corresponding peak voltage value is treated as the base voltage. Next, the wavelet coefficients are determined for each event; where the DWT decomposition level is adapted as per the corresponding sampling frequency and following the rule of thumb in~eq~(\ref{eq:ThumbRule}), \textit{e.g.}, $D=8$ for the UoC database. The subsequent energy of wavelet coefficients is used to estimate the severity of each event using the proposed index, $\mathcal{ENI}$ (see~Algorithm~\ref{al:propIndex}), and the \emph{wavelet deviation index}, $\Delta WD_e$, see~\cite{Naik:Kundu:2013}.

To facilitate comparative evaluation, the characteristic parameters of each real sag event are estimated, \textit{i.e.}, sag magnitude, $\alpha \ (pu)$; and duration, $t_d \ (cycles)$. Fig.~\ref{f:realSags} shows the estimated sag parameters and the corresponding variations in $\mathcal{ENI}$ and $\Delta WD_e$. The results clearly outline the desired monotonic relationship among the sag parameters $\{ \alpha, \ t_d\}$ and $\mathcal{ENI}$; for both databases, UoC (Fig.~\ref{f:UoC+ENI}) and EPRI (Fig.~\ref{f:EPRI+ENI}). In contrast, a non-monotonic behavior is observed in $\Delta WD_e$ results, Fig.~\ref{f:UoC+WDe}-\ref{f:EPRI+WDe}. For instance, consider the values of $\Delta WD_e$ corresponding to EPRI sags with $\{ \alpha \leq 0.35 , \ t_d \leq 5 \ cycles$\}, which are contained inside black dotted lines in Fig.~\ref{f:EPRI+WDe}. For convenience, a subset of results in Fig.~\ref{f:realSags} are reproduced in Table~\ref{t:realSag}, which shows the estimated sag parameters along with the values of indices for a pair of sag events from both databases. These results clearly demonstrate that the non-monotonic behavior of $\Delta WD_e$ can lead to an incorrect severity estimation of sags. These limitations, in major part, can be attributed to the over-emphasis on the higher frequency bands in the calculation of $\Delta WD_e$, see~\cite{Naik:Kundu:2013} for details. %Consider the first pair of UoC sags in Table~\ref{t:realSag}; it is clear that given that the duration of \texttt{sag-1} and \texttt{sag-2} do not differ significantly, ($t_{d,1} \approx t_{d,2} \approx 3.9 \ cycles$

%-----------------------------------------
%%%===========================================
\begin{figure}[!t]
% \vspace{-5mm} 
    \centering
    \begin{subfigure}{0.22\textwidth}
        \includegraphics[width=\textwidth]{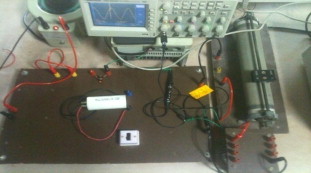}%
        \caption{Laboratory Setup}%
        \label{f:expsetup}
    \end{subfigure}
    \hfill
    \begin{subfigure}{0.22\textwidth}
        \includegraphics[width=\textwidth]{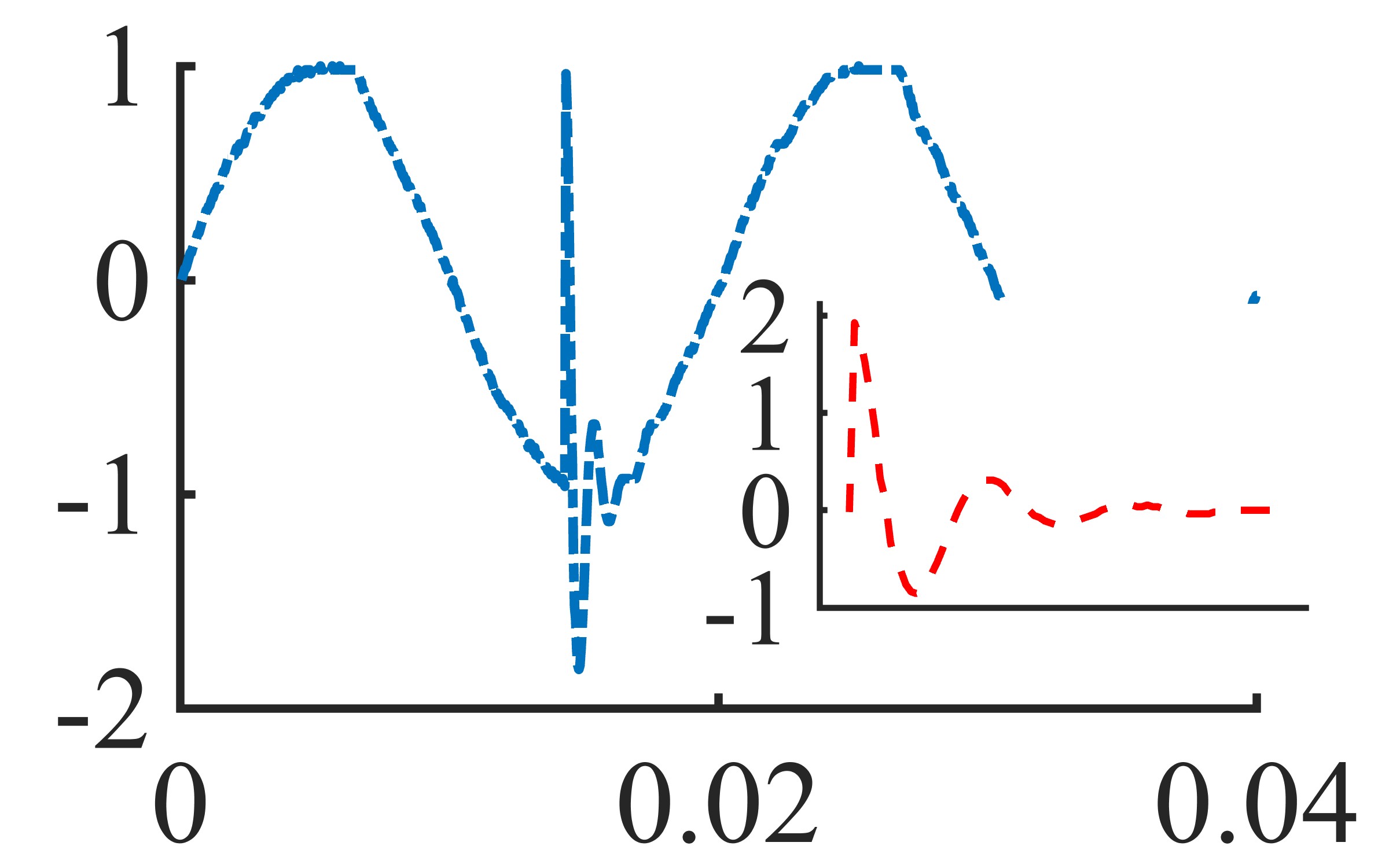}%
        \caption{Illustrative Transient}
        \label{f:illTrans}
    \end{subfigure}
    \caption{The laboratory setup to emulate capacitor bank switching and an illustrative experimental transient. The x-axis and y-axis respectively represent \textit{time} (s) and \textit{voltage} (pu). The transient parameters are determined to be: magnitude, $\beta = 1.93$; duration, $t_{tr} = 3.32 \ (ms)$; and $f_{tr}=850 \ Hz$. The severity of this event is estimated to be: $\mathcal{WNI}=7.80$ and $\Delta WD_e=1.09$.}
    \label{f:RealTran}
    % \vspace{-5mm}
\end{figure}
%%%=============================================
%======================================================
\vspace{-2mm}
\subsection{Experimental Disturbances}
\label{subsec:realtrans}

A set of capacitor switching transients is used next for the comparative evaluation of indices. Such events were generated using a simple laboratory setup, see~Fig~\ref{f:expsetup}. The setup essentially emulates a single phase distribution system with a power factor correction capacitor bank and a constant inductive load. The manual switch in the setup mimics the switching of a power factor correction bank, and the consequent transients are captured using a digital oscilloscope at the sampling frequency of $25 \ kHz$. Fig.~\ref{f:illTrans} shows an illustrative $850$ Hz switching transient generated using this setup. The manual switch operations allowed for the collection of seven distinct transients which are used for further analysis.

\begin{figure}[!t]
% \vspace{-9mm}
   \centering
   \begin{subfigure}{0.21\textwidth}
        \includegraphics[width=\textwidth]{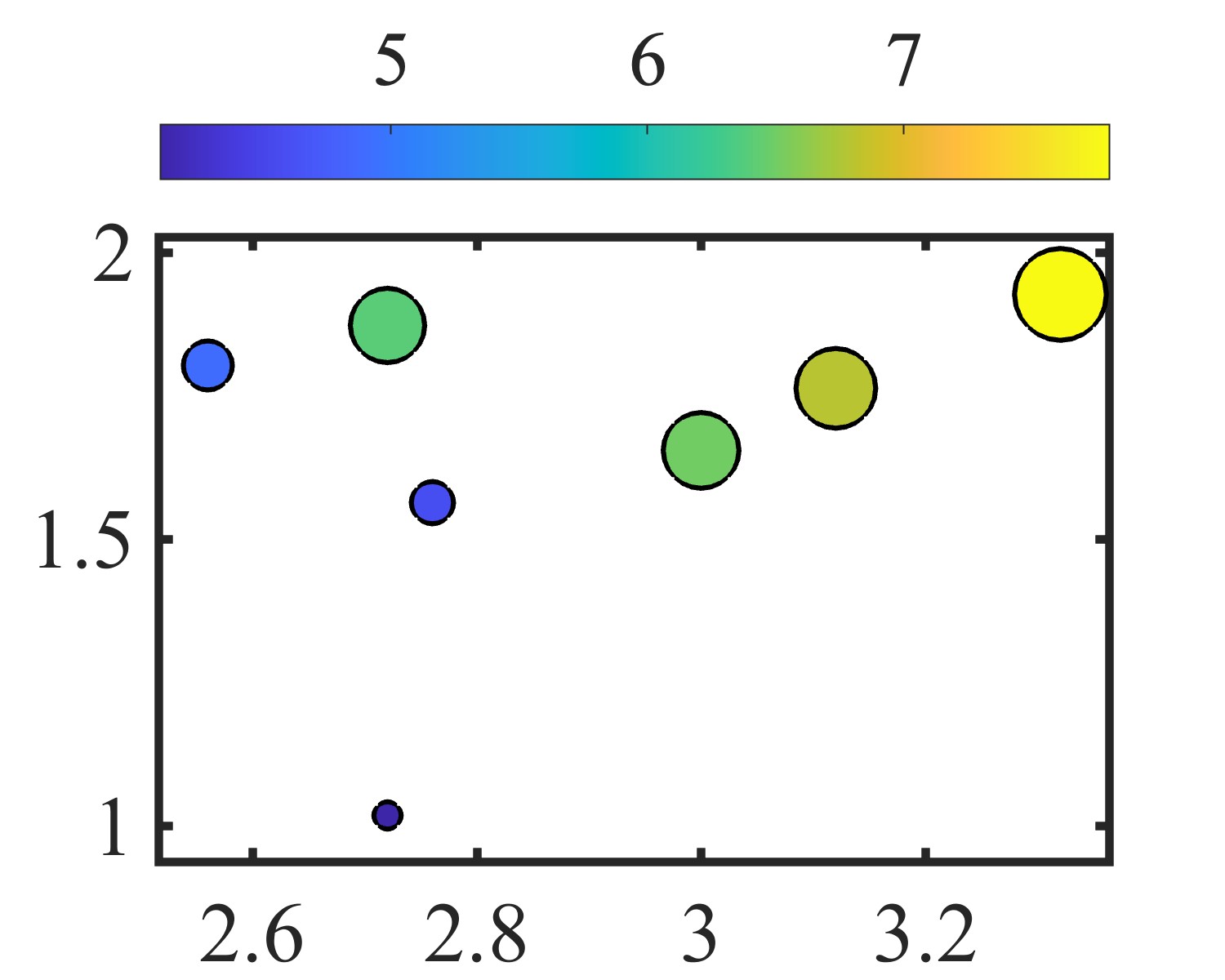}%
        \caption{$\mathcal{ENI}$}%
        \label{f:expTrENI}
    \end{subfigure}
    \hfill
    \begin{subfigure}{0.21\textwidth}
        \includegraphics[width=\textwidth]{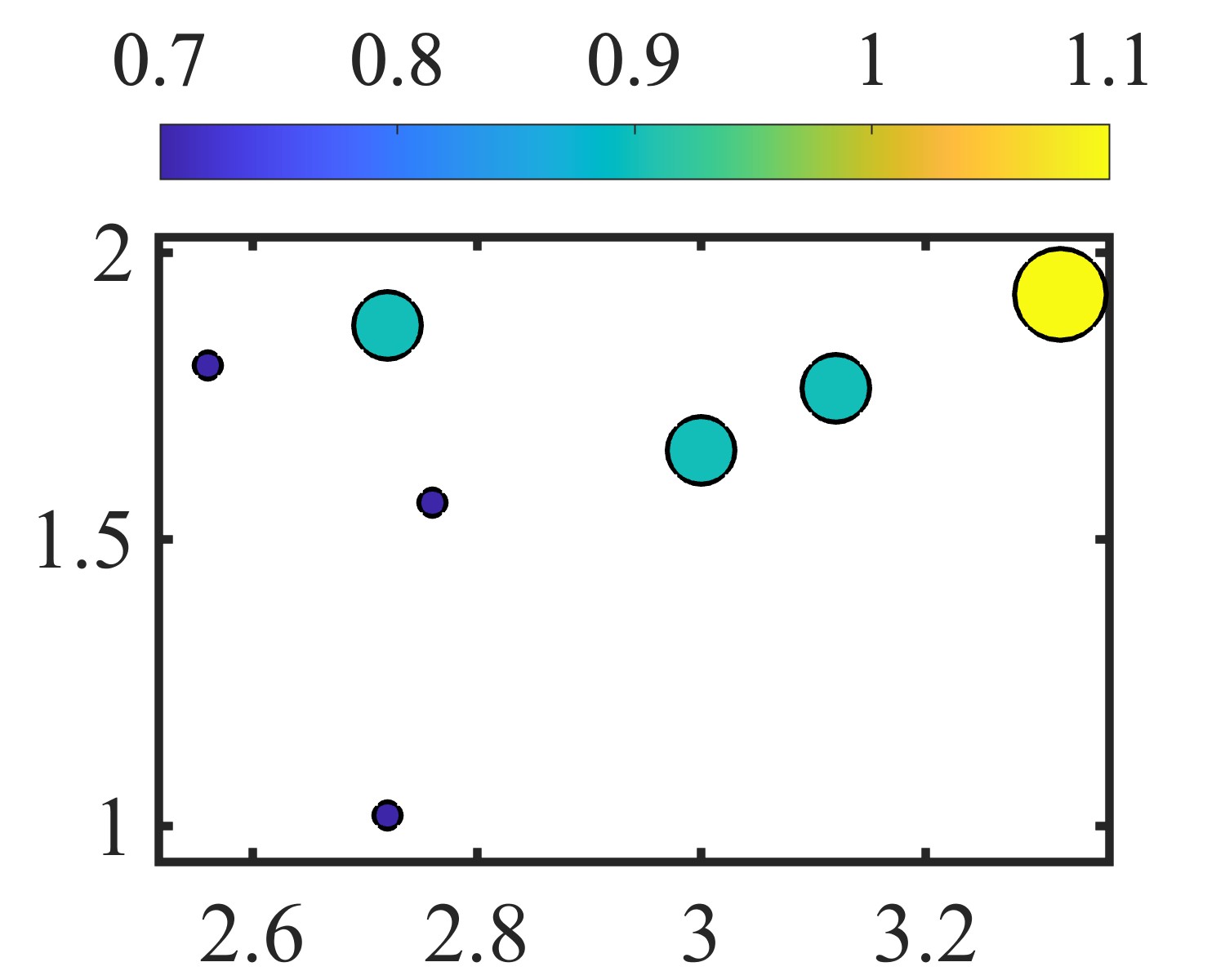}%
        \caption{$\Delta WD_e$}
        \label{f:expTrWDe}
    \end{subfigure}
   \caption{Comparative evaluation of indices over experimental transient events. \texttt{x} and \texttt{y-axes} respectively denote the \texttt{duration}, $t_{tr} \ (cycles)$, and the estimated \texttt{transient magnitude}, $\beta \ (pu)$. The size of each bubble is proportional to a particular index; a higher index value is depicted by a larger radius and bright yellow color.}
  \label{f:compEval_expTr}
\end{figure}

For comparative evaluation purposes, the severity of the experimental transients is estimated using the proposed weighted index ($\mathcal{WNI}$, Section~\ref{s:WNI} and~\ref{s:indexselect}) and $\Delta WD_e$~\cite{Naik:Kundu:2013}. Each transient is decomposed to 8 levels and subsequent energy of wavelet coefficients is used to calculate both indices. Further, the characteristic  parameters of each transient are also estimated, \textit{e.g.}, magnitude, $\beta \ (pu)$; duration, $t_{tr} \ (cycles)$. Given that the characteristic parameters, $\{\beta,t_{tr}\}$, exert a clear influence on transient severity, the objective of the comparative evaluation is to determine whether the indices display monotonic behavior with these parameters. Fig.~\ref{f:compEval_expTr} shows the estimated transient parameters and severity indices for all experimental transients. These results highlight a systematic rise in both indices corresponding to increasing values of $\beta$ and $t_{tr}$. To summarize, the comparative evaluation reaffirms the severity estimation capabilities of $\mathcal{ENI}$ on both fundamental and high frequency non-stationary events (cf. $\Delta WD_e$ behavior in Section~\ref{subsec:realsag}).
%-------------------------------------------------
%%%===================================================
\section{Discussions}
\label{s:discuss}
%%%============================================
\subsection{Sensitivity to Wavelet Selection}
\label{s:wavsense} 

Given that the proposed index primarily relies on the temporal-frequency energy distribution of wavelet coefficients, it is pertinent to determine whether the index retains monotonic behavior across different wavelets. To this end, a total of 70 distinct wavelets from three different wavelet families (\emph{Daubechies}, \emph{Symlets}, and \emph{Coiflets}) are considered. For each wavelet, the index values are determined for both fundamental (\emph{sags}, \emph{swells}, and \emph{interruptions}) and high frequency (\emph{transients}) events with the parametric variations given in Table~\ref{t:events}.

%--------------------------------------------
% \vspace{-5mm}
%-----------------------------------------

\begin{figure}[!t]
    \centering
    \begin{subfigure}{0.23\textwidth}
        \includegraphics[width=\textwidth]{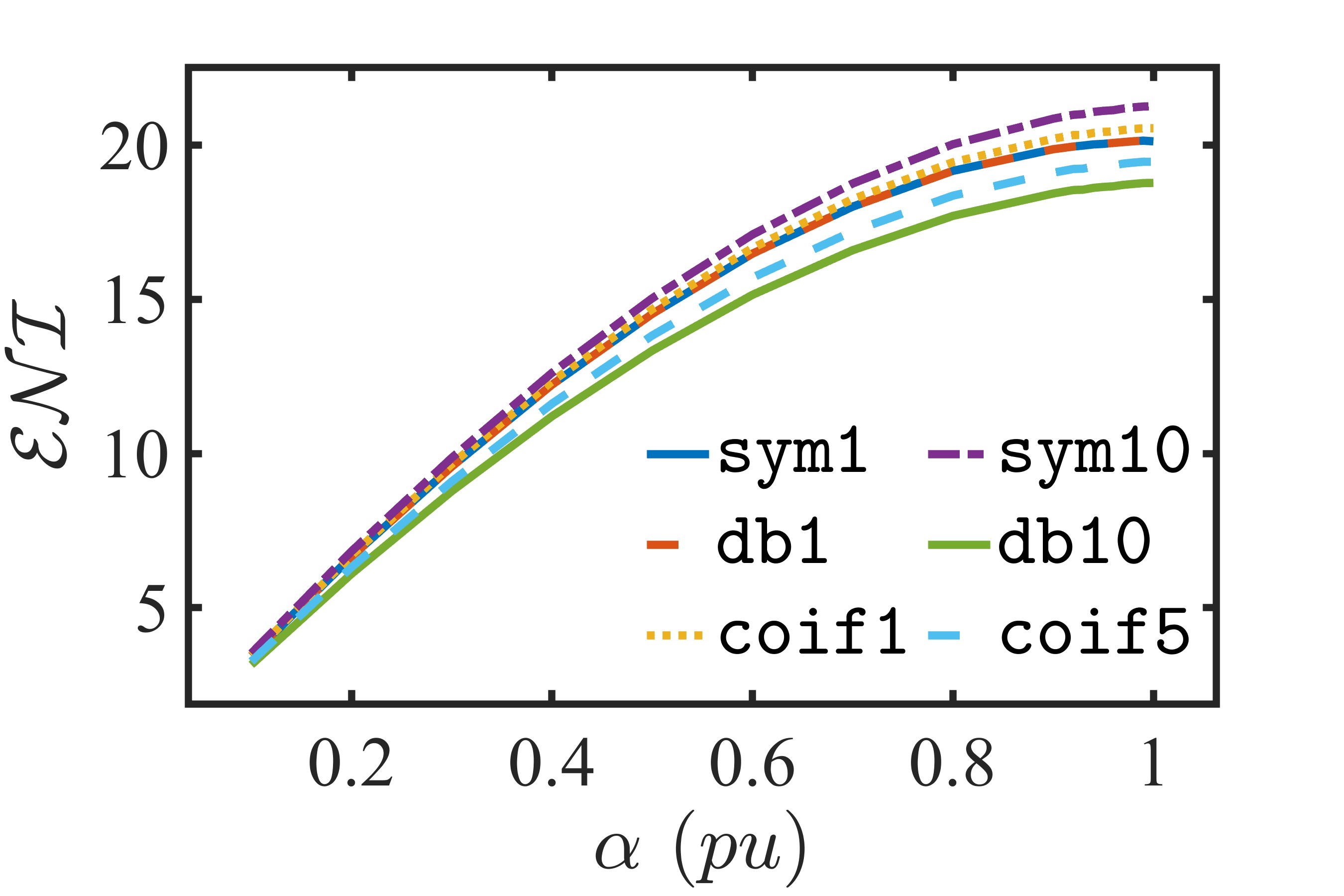}
        \caption{sags with $t_d = 10 \ cycles $}
        \label{f:sens_sg_a}
    \end{subfigure}
    \hfill
    \begin{subfigure}{0.23\textwidth}
        \includegraphics[width=\textwidth]{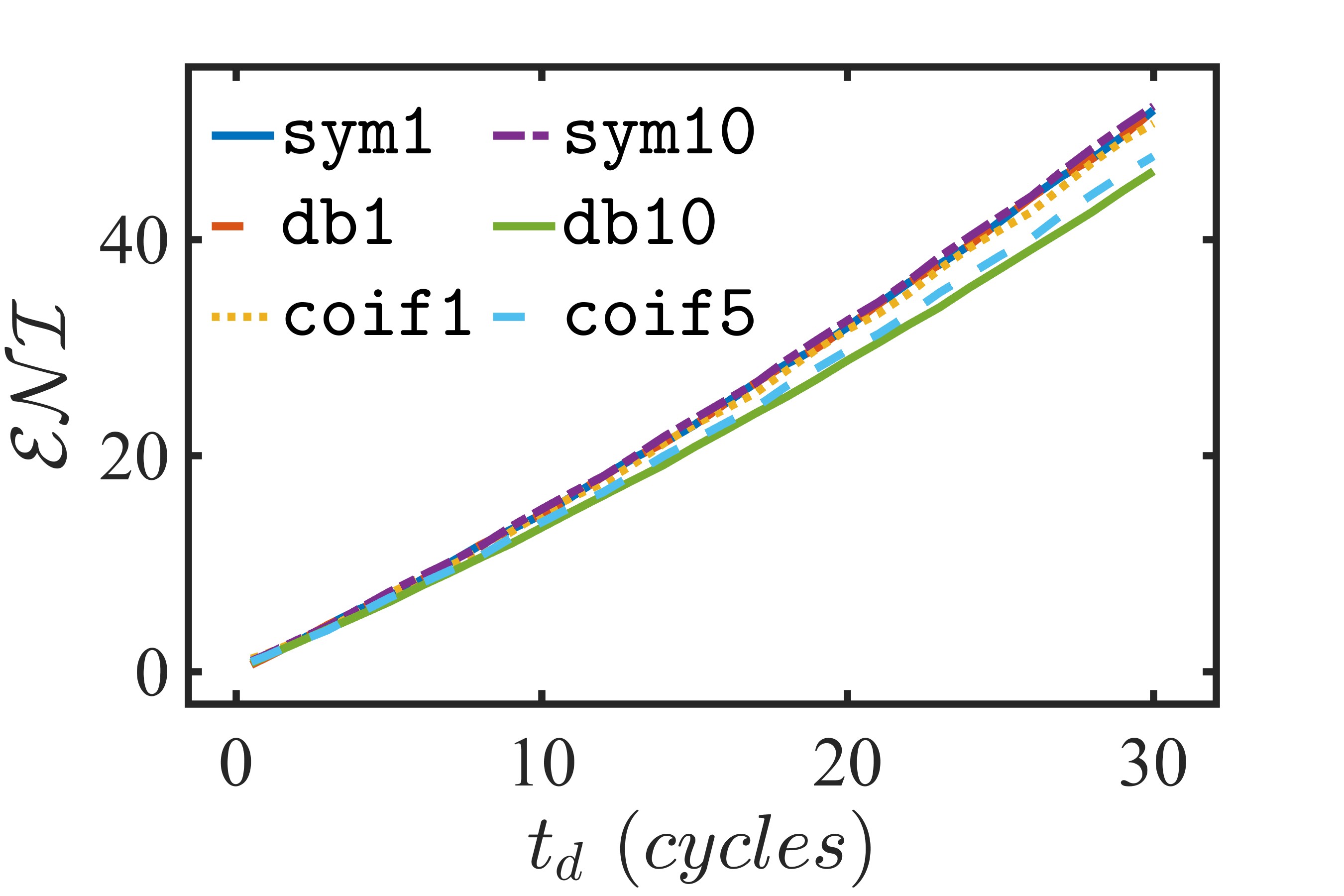}%
        \caption{sags with $\alpha = 0.5 \ pu $}
        \label{f:sens_sg_td}
    \end{subfigure}
    \hfill
    \begin{subfigure}{0.23\textwidth}
        \includegraphics[width=\textwidth]{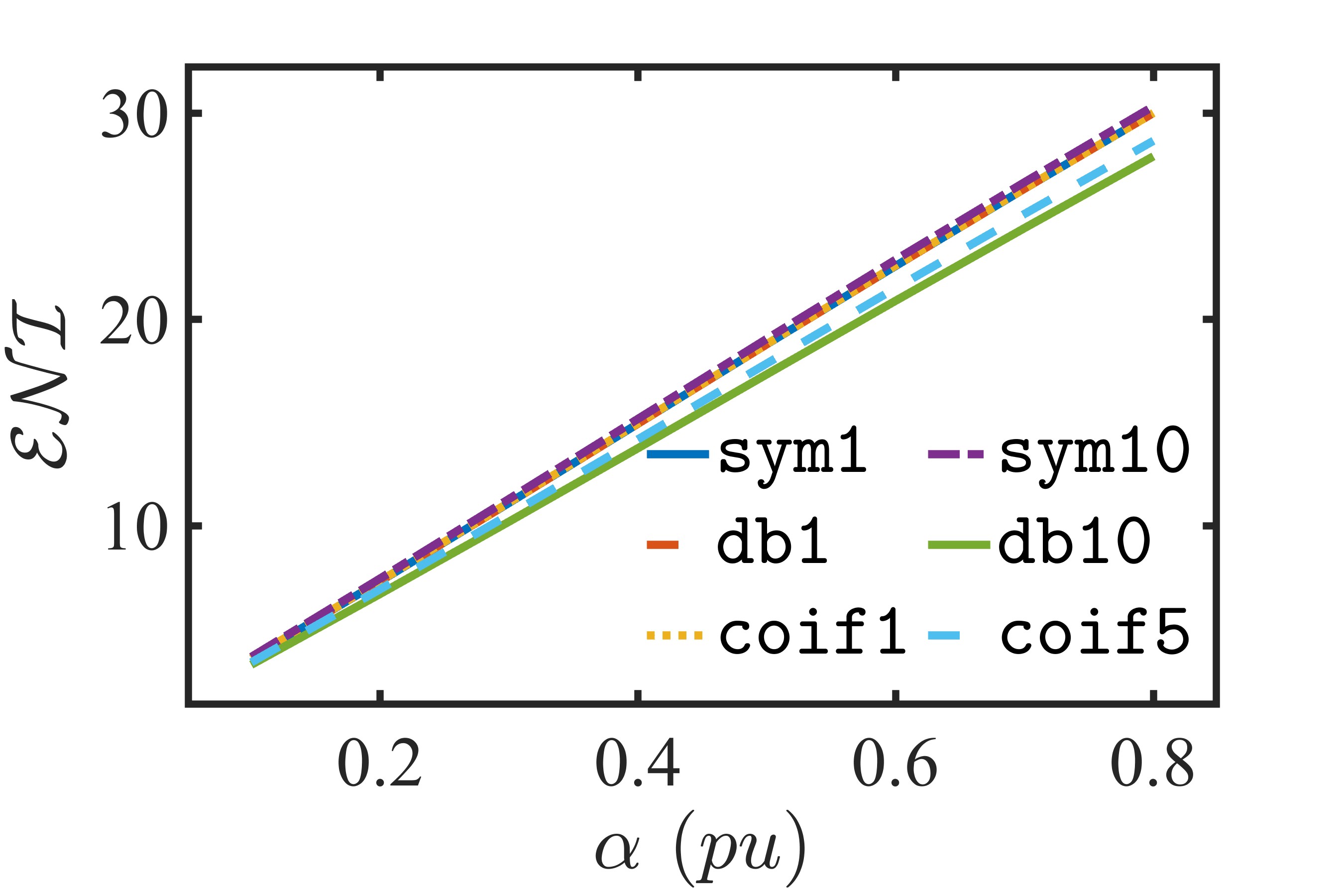}
        \caption{swells with $t_d = 10 \ cycles $}
        \label{f:sens_sw_a}
    \end{subfigure}
    \hfill
    \begin{subfigure}{0.23\textwidth}
        \includegraphics[width=\textwidth]{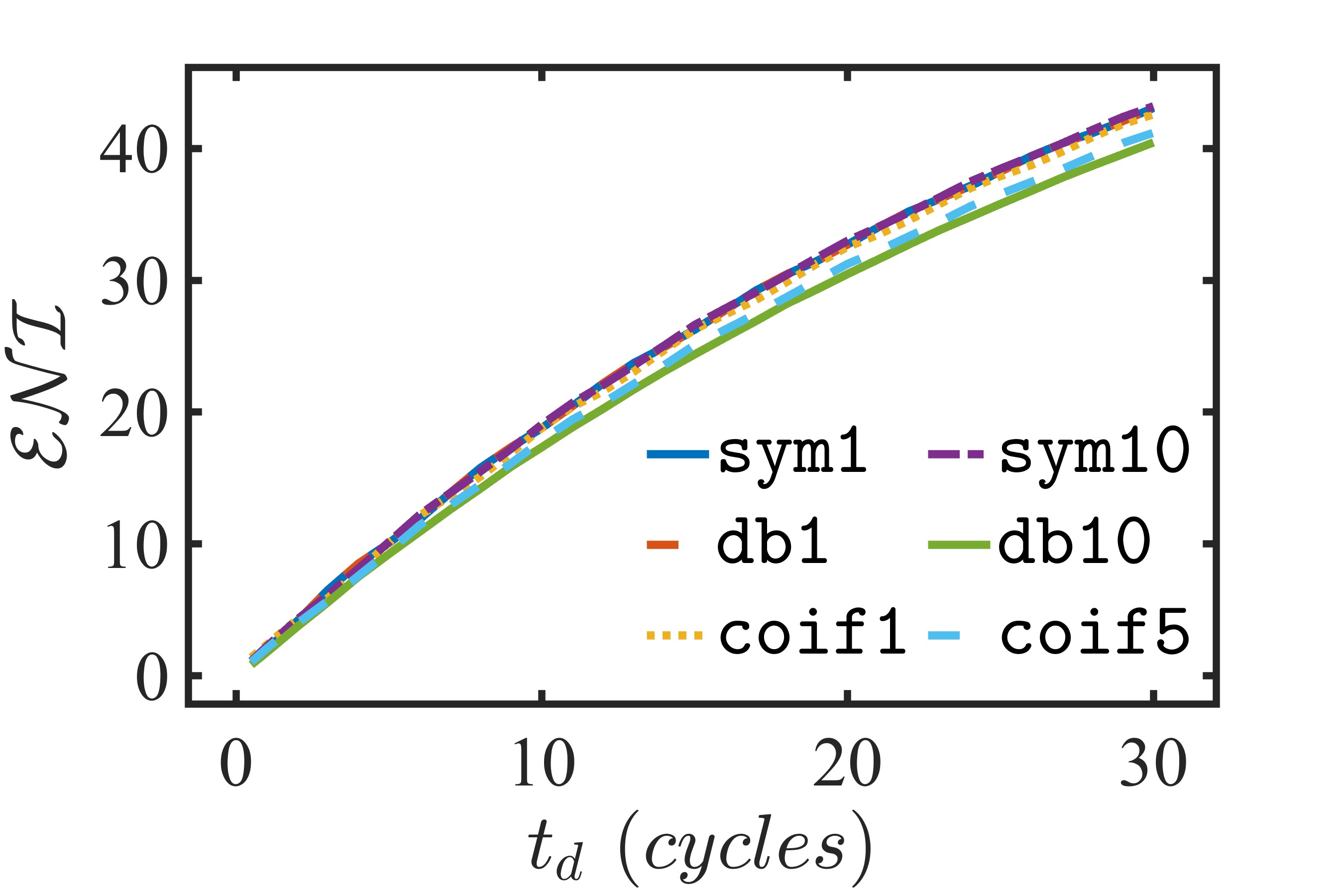}%
        \caption{swells with $\alpha = 0.5 \ pu $}
        \label{f:sens_sw_td} 
    \end{subfigure} \\ 
    % \hfill
    \begin{subfigure}{0.239\textwidth}
        \includegraphics[width=\textwidth]{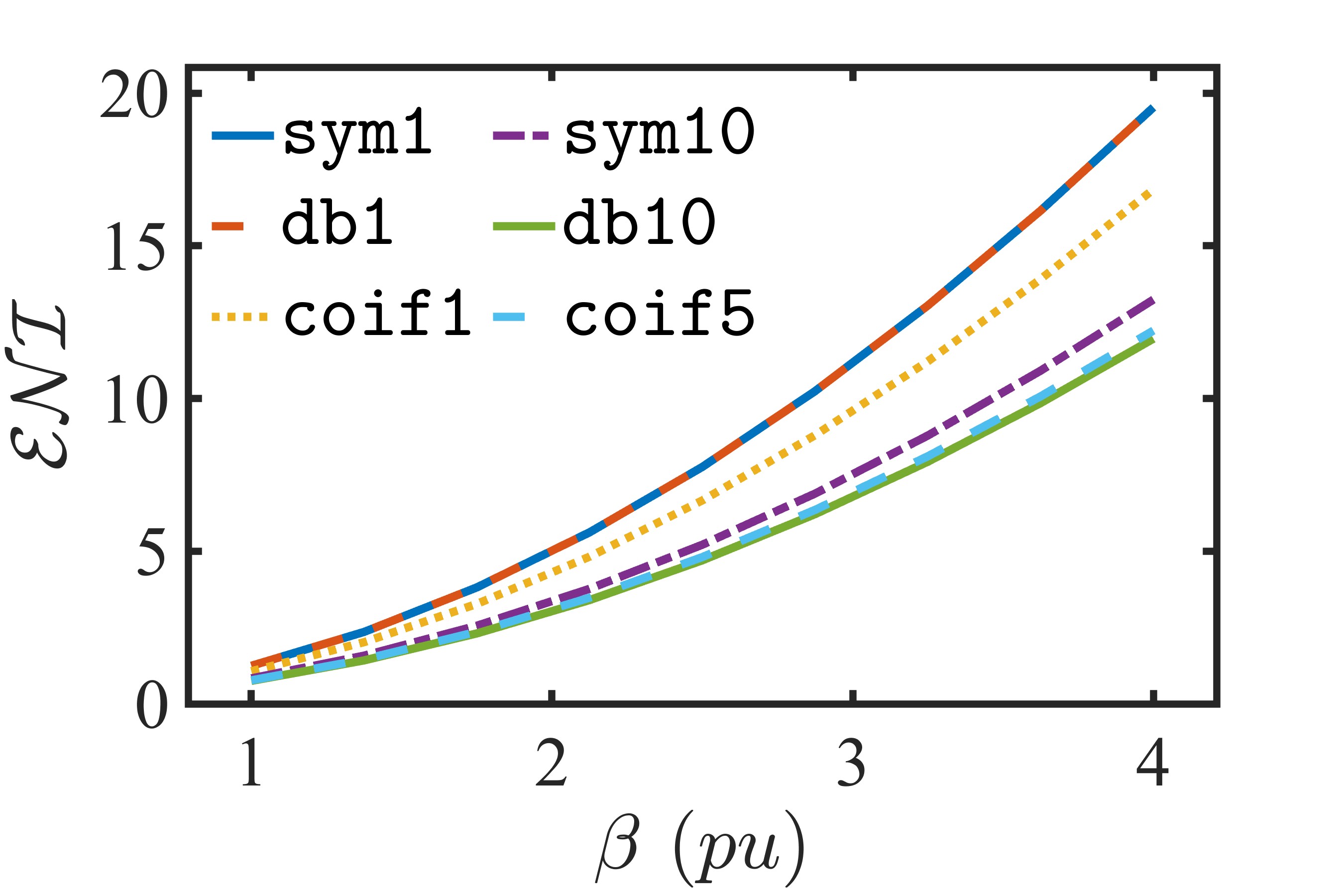}
        \caption{transients,$\{\gamma, f_{tr} \} = \{-55, 800\}$}
        \label{f:sens_tr_a}
    \end{subfigure}
    \hfill
    \begin{subfigure}{0.235\textwidth}
        \includegraphics[width=\textwidth]{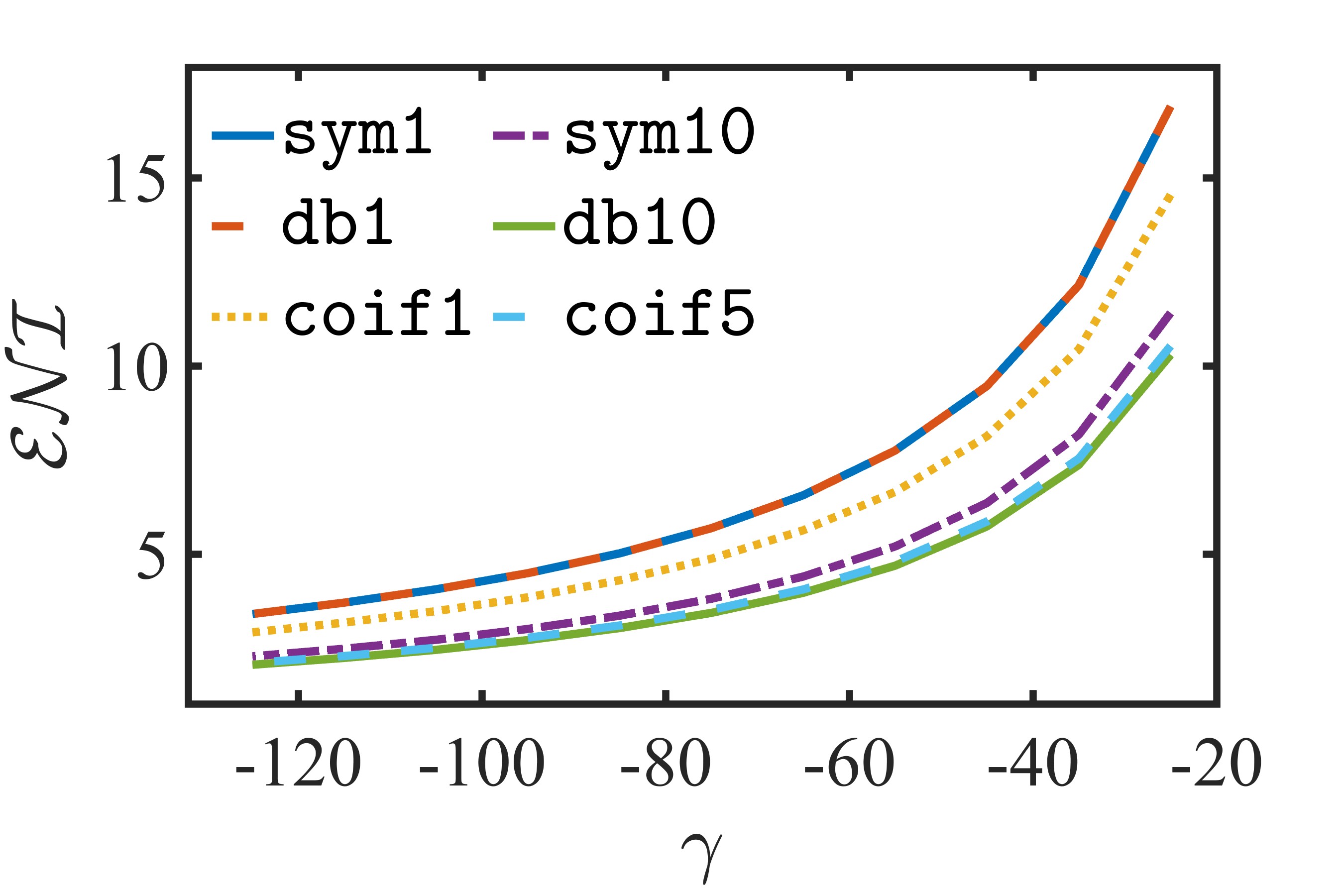}%
        \caption{transients, $\{\beta, f_{tr} \} = \{2.5, 800\}$}
        \label{f:sens_tr_td}
    \end{subfigure}
    
    \caption{The variation in the index, $\mathcal{ENI}$, values under different wavelets. A monotonic variation in $\mathcal{ENI}$ irrespective of wavelets is noteworthy. Further, a higher slope in the index variation is desirable as it indicates a higher sensitivity to an event severity parameter.}
    \label{f:wavesens}
\end{figure}

The results of this analysis are shown for a few wavelets in Fig.~\ref{f:wavesens}. It is clear that the index retains the monotonic behavior for all events under consideration, irrespective of the selected wavelet. For instance, the results for the fundamental frequency events (Fig.~\ref{f:sens_sg_td}-\ref{f:sens_sw_td}) suggest that the index is relatively robust to the wavelet selection. Further, the results for transients, in Fig.~\ref{f:sens_tr_a}~and~\ref{f:sens_tr_td}, indicate that while the index remains monotonic across different wavelets, it is more sensitive to transient severity variations with lower order wavelets, \textit{i.e.}, for a specific transient severity, the index values are higher with the lower order wavelets such as \texttt{db1} and \texttt{sym1}. This can be explained by the shorter support and fewer vanishing moments associated with low order wavelets (\textit{e.g.}, \texttt{db1} or \texttt{sym1}) that can detect rapid transitions, and, therefore, are more sensitive to transient events~\cite{Darwish:Hesham:2010}. However, such wavelets often suffer from an \emph{energy leakage} issue, \textit{i.e.}, poor frequency localization stemming from a higher band overlap between quadrature analysis filters; see~\cite{Peng:Jackson:2009} for details. The energy leakage issue can be alleviated by using higher order wavelets (\textit{e.g.}, \texttt{db20} or \texttt{sym20}), which provide improved frequency localization and, therefore, are often more suitable for the fundamental frequency events~\cite{Darwish:Hesham:2010}, \textit{e.g.}, \emph{sags}. This improvement in frequency localization, however, comes at the expense of poor time localization~\cite{Peng:Jackson:2009}. The selection of wavelet for PQ analysis requires a careful balancing of time-frequency localization capabilities for high frequency transients and fundamental frequency sags/swells.

The overall results of our empirical sensitivity analysis suggest that mid-order wavelets (\textit{e.g.}, \texttt{sym4}, \dots, \texttt{sym8} or \texttt{db4}, \dots, \texttt{db8}) can balance the time-frequency localization requirements of sags/swells and transients. It is worth noting that these results agree with the earlier investigations in~\cite{Hafiz:Swain:IET:2018,Darwish:Hesham:2010}. Finally, it is worth highlighting that the monotonic behavior of the index across different wavelets can be explained by the fact that the monotonicity conditions of Proposition 2 (see Section~\ref{s:ENI}) are satisfied irrespective of the selected wavelet.

%%%===========================================
% \vspace{-5mm}
%%%=============================================

%%%===================================================
\vspace{-4mm}
\subsection{Utility Significance}
\label{s:utilityapp} 

\begin{SCfigure}[1.0][!t]
% \begin{figure}[!t]
   \centering
   \includegraphics[width=0.3\textwidth]{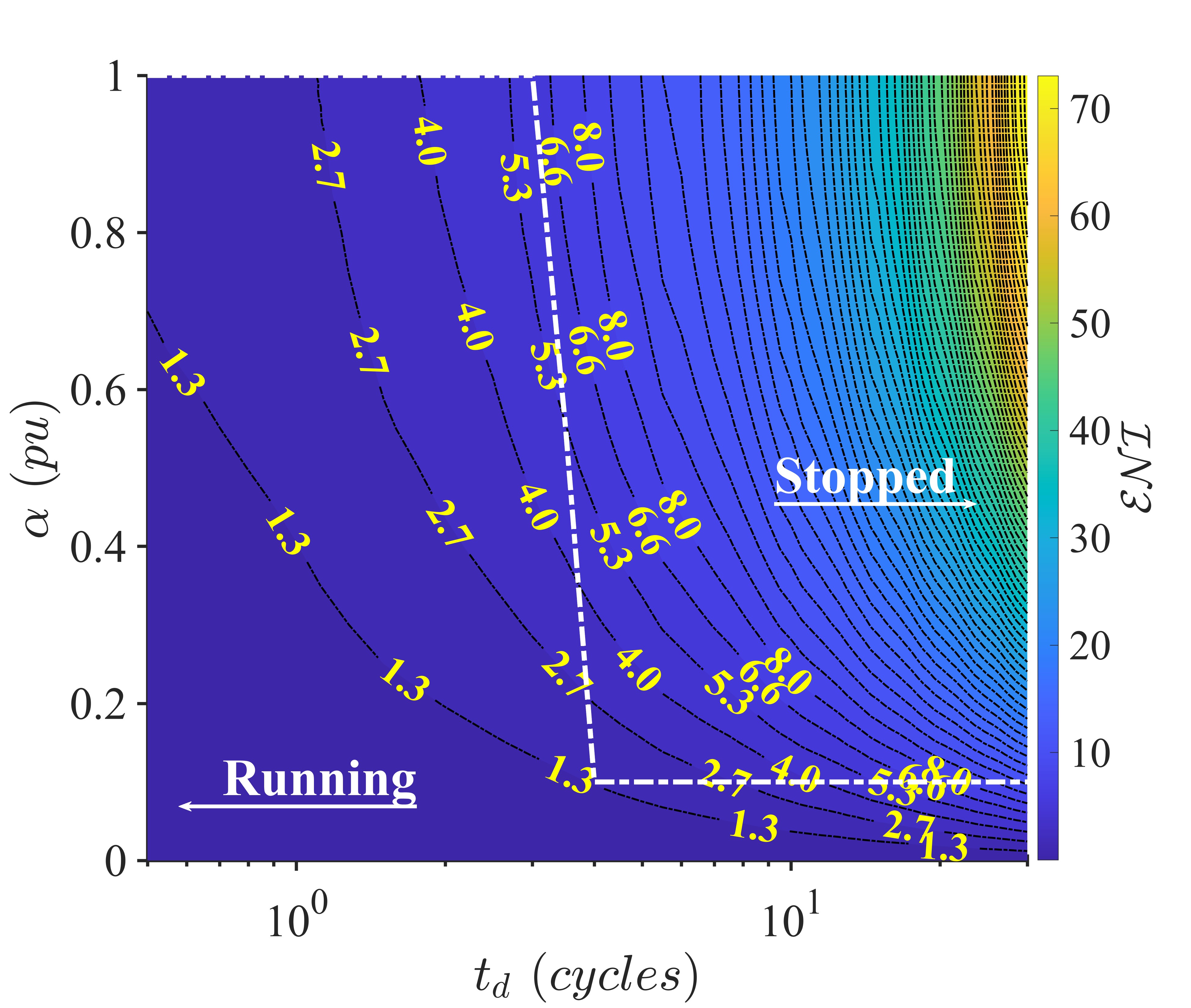}
   \caption{Voltage sag ride through of adjustable speed drives (dotted white line) and the corresponding variations in $\mathcal{ENI}$. $\alpha$ denotes the sag magnitude, \textit{i.e.}, $\alpha: 0\rightarrow 1$ indicates \emph{nominal voltage $\rightarrow$ interruption}. The values along \emph{contour} lines indicate the values of $\mathcal{ENI}$.}
  \label{f:asd}
% \end{figure}
\end{SCfigure}

In a distribution network, the issues related to PQ are usually approached in terms of consumer equipment sensitivity~\cite{Dugan:Book:2012,Herath:Gosbell:2005}. For instance, equipment susceptibility is often characterized in terms of voltage \emph{sag ride-through} capability, which provides safe operating ranges in terms of magnitude and duration for the given equipment. It is, therefore, pertinent to explore the possible applications of the proposed severity indices to assess such capabilities. In particular, we consider adjustable speed drives (ASD), which are widely used in industries but are typically susceptible to voltage sags. Such ASD failures are often associated with significant economic burdens as they may lead to unscheduled downtime and process loss. Typical ASDs can sustain all sag events with duration limited to $t_d\leq3 \ cycles$, with a linear decrease in \emph{tolerable} sag magnitude $\alpha: 1\rightarrow0.1$ as the duration increases, $t_d: 3 \rightarrow 4 \ cycles$, and finally, $\alpha \leq 0.1$ can be tolerated in a sustained operation, \textit{i.e.}, $t_d\geq4 \ cycles$~\cite{Dugan:Book:2012}. This sag ride-through behavior for different values of sag magnitude, $\alpha \ (pu)$, and duration, $t_d \ (cycles)$, is depicted by a white dotted line in Fig.~\ref{f:asd}, with `\emph{running}' and `\emph{stopped}' regions marked for convenience.

Further, the variations in $\mathcal{ENI}$ for the same values of $\{\alpha, t_d\}$ are also shown as \emph{contour} plots in Fig.~\ref{f:asd}. The minimum and maximum values of $\mathcal{ENI}$  are respectively determined to be $0$ (\emph{corresponding to $\{\alpha, t_d\} = \{0,0\}$}) and $73\%$ (\emph{corresponding to $\{\alpha, t_d\} = \{1 \ pu, \ 30 \ cycles\}$}). The \emph{contour} analysis of $\mathcal{ENI}$ delineates three regions corresponding to ASD sag susceptibilities, as seen in Fig.~\ref{f:asd} and outlined in the following: all sag events with severity $\mathcal{ENI}\leq1.3\%$ fall in the \emph{operating} region; whereas $\mathcal{ENI}>8\%$ indicate relatively \emph{severe} sags which lead to loss of ASD operations. The intermediate region, $1.3\%<\mathcal{ENI}\leq8\%$, requires further analysis as the values in this region can correspond either to \emph{tolerable} or \emph{severe} sag category depending on the duration $t_d$, see Fig.~\ref{f:asd}. The severity estimation capabilities of the proposed indices are, thus, envisaged as continuous monitoring tools at a utility PCC. A historical analysis of $\mathcal{ENI}$ values (\textit{e.g., histogram}) can reveal the severity distribution of prevailing sag events, which can aid further decision-making. If, for instance, such severity distribution is skewed towards $\mathcal{ENI}>1.3$, then subsequent mitigation steps may be required.
%--------------------------------------------
\vspace{-5mm}
%-----------------------------------------
%%%=======================================
\section{Conclusions}
\label{s:conclusion}

A new approach is proposed to estimate the severity of non-stationary power quality events. In particular, the idea of temporal-frequency energy distribution as an estimator of the severity is put forward. Three energy-norm indices are developed based on this notion to accommodate different classes of non-stationary events. Additionally, the issue of frequency sensitivity is approached as a \emph{preference}. It is shown that the index can easily be made sensitive to a frequency band of interest. Moreover, the behavior of these indices is analyzed theoretically and empirically by considering a large set of synthetic and real-life recorded events, which convincingly demonstrate that regardless of the nature of the non-stationary event, the proposed indices can be used to ascertain the severity of events. Additionally, a detailed sensitivity analysis highlights that the index retains the desired monotonic behavior regardless of the selected wavelet. Moreover, the ASD case study clearly illustrates how the proposed index can easily be linked with the equipment sensitivity through corresponding \emph{tolerance curves} and can clearly delineate different operating regions. Finally, from a utility perspective, the energy norm indices are not only valuable for determining acceptable disturbance thresholds but can also be used to identify the source for a particular disturbance at the PCC, which presents an intriguing direction for future research.

% \textcolor{blue}{}
\ifCLASSOPTIONcaptionsoff
  \newpage
\fi

\bibliographystyle{IEEEtran}

\end{document}